\documentclass[aps,prd,showpacs,nofootinbib,twocolumn]{revtex4}
\usepackage{latexsym}
\usepackage{amsmath,amsfonts}
\usepackage{amsbsy}
\usepackage{mathrsfs}
\usepackage{color}
\usepackage{float}
\usepackage{dsfont}
\usepackage{psfrag}
\usepackage{enumerate}
\usepackage{soul}
\usepackage{gensymb}
\newtheorem{theorem}{Theorem}

\usepackage{amsmath,amssymb,calc,amsfonts}
\usepackage{latexsym}
\usepackage{hyperref}

\usepackage{graphicx,calc,epsfig}

\begin{document}

\title{Finite distance corrections to the light deflection in a gravitational field with a plasma medium}

\date{\today}

 \author{Gabriel Crisnejo$^1$\footnote{gcrisnejo@famaf.unc.edu.ar}, Emanuel Gallo$^1 \, ^2$\footnote{egallo@unc.edu.ar} and Adam Rogers$^3$\footnote{rogersa@brandonu.ca} } 

 \affiliation{ 
 $^1$FaMAF, UNC; Ciudad Universitaria, (5000) C\'ordoba, Argentina. \\ 
 $^2$Instituto de F\'isica Enrique Gaviola (IFEG), CONICET, \\
 Ciudad Universitaria, (5000) C\'ordoba, Argentina. \\
 $^3$ Department of Physics and Astronomy, Brandon University, \\ 
 Brandon, MB, Canada, R7A 6A9.}

\begin{abstract}

The aim of the present work is twofold: 
first, we present general remarks about 
the application of recent procedures to compute the 
deflection angle in 
spherically symmetric and asymptotically 
flat spacetimes, taking into account 
finite distance corrections based on the 
Gauss-Bonnet theorem. Second, and as the 
main part of our work, we apply this 
powerful technique to compute corrections 
to the deflection angle produced by 
astrophysical configurations in the weak 
gravitational regime when a plasma medium 
is taken into account. For applications, 
we use these methods to introduce new 
general formulae for the bending angle of 
light rays in plasma environments in 
different astrophysical scenarios, 
generalizing previously known results.  We also present new and useful formulae for 
the separation angle between the images of two sources when they are lensed by an 
astrophysical object surrounded by plasma. In particular, for the case of an 
homogeneous plasma we study these 
corrections for the case of light rays 
propagating near astrophysical objects 
described in the weak gravitational regime by a Parametrized-Post-Newtonian (PPN) 
metric which takes into account the mass 
of the objects and a possible quadrupole 
moment. Even when our work concentrates on finite distances corrections to the 
deflection angle, we also obtain as 
particular cases of our expressions new 
formulae which are valid for the more 
common assumption of infinite distance 
between receiver, lens and source. We also consider the presence of an inhomogeneous 
plasma media introducing as particular 
cases of our general results explicit 
expressions for particular charge number 
density profiles. 
\end{abstract}


  \maketitle


%
\section{Introduction}\label{section1}
Gravitational lensing is a crucial tool to study the dynamics, evolution and distribution of matter in the Universe \cite{Hoekstra:2013via,2015IAUS..311...86M,Giocoli:2013tga,Lewis20061}. The response of electromagnetic radiation to gravitational fields occurs at all size scales in the Universe, ranging from the size of individual black holes \cite{AR-EHT} to clusters containing many individual galaxies \cite{AR-CLUSTER}. In fact, recently the first observational test of Einstein's general relativity confirmed the theory to high precision on extragalactic scales \cite{Collett:2018gpf}. At all scales, the study of the strong gravitational lensing regime gives us information about the response of electromagnetic radiation to gravitational fields and will be crucial in providing tests of gravitational theory under strong field conditions. 

Typically, gravitational lens effects are considered in the vacuum. However, many compact objects are surrounded by dense, plasma-rich magnetospheres \cite{AR-PSR1, AR-PSR2}, and even galaxies and galaxy clusters \cite{AR-HITOMI} are in general immersed in a plasma fluid. In the visible spectrum, the modification of gravitational lensing quantities due to the presence of the plasma is negligible. The same cannot be said of observations in the radio frequency spectrum where the index of refraction of the plasma causes strong frequency-dependent modifications of the usual gravitational lensing behavior.

The effect of plasma on light propagation has been studied since the 1960s. In 1966, Muhleman and Johnston studied the influence on the time delay by the solar corona electron plasma at radio frequencies in the gravitational field of the Sun \cite{Solar-Radio}. In 1970, Muhleman, Ekers and Fomalont calculated for the first time the light deflection in the presence of a plasma in the weak-field approximation \cite{Muhleman-1970}. A variety of studies focusing on the solar wind and the electron density profile in the outer corona were also performed, using analysis of  light propagation in different spatial missions such as Viking, Mariner 6 and 7 and the Cassini mission  \cite{1977ApJ...211..943M,Tyler-1977}. In navigating interplanetary spacecraft, the plasma contribution to light propagation is also routinely considered (see for example the review article of Turyshev and Toth \cite{Turyshev2010}.
A rigorous derivation of a Hamiltonian for light rays including a magnetized plasma and curved background was performed by Breuer and Ehlers in 1980 \cite{Breuer-1980, Breuer-1981a, Breuer-1981b}. The light deflection in a plasma was calculated for the first time in the Schwarzschild spacetime (and in the equatorial plane of the Kerr spacetime) without the weak-field approximation by Perlick in 2000 \cite{Perlick-book}.

 Presently there exist some radio-telescope projects which operate in frequency bands where such plasma effects must be taken into account \cite{1991CuSc...60..106S,2013AA...556A...2V,2009IEEEP..97.1497L,Budianu-2015,Bentum:2016ekl}. For that reason, in the last years the study of the influence of plasma media on the trajectory of light rays in a external gravitational fields associated with compact bodies has continued and become a very active research area \cite{BisnovatyiKogan:2008yg,
BisnovatyiKogan:2010ar,
Tsupko:2013cqa,Tsupko:2014sca,
Tsupko:2014lta,Perlick:2015vta,
Bisnovatyi-Kogan:2015dxa,
Perlick:2017fio,
Bisnovatyi-Kogan:2017kii, 
2013ApSS.346..513M,
2016ApSS.361..226A, 
2017IJMPD..2650051A,
2017IJMPD..2641011A,
2017PhRvD..96h4017A,
2018arXiv180203293T,
Rogers:2015dla,
Rogers:2016xcc,
Rogers:2017ofq,
2018MNRAS.475..867E, 
Crisnejo:2018uyn,Er:2013efa}. 
    
One of the main quantities in the study of gravitational lensing is the deflection angle. In general, expressions for the deflection angle are written in terms of derivatives of the various metric components. However, in \cite{PhysRevD.83.083007}, we introduce an expression for the deflection angle in the weak lensing regime which is written in terms of the curvature scalars. It was generalized to the cosmological context by Boero and Moreschi \cite{Boero:2016nrd} and recently by us to take into account second order corrections in perturbations of a flat metric \cite{PhysRevD.97.084010}. 
On the other hand, Gibbons and Werner have also established a new geometrical (and topological) way of studying gravitational lensing using the Gauss-Bonnet theorem and an associated optical metric \cite{Gibbons:2008rj}. It is worthwhile to mention that the concept of the optical metric and the related Fermat principle for light rays in general relativity was introduced first by Weyl in 1917 \cite{Weyl-1917}. More precisely, Gibbons and Werner obtained an elegant relation between the deflection angle, the Gaussian curvature of the associated optical metric and the topology of the manifold. In this approach the deflection angle can be obtained by integrating the Gaussian curvature of the metric in an appropriate two-dimensional integration region $D$.

Since the seminal work of Gibbons and Werner \cite{Gibbons:2008rj}, many applications of this method for purely gravitational lensing in astrophysical situations have emerged. In particular, this new technique is being used to compute gravitational lensing quantities in a variety of spacetimes including vacuum, electro-vacuum, and with a vast array of scalar fields or effective fluids at both finite and infinite distances \cite{Jusufi:2015laa,Jusufi:2016wiz,Jusufi:2016sym,Jusufi:2017gyu,Jusufi:2017hed,Jusufi:2017vta,Jusufi:2017drg,Jusufi:2017xnr,Jusufi:2018kmk,Jusufi:2018waj,Sakalli:2018nug,Arakida:2017hrm,Ishihara:2016vdc,Ishihara:2016sfv,Ovgun:2018xys,Ovgun:2018ran,Jusufi:2018kry,Ovgun:2018prw,Ovgun:2018oxk,Ovgun:2018fte,Ono:2018ybw,Werner:2012rc,Jusufi:2017lsl,Jusufi:2017vew,Jusufi:2017mav,Jusufi:2017uhh,Jusufi:2018jof,Ono:2017pie}.
 More recently, in \cite{Crisnejo:2018uyn}, we have shown for the first time how the Gibbons-Werner method can also be applied to the study of light rays simultaneously interacting with gravitational fields and a plasma medium. It is worth noting that in this case the light rays do not propagate along null geodesics of the underlying physical spacetime. Despite this apparent difficulty we have shown how the Gauss-Bonnet method, originally designed to study null geodesics in pure gravitational fields, can also be applied through judicious choice of optical metric to the study of timelike curves followed by light rays in a plasma environment. In fact, these results also apply to timelike geodesics followed by massive particles in pure gravitational fields. Thus, our results highlight the elegance and power of the Gibbons-Werner method by demonstrating the beautiful relationship that exists between the deflection angle, geometric and topological quantities associated with spacetime, and the implications of these relationships for both optics and mechanics.

Due to the deep connections between geometry and topology exposed by the Gibbons-Werner method, several authors have proposed alternative extensions to the Gauss-Bonnet theorem in situations where the source or the observer can not be considered to be at infinite distance to the lens. The first alternative is presented in references \cite{Ishihara:2016vdc,Ishihara:2016sfv,Ono:2017pie,Ono:2018ybw} and the second in \cite{Arakida:2017hrm}. 
Some remarks are in order with respect to these alternative formulations. First, even when these proposals are based in the Gauss-Bonnet theorem, they do not agree in their predictions. In particular for asymptotically flat spacetimes, the proposal given in \cite{Arakida:2017hrm} is different and not generally compatible with the proposal given and used in the others \cite{Ishihara:2016vdc,Ishihara:2016sfv,Ono:2017pie,Ono:2018ybw}.
It can be easy checked by comparison of the expressions for the deflection angle in a Schwarzschild spacetime that these authors obtained using their respective definitions. More precisely, even when different authors use the same coordinate system (the usual Schwarzschild coordinates), in \cite{Arakida:2017hrm} Arakida obtained an expression for the deflection angle (at linear order in the mass) with some extra terms that are missing in the Ishihara \emph{et.al.} definition \cite{Ishihara:2016vdc} (see eq. (54) of reference \cite{Arakida:2017hrm} and the particular case of eq. (A.3) with $a=0$ of reference \cite{Ono:2017pie}).
\footnote{Note that eq. (A.3) of reference \cite{Ono:2017pie} refers to the deflection angle computed for a Kerr metric as expressed in Boyer-Lindquist coordinates. However, by taking $a=0$ this metric agrees with the Schwarzschild metric in the standard Schwarzschild coordinates and in that situation the comparison is explicit. At 
linear order in the mass these 
extra terms are given by 
\begin{equation}
\delta\alpha=-\frac{m}{b}(\sin^2(\hat\varphi_R)\cos(\hat\varphi_R)
-\sin^2(\hat\varphi_S)\cos(\hat\varphi_S))
\end{equation}
where $\hat\varphi_S$, and 
$\hat\varphi_R$ represent the angular 
coordinate of the position of the 
source and the receiver, 
respectively, and $\delta\alpha$ 
is the difference between the 
expressions given by Arakida 
and Ishihara \emph{et.al.} More details, in Sec.\eqref{sec:remark} } 

Before continuing, it is 
important to note the following caveat. The two separate groups of authors mentioned above have both considered the
interesting effect of the 
cosmological constant on the bending 
angle. When a cosmological constant is included, these groups also obtain different expressions for the 
deflection angle. An interesting question is whether this difference introduces incompatible
predictions for the related observable 
quantities. However, we do 
not address this case in the present work, and restrict all discussion to 
asymptotically flat spacetimes. Therefore, it remains an open question if the two alternative definitions give incompatible results for that situation.

Returning to the discussion for asymptotically flat spacetimes, since these two alternative 
definitions use two different integration 
regions $D$ and $D'$ for the integration of the 
Gaussian curvature, it is difficult to see the 
reason for that difference in the original 
presentations. In particular, even when both of 
these authors use quadrilateral regions, in the 
case of Arakida it is a finite region and in the 
case of Ishihara \emph{et.al.} it is an 
unbounded one. We will show below how the difference 
in the results of these expressions for asymptotically flat spacetimes can be 
easily understood by presenting 
an alternative integration region in the 
Ishihara \emph{et.al.} definition.
Second, even when finite-distance corrections to 
the deflection angle are derived by these two 
alternative definitions, the authors of 
\cite{Arakida:2017hrm,Ishihara:2016vdc,Ishihara:2016sfv,Ono:2017pie,Ono:2018ybw} or 
\cite{Arakida:2017hrm} do not attempt to make a 
comparison with known (and compatible) expressions
from the literature that were obtained using 
different techniques 
\cite{1970ApJ...162..345W,Shapiro67,Richter:1982zz,DeLuca:2003zz,Misner1973,Will:1993ns,He:2016vxc,deFelice:2006xm,deFelice:2004nf}. These expressions have been tested in observations reaching back several decades ago\cite{PhysRevLett.75.1439,Shapiro-2004,Fomalont-1975,Robertson91, Lambert-2009,2017arXiv170206647T,1997froeschle,2010hobbs}. Moreover, they
are also routinely used and needed for high precision relativistic astrometry 
\cite{deFelice:1998wk,deFelice:2001wr,Crosta:2003zh,kaplan115,Turyshev:2008tm}. Due to the existence of these two 
incompatible definitions (as previously 
mentioned, they do not agree even at linear 
order in a Schwarschild spacetime), the 
comparison between their predictions and the 
known quantities can be used as a good test of 
their validity. We will carry out this 
comparison and we will shown that the Ishihara 
\emph{et.al.} definition is in complete 
agreement with known expressions. 

In addition to dealing with the technical issues 
around the calculation, our main motivation for 
the present work is to study how the 
consideration of finite distances between the 
source, lens and observer can affect the 
expression for the deflection angle in 
astrophysical situations where a plasma medium 
is present. The usual way to study lensing due to plasma is through 
the Hamiltonian equations for the timelike 
curves followed by light rays in the plasma 
environment. On the other hand, recently we 
presented a geometrical formulation of the 
problem using the Gibbons-Werner method \cite{Crisnejo:2018uyn}. 
Therefore it is natural to use this powerful 
technique to study the corrections in the known 
expressions for the deflection angle in these 
situations. 

Motivated by these issues, we propose a number 
of points to contribute to the discussion of 
this subject: First, we present an alternative 
formulation of the definition given in 
\cite{Ishihara:2016vdc} for the bending angle at 
finite distances {for static and spherically symmetric asymptotically flat spacetimes}\footnote{{As we will see, the definition can also be applied to the equatorial plane of more general static metrics with SO(2) symmetry.}}. We remark that it is not a new 
definition, but an equivalent formulation. Our 
approach is based on a finite region and allows 
us to compare with the expression given by
Arakida in \cite{Arakida:2017hrm} in that situation. 
Furthermore, this work fills the existent gap in the 
comparison with known expressions for the 
bending angle at finite distance and the results 
obtained using the definition given in 
\cite{Ishihara:2016vdc}. This comparison 
provides confidence in the veracity of the 
region definitions in that work. Finally, as the 
focus of our work, we apply this powerful 
technique to compute corrections to the 
deflection angle produced by astrophysical 
configurations in the weak gravitational field 
regime when a plasma medium is taken into 
account. In particular, for the case of a 
homogeneous plasma we study finite distance 
corrections for the case of light rays 
propagating through astrophysical objects 
described in the weak gravitational region by a 
Parametrized-Post-Newtonian (PPN) expansion 
which takes into account the mass of the objects 
and a possible quadrupole moment. Even when our 
work concentrates in finite distance corrections 
to the deflection angle, we also obtain as 
particular cases of our expressions new formulas 
which are valid for the more common assumption 
of infinite distance between receiver, lens and 
source generalizes previous 
ones.  

This work is organized as follows. In 
Sec.\eqref{sec:remark} we review the definition 
of bending angle given by Ishihara \emph{et.al.} 
in \cite{Ishihara:2016vdc} and we propose 
an alternative presentation by 
using a finite quadrilateral region
which allows us to make a comparison and remark on the difference with the Arakida definition\cite{Arakida:2017hrm}. We also present a review of known finite distance expressions for the bending angle in order to prepare for later comparison with what is obtained by the use of the Gauss-Bonnet method. In Sec.\eqref{section3} we review the theory of light rays in cold non-magnetized plasma and the associated optical metric. We also present a simple formula which relates the separation angle between the images of two sources when they are lensed by a gravitational field surrounded by a plasma medium. In Sec.\eqref{section4} we study finite distance corrections to the deflection angle in astrophysical situation where the gravitational field can be represented by a PPN metric and where a homogeneous plasma medium is present. We also carry out detailed comparisons between known expressions for the bending angle and results obtained using the Ishihara \emph{et.al} definition. As a by-product, we obtain several new formulae for the bending angle which generalizes previous known results in several ways. Finally in section \eqref{section5} we briefly discuss the situation where the plasma is non-homogeneous presenting the study of a Schwarzschild spacetime surrounded by some particular cases of inhomogeneous plasma media.  In particular, we study the relevance of finite distance corrections in a model for the plasma density of the solar corona. We conclude with final remarks. For completeness, in Appendix \ref{Ap} we show how three different versions of the deflection angle calculation give the same result using the finite quadrilateral region (as defined in Sec.\eqref{sec:remark} ). In appendix \ref{Bnueva} we study the relationship between different angular coordinates and the elongation angle. In Appendix \ref{Bp} we give explicit finite distance expressions for the contribution to the deflection angle in a Schwarzschild spacetime produced by a model for the electronic charge distribution of the extended solar corona.

\section{Finite distance corrections to the deflection angle using the Gauss-Bonnet theorem}\label{sec:remark} 
\subsubsection{General remarks} 

{The Gauss-Bonnet theorem provides a powerful framework to describe finite distance corrections to the gravitational lens deflection angle. A thorough discussion of this topic first requires some general discussion of definitions recently used in the literature \cite{Ishihara:2016vdc,Ishihara:2016sfv,Ono:2017pie,Arakida:2017hrm}.}

Let us {recall the application of the Gauss-Bonnet theorem to a two-dimensional riemannian manifold.} Let $D\subset S$ be a regular domain of an oriented 2-dimensional surface S with Riemannian metric $\hat{g}_{ij}$, whose boundary is formed by a closed, simple, piece-wise, regular and positive oriented curve $\partial D : \mathbb{R} \supset I \to D$. Then{,} the Gauss-Bonnet theorem states
\begin{equation}
\int\int_{D}\mathcal{K} dS + \int_{\partial D} \kappa_g\;d\sigma+ \sum_{i} \Theta_{i}=2\pi\chi(D), \ \ \sigma\in I;
\end{equation}
where $\chi(D)$ and $\mathcal{K}$ are the Euler characteristic and Gaussian curvature of $D$, respectively; $\kappa_g$ is the geodesic curvature of $\partial D$ and $\Theta_{i}$ is the exterior angle defined in the $i^{\text{th}}$ vertex, in the positive sense. {Given} a smooth curve $\gamma$ with tangent vector $\dot{\gamma}$ such that
\begin{equation}
\hat{g}(\dot{\gamma},\dot{\gamma})=1,
\end{equation}
and acceleration vector $\ddot{\gamma}$, the geodesic curvature $\kappa_{g}$ of $\gamma$ can be computed as
\begin{equation}
\kappa_g=\hat{g}(\nabla_{\dot{\gamma}}\dot{\gamma},\ddot{\gamma}),
\end{equation}
which is equal to zero if and only if $\gamma$ is {a} geodesic, because $\dot{\gamma}$ and $\ddot{\gamma}$ are orthogonal.

Let us consider a static spherically symmetric and asymptotically flat spacetime,
\begin{equation}\label{eq:spherimetr}
ds^2=-A(r)dt^2+B(r)dr^2+C(r)(d\theta^2+\sin^2(\vartheta)d\varphi^2),
\end{equation}\footnote{As we have mentioned, the restriction to spherical symmetry is not at all necessary: the following discussion can also by applied to the equatorial plane of static spacetimes with $SO(2)$ symmetry.}
and a light ray propagating from a source S to a receiver $R$ on a null geodesic{, which can be taken as lying in the plane defined by $\theta=\pi/2$ without a loss of generality.} This null geodesic can be put in {one-to-one correspondence} with {a} spatial geodesic of the associated optical metric given by  \cite{book:75670,Gibbons:2008rj,Weyl-1917}
\begin{equation}
d\sigma^2=\frac{B(r)}{A(r)}dr^2+\frac{C(r)}{A(r)}d\varphi^2.
\end{equation}
{Ishihara \emph{et.al.} \cite{Ishihara:2016vdc} proposed a new definition for the deflection angle at finite distance using the Gauss-Bonnet theorem,} which can be written as
\begin{equation}\label{eq:asad}
\alpha=-\int\int_{{}^\infty_R\square^\infty_S}\mathcal{K}dS.
\end{equation}
In order to define the integration region 
${}^\infty_R\square^\infty_S$ one start{s} with 
a region $D_{r}${, bounded} by the geodesic 
$\gamma_\ell$ {with its} origin at a point $S$ 
and end at $R$. {Let us consider} two radial 
geodesics {$\gamma_S$ and $\gamma_R$}, defined 
by respective constants $\varphi_S$ and 
$\varphi_R${, pass through the points $S$ and 
$R$ respectively. Then, let a} circular arc 
segment defined by $r=r_C=\text{constant}$ 
{close the region. The arc segment is} chosen 
{to be} orthogonal to the radial geodesics 
$\gamma_R$ and $\gamma_S$.  {The region} 
${}^\infty_R\square^\infty_S$ is {then} obtained 
as the limit of the region $D_{r}$ as $r_C$ goes 
to infinity. For a motivation 
of the choice of this region see the original 
references 
\cite{Ishihara:2016vdc,Ishihara:2016sfv,Ono:2017pie,Ono:2018ybw}. Since we are interested in the comparison of this formula with the Arakida proposal which is based in a different quadrilateral (and finite) region\cite{Arakida:2017hrm}, we will give an alternative presentation of \eqref{eq:asad} which also makes use of a finite quadrilateral region. Of course, when we talk about the bending angle, we are referring to how the path of light rays are curved with respect to a flat spacetime. Therefore, it is natural that we relate the behavior of null geodesics in the two spacetimes. 

\begin{figure}[H]
\centering
\includegraphics[clip,width=75mm]{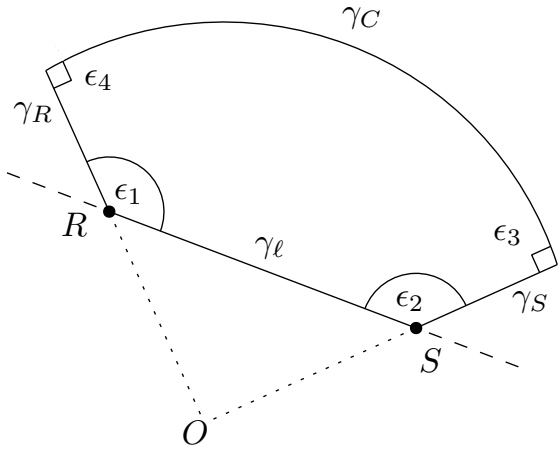}
\caption{The region $D_r$ in {a E}uclidean two-dimensional space as described in the text. It is bounded by 4 curves: a straight line geodesic connecting the points $R$ and $S$, two radial geodesics $\gamma_R$ and $\gamma_S$ and a circular curve $\gamma_C$ which intersects $\gamma_R$ and $\gamma_S$ {orthogonally}.}
 \label{grafico-flat}
\end{figure}

Consider a two dimensional space with a Euclidean metric written in a standard polar coordinate system $\{r,\varphi\}$. In this space let $D_r$ be a region with boundaries formed by two straight line segments defined by $ \varphi=\varphi_S=\text{constant}$ and $ \varphi=\varphi_R=\text{constant} $ and such that their ends farthest from the origin are connected by a circular arc segment $\gamma_C$ defined by $r=r_C=\text{constant}$ and the two ends nearest the origin connected by a straight line segment $\gamma_\ell$ (see Fig.\eqref{grafico-flat}. For all the following discussion, the azimuth angular coordinate $\varphi$ is measured from a given polar axis. But the choice of this axis is arbitrary for the moment, and this is the reason that we have not plotted any axis or azimuth angular coordinate in Fig.\eqref{grafico-flat}. We have only plotted geometrical quantities adapted to the rotational azimuth Killing symmetry of the metric. Later we will introduce  particular azimuth angular coordinates (See also Appendix \eqref{Bnueva}).
If we apply the Gauss-Bonnet theorem in this region, we obtain the following relation for the sum of the interior angles $\epsilon_i$ of the region $D_r$ (which are related to the exterior angles $\Theta_i$ by $\Theta_i=\pi-\epsilon_i$):
\begin{equation}
\sum_{i}\epsilon_{i}=\int_{\gamma_C} \kappa\;d\sigma +2\pi, \ \ \sigma\in I.\label{S}
\end{equation}
Of course, 
\begin{equation}\label{eq:kappaeuc}
\int_{\gamma_C} \kappa\;d\sigma=\varphi_R-\varphi_S,
\end{equation}
but it will be not relevant for us.

In a similar way, let us consider a Riemannian two-dimensional space defined in a region $\mathcal{R}^2/B$ with $B$ a compact set, such that it allows a SO(2) symmetry group and it is also asymptotically Euclidean. The metric associated with this Riemannian manifold can be written as
$d{\tilde\sigma}^2=a(\tilde{r})d\tilde{r}^2+r'^2b(\tilde{r})d\tilde{\varphi}^2$, with $a(\tilde{r})$ and $b(\tilde{r})$ going to 1 as $\tilde{r}$ goes to infinity. As this metric is asymptotically Euclidean and therefore tends to the Euclidean metric $d\tilde{r}^2+\tilde{r}^2d\tilde\varphi^2$ as $r$ goes to infinity, we can make an identification between the coordinates 
$\{r,\varphi\}$ used in the polar coordinates system of the Euclidean space where the region $D_r$ was defined and the new coordinate $\tilde{r},\tilde\varphi$ of the Riemannian manifold.
\begin{figure}[H]
\centering
\includegraphics[clip,width=75mm]{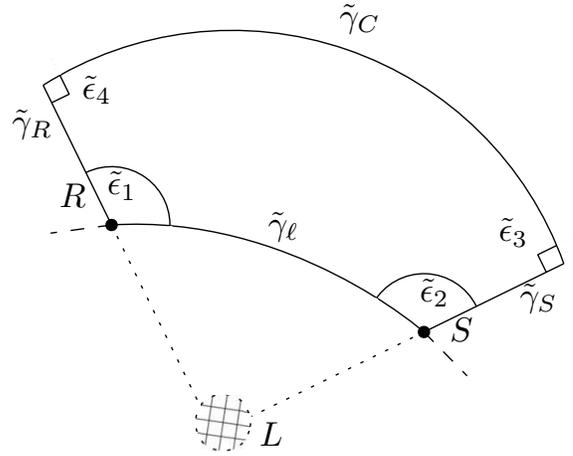}
\caption{The region $\tilde{D}_r$ in {a 
R}iemannian two-dimensional space as described 
in the text. It is bounded by 4 curves: a 
spatial geodesic $\tilde{\gamma}_\ell$ 
connecting the points $R$ and $S$, and three 
curves $\tilde\gamma_R${,} $\tilde\gamma_S$ and 
a circular curve $\tilde\gamma_C$ identified 
with the respective curves in the  Euclidean 
space. By construction the curve 
$\tilde{\gamma}_C$ also intersects 
$\tilde\gamma_R$ and $\tilde\gamma_S$ 
orthogonally. The circular area plotted 
with reticulated lines in the interior 
represents the region where an astrophysical 
object that acts as a lens is contained. This 
region is not necessary covered by the polar 
coordinate system described in the text.}
\label{grafico-no-flat}
\end{figure}

Now let $\tilde{D}_r$  be a slightly modified 
region in this Riemannian manifold chosen such 
that three of its sides are defined in a similar 
way as were  $\gamma_R$, $\gamma_S$ and 
$\gamma_C$ and with the remaining boundary 
chosen as the geodesic $\tilde\gamma_\ell$ which 
coincides with the spatial geodesic associated 
with the spatial orbit of the null geodesic 
followed by a light ray connecting $S$ with $R$ 
in the physical curved spacetime. See 
Fig.\eqref{grafico-no-flat}. Therefore, for this 
region we obtain
\begin{equation}
\sum_{i}\tilde\epsilon_{i}=\int\int_{\tilde{D}_r}\mathcal{K}dS+\int_{\tilde\gamma_C} \tilde\kappa\;d\tilde\sigma +2\pi, \ \ \tilde\sigma\in I.\label{S'}
\end{equation}
Note that by construction the following crucial 
property is satisfied 
$\epsilon_3=\tilde\epsilon_3=\epsilon_4=\tilde\epsilon_4=\pi/2$, and therefore the difference between the sum of inner angles for the regions $D_r$ and $\tilde{D}_r$ is only related to the difference in the angles that the geodesic $\gamma_\ell$ and $\tilde{\gamma}_\ell$ make with the radial geodesics $\gamma_R$ and $\gamma_S$. 
Motivated by the last fact, we propose the following expression as the definition of the deflection angle $\alpha$: 
\begin{equation}\label{def1}
\alpha=\sum_{i}(\epsilon_{i}-\tilde\epsilon_{i}).
\end{equation}
Therefore, taking into account the equations \eqref{S} and \eqref{S'} we obtain the alternative expression:
\begin{equation}
\alpha=-\int\int_{\tilde{D}_r}\mathcal{K}dS-\int_{\tilde{\gamma}_{C (S\to R)}}  \tilde\kappa\;d\tilde\sigma+\int_{\gamma_{C(S\to R)}} \kappa\;d\sigma.\label{eqmain}
\end{equation}     
Where the notation $\tilde{\gamma}_{C (S\to R)}$ is to recall that the integration must be done on the circular arc segment $\gamma_C$ in the direction from $S$ to $R$.   
Alternatively, as the other three curves in the quadrilateral region are geodesics,  Eq.\eqref{eqmain}
can be written as
\begin{equation}
\alpha=-\int\int_{\tilde{D}_r}\mathcal{K}dS-\oint_{\partial\tilde{D}_r}   \tilde\kappa\;d\tilde\sigma+\oint_{\partial D_r} \kappa\;d\sigma,\label{eqmain2}
\end{equation} 
with the line integrals made on the respective boundaries $\partial\tilde{D}_r$ an $\partial\tilde{D}_r$ of the regions $\tilde{D}_r$ and $D_r$ in a counter-clockwise direction.
By construction the right hand side of Eq.\eqref{eqmain}  gives the same result for \emph{any} curve $\gamma_C$ defined by $r_C=\text{constant}$. This definition is an alternative presentation of the proposed definition of Ishihara \emph{et.al.}\cite{Ishihara:2016vdc}. In particular, as the metric is assumed to be asymptotically Euclidean, we can take the limit of $r_C$ going to infinity, in which case $\int_{\tilde\gamma_C} \tilde\kappa\;d\tilde\sigma\to\int_{\gamma_C} \kappa\;d\sigma$, resulting in an expression for the angle $\alpha$ which reduces to the formula \eqref{eq:asad} as given by Ishihara \emph{et.al.}\cite{Ishihara:2016vdc}. 

In fact, we can repeat the same procedure but without the assumption that the curve $\tilde{\gamma}_\ell$ is geodesic. In this case, even when the region $D_r$ remains unchanged, we obtain a new region $\tilde{D}^*_r$ and Eq.\eqref{eqmain} is modified to:
\begin{equation}
\begin{aligned}
\alpha=&-\int\int_{\tilde{D}^*_r}\mathcal{K}dS-\int_{\tilde{\gamma}_{\ell (R\to S)}} \tilde\kappa\;d\tilde\sigma -\int_{\tilde{\gamma}_{C  (S\to R)}} \tilde\kappa\;d\tilde\sigma\\
&+\int_{\gamma_{C  (S\to R)}} \kappa\;d\sigma\\
=&-\int\int_{\tilde{D}^*_r}\mathcal{K}dS-\oint_{\partial\tilde{D}^*_r}   \tilde\kappa\;d\tilde\sigma+\oint_{\partial D_r} \kappa\;d\sigma.
\label{eqmain3}
\end{aligned}
\end{equation}
If we assume a region $\tilde{D}\equiv 
{}^\infty_R\square^\infty_S$ obtained from 
$\tilde{D}^*_r$ in the limit of $r_C$ going to 
infinity, it is easy to see that the relation 
\eqref{eqmain3} reduces formally to the 
expression found in Ref.\cite{Ono:2017pie} for 
the deflection angle at finite distances valid 
for a general stationary and axially-symmetric 
spacetime (Note that in such cases, as explained 
in detail in \cite{Ono:2017pie} a modification 
of the form of the optical metric is needed).

With the expression \eqref{eqmain}, we are ready to compare with the Arakida definition \cite{Arakida:2017hrm}. In that reference the author also takes a finite quadrilateral region but instead of using the circular curve $\gamma_C$, a new curve $\gamma_\Gamma$ is chosen which is identified with the spatial geodesic associated to a light ray connecting $R$ and $S$ if the spacetime were flat, that is, in the Euclidean space it is in fact a straight line. 

Keeping the definition \eqref{def1} for the deflection angle with these new regions, and noting that for a quadrilateral trapezoid in Euclidean space, the sum of interior angles always is equal to $2\pi$, we obtain a new  deflection angle, 
\begin{equation}
\tilde\alpha=2\pi-\sum_i\tilde\epsilon_i;\label{def:Arak}
\end{equation}
which exactly agrees with the definition of Arakida \cite{Arakida:2017hrm} (in that reference the interior angles are denoted $\beta_i$.) Equivalently, for this new choice 
of regions, 
the integration around the 
curve $\gamma_\Gamma$ (which replaces the curve $\gamma_C$) in the
last term in \eqref{eqmain} is exactly zero, because it is computed
in the Euclidean background spacetime and $\gamma_\Gamma$ is a geodesic of the Euclidean space by construction, and therefore only the first
two terms in \eqref{eqmain} survive, and we arrive at an expression 
with exactly the same form as found in \cite{Arakida:2017hrm} (Equation (35) of that reference). 
Therefore, it should appear at first sight that 
the definition \eqref{def1} 
also contains as particular case the definition for the deflection 
angle given by Arakida. However, note that in the motivation for 
\eqref{def1} the equality between the interior angles $\epsilon_3$ 
and $\tilde{\epsilon}_3$ and between  $\epsilon_4$ and 
$\tilde{\epsilon}_4$ was crucial. 
Note, that we could have 
written the expression for the deflection angle as the difference between the sum of $\epsilon_1$ and $\epsilon_2$ and their tilde 
version, emphasizing in this way that it only depends on the angles 
that the null geodesic connecting $S$ with $R$ makes with the radial curves in 
the curved space as compared  to the similar angles defined in the 
background. More precisely, we could also have written the deflection angle without any reference to closed regions as 
\begin{equation}
\alpha=(\epsilon_1-\tilde\epsilon_1)+(\epsilon_2-\tilde\epsilon_2),
\end{equation}
 For the use of the Gauss-Bonnet theorem we need a closed region. Since the deflection angle only depends on the difference of angles formed by the actual null geodesic and the radial geodesics in both spaces, the new curve used to close the region must be chosen such that the angles between the new curve and the radial directions are the same in both curved and flat space. This can only be the case if the new curve is chosen as the circular arc segment as $\gamma_C$ whose tangent vectors are the rotational Killing vectors $\frac{\partial}{\partial \varphi}$, and therefore are always orthogonal to the radial geodesics in both spaces. However, this will not be the case if instead of the curve $\gamma_C$ we choose as the new curve $\gamma_\Gamma$ (as used by Arakida). In such a situation, the curve $\gamma_\Gamma$ forms different inner angles with the radial directions in both spaces, and therefore the new deflection angle as given by \eqref{def:Arak} has not only information concerning to the bending of the light ray which connects the source with the receiver but also of the newly introduced curve $\gamma_\Gamma$. Therefore, the use of the equation \eqref{def:Arak} seems not to be well motivated. In fact, as we mentioned in the introduction, 
comparison of the expressions found in \cite{Ishihara:2016vdc} and 
\cite{Arakida:2017hrm} for the deflection angle at finite distances 
in a Schwarzschild background do not agree even at first order in the 
mass. {For this example, it is easy to check  that 
the origin of the difference between the Ishihara \emph{et.al.} 
expression for the deflection angle and the Arakida expression does indeed originate in the difference between the values of the inner angles that the curve $\gamma_\Gamma$ makes with the radial curves in both the Euclidean and the curved spaces. More precisely, as follows
from eq.(44) of \cite{Arakida:2017hrm} at linear order in the mass the difference 
between $\epsilon_3$ and $\tilde\epsilon_3$} 
\footnote{In the Arakida notation our inner angle $\epsilon_3$ is 
denoted as $\beta_2$, and in particular $\beta_2=E$ in his 
notation. See Eq.(44) of \cite{Arakida:2017hrm}.} 
{is 
\begin{equation}\label{eq:eps3epstilde3}
\epsilon_3-\tilde\epsilon_3=\frac{m}{b}\sin^2(\hat\varphi_S)\cos(\hat\varphi_S).
\end{equation}
The hat symbol above the angular coordinate ($\hat\varphi$) is to differentiate it from other azimuth 
angular coordinate associated to a 
different polar axis that we will choose 
later. They are chosen such that the 
point of closest approach  of the light 
ray to the to the central lens is at an azimuth angle of $\hat\varphi =\pi /2$. It was the choice used by Ishihara \emph{et.al}. and Arakida in their respective works. 
A similar expression follows for the difference of the angles 
$\epsilon_4$ and the tilde version.
This difference will contribute to the deflection angle (even at linear order in $m$). Hence, from the two definitions, we arrive at a difference between the formulas for the deflection. In particular, using the definition of \cite{Ishihara:2016vdc}, the deflection angle at linear order in the mass, denoted as $\alpha_{\text{\cite{Ishihara:2016vdc}}}$ reads (See Eq.(A.3) of Ref.\cite{Ono:2017pie} in terms of the angular coordinates with $a=0$, or Eq.(37) of Ref.\cite{Ishihara:2016vdc} with $\Lambda=0$):
\begin{equation}\label{eq:isexp}
\alpha_{\text{\cite{Ishihara:2016vdc}}}=\frac{2m}{b}\bigg(\cos(\hat{\varphi}_S)-\cos(\hat{\varphi}_R)\bigg).
\end{equation}
In comparison, the expression $\alpha_{\text{Arakida}}$ given by Arakida(Eq.(54) of \cite{Arakida:2017hrm}) reads:
\begin{equation}\label{eq:arakexp}
\alpha_{\text{Arakida}}=\alpha_{\text{\cite{Ishihara:2016vdc}}}+\delta\alpha;
\end{equation}
with
\begin{equation}\label{eq:deltaalphaarak}
\delta\alpha=-\frac{m}{b}\bigg(\sin^2(\hat{\varphi}_R)\cos(\hat{\varphi}_R)-\sin^2(\hat{\varphi}_S)\cos(\hat{\varphi}_S)\bigg).
\end{equation}
As anticipated, these extra terms are originated by the relations like \eqref{eq:eps3epstilde3} and a similar formula for $\epsilon_4-\tilde\epsilon_4$.

With respect to the expressions \eqref{eq:isexp} and \eqref{eq:arakexp} some remarks are in order. First, let us note that these differences are more than relevant 
with respect to the actual observability of the 
finite distance corrections. Let us consider, for example, the deflection produced by our Sun when the 
light rays of a far away source graze the surface and reach us on Earth. For such 
situation we can make the following approximations: 
$\hat\varphi_S=0+\mathcal{O}(m)$\footnote{
More precisely as the 
source goes to infinity, 
$\hat\varphi_S\to-\alpha_\infty/2$ where 
$\alpha_\infty$ should be the total 
deflection angle if both the source and 
the receiver were at a great distance from the lens. See Appendix \eqref{Bnueva} for 
more details between different azimuthal 
angular coordinate systems.}, 
$\hat\varphi_R=\pi-\delta\hat\varphi$, 
with $\delta\hat\varphi=\arcsin(b/r_o)\approx b/r_o\approx 4\times 10^{-3}$, 
where $r_o$ is the distance from the Sun 
to the 
Earth, and $b$ equal to the radius of the 
Sun.  
Then, as proved by Ishihara \emph{et.al.}, the 
difference between the infinite distance 
expression 
and $\alpha_{\text{\cite{Ishihara:2016vdc}
}}$ is of 
the order of $10^{-5}\text{arcsec}$. More precisely, by 
doing a Taylor expansion of \eqref{eq:isexp} we 
obtain 
\begin{equation}\label{tayara1}
\alpha_{\text{\cite{Ishihara:2016vdc}}}\approx 
\frac{4m}{b}-\frac{m\delta\varphi^2}{b}-
\frac{m\delta\varphi^4}{4b}+\mathcal{O}
(\delta\varphi^5).
\end{equation}
Hence, the first correction to the infinite 
distance expression is approximately given by 
$\frac{m\delta\varphi^2}{b}\approx10^{-5}\text{arcsec}$, 
which is within the capabilities of actual observations.
Even when Arakida do not compute the numerical 
correction for this example, we can do the 
same exercise. The new terms contribute as
\begin{equation}\label{tayara2}
\delta\alpha\approx\frac{m\delta\varphi^2}{b}-
\frac{m\delta\varphi^4}{2b}+\mathcal{O}
(\delta\varphi^5).
\end{equation}
Surprisingly, as the Arakida expression is obtained 
by the addition of \eqref{tayara1} and 
\eqref{tayara2}, we note that there exists a 
cancellation between the quadratic terms in 
$\delta\varphi$, resulting in a final expression given by 
\begin{equation}
\alpha_{\text{Arakida}}\approx \frac{4m}{b}-
\frac{3m\delta\varphi^4}{4b}.
\end{equation}
Hence, the correction to the usual Schwarzschild 
expression is of the order of $10^{-5}\mu$sec, a 
value undetectable with the actual 
technology.} Let us recall that the 
deflection angle is routinely measured by 
using the Eddington procedure on 
observations of the same source at two 
different times: when the Sun is present 
between the source and observer and when 
it is not. More generally, the change in 
the angular position of the images are 
usually compared with respect to some 
references objects using differential 
astrometry \cite{Will:1993ns,Turyshev:2008tm}. Alternatively, observations in one 
only session are made by observing the 
passing of the Sun around the line of 
sight of radio sources \cite{refId0}. A 
similar procedure is used to study the 
deflection of light by planets when they 
pass over the star background field 
\cite{2010IAUS..261..306M}. Moreover, as 
we will see below, even considering rays 
coming from far away sources whose images 
form an elongation angle (that is, the 
angle between the Sun, the Earth and the 
image) of $\theta_I=45\degree$  
($\hat\varphi_R=3\pi/4+\mathcal{O}(m)$), 
the expression of Arakida differs with the Ishihara et.al. alternative up to as much 
as 1mas (one milliarcsecond).
Therefore, the ramifications of these two 
formulae are not only of academic 
interest, but also practical.

Second, as we will show below (Secs. \eqref{subsubpreli} and \eqref{subsection:thetaI}), Eq. \eqref{eq:asad} or 
its equivalent \eqref{eqmain} yield the same results when they are 
compared with other well-known expressions obtained using standard post-
Newtonian techniques, even in more general situations such as the inclusion of a possible quadrupole moment of the lens and second order corrections in the mass. In fact, even when Ishihara \emph{et.al.} \cite{Ishihara:2016sfv} studied the bending of light rays from far away sources ($\hat\varphi_S=0+\mathcal{O}(m)$) due to the Sun, the expression obtained is a particular case of a  well-known result found by Shapiro in 1967 \cite{Shapiro67}  (see also Ward \cite{Ward-1970}), namely
\begin{equation}\label{eq:shapward}
\alpha=\frac{(1+\gamma)m}{b}(1+\cos(\theta_I)),    
\end{equation} 
where $\theta_I$ is again the elongation 
angle between lens, observer and source 
related to $\hat\varphi_R$ by 
$\theta_I=\pi-\hat\varphi_R+\mathcal{O}(m)$\footnote{See Section 
\eqref{subsubpreli} and Appendix 
\eqref{Bnueva}. In terms of 
$\hat{\varphi}_R$ and the radial 
coordinate of the receiver $r_o$ (related 
to $b$ by $b=r_o\sin(\hat{\varphi}_R)+\mathcal{O}(m)$) the expression \eqref{eq:shapward} reads 
$\alpha=\frac{(1+\gamma)m}{r_o}\tan(\frac{\hat{\varphi}_R}{2})$.}. In this 
expression  $\gamma$ is the Eddington 
post-Newtonian parameter (See 
Fig.\eqref{finitedistance}. More details 
will be discussed in the following Section \eqref{subsubpreli}). The usual infinite 
distance expression is recovered by taking $\theta_I\to 0$. The expression 
\eqref{eq:shapward} is not only well 
discussed in several textbooks 
\cite{Misner1973,Will:1993ns,Poisson-2014}
, but also it has been continuously tested experimentally using distant sources whose images form different elongation angles 
with the Sun. These elongation angles vary from $\arcsin(R_{\odot}/r_o)$ (with 
$R_{\odot}$ the solar radius) to 
$180\degree$. Even more, these 
observations are used to constrain the 
value of the $\gamma$ factor 
\cite{Robertson91, Shapiro-2004, Lambert-2009,2017arXiv170206647T,1997froeschle,2010hobbs}; testing the equivalence 
principle  by observing the apparent 
shifts of active galactic nuclei (AGN) 
core positions by using our own galaxy as 
a lens\cite{ZHANG201639}; the deflection 
of light produced by the Earth using data 
of Hipparcos\cite{gould93} or to reduce 
geodetic Very Long Baseline Interferometry (VLBI) data\cite{titovreduction}.

\begin{figure}[H]
\centering
\includegraphics[clip,width=90mm]{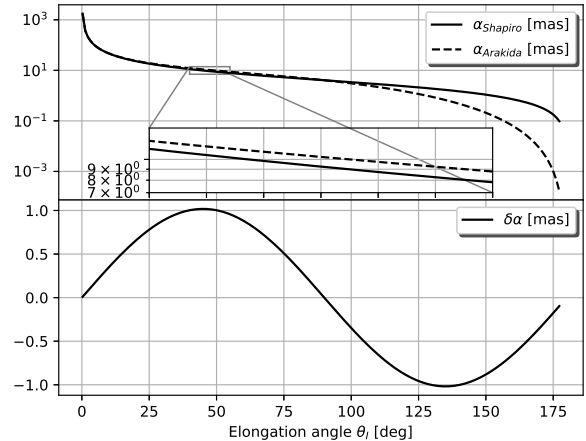}
\caption{Comparison between the Shapiro and Arakida expression for the deflection angle in terms of the elongation angle $\theta_I$ for an observer placed at 1 AU of the Sun ($\gamma=1$). Around $\theta_I=45\degree$ and $135\degree$ the difference can be as big as 1mas.}
 \label{anglecompS-A}
\end{figure}

In comparison with \eqref{eq:shapward}, the Arakida expression has an extra term obtained from \eqref{eq:deltaalphaarak} by taking $\hat{\varphi}_S=0+\mathcal{O}(m)$ and $\hat{\varphi}_R=\pi-\theta_I+\mathcal{O}(m)$ given by (see Appendix \eqref{Bnueva})
\begin{equation}\label{eq:difdektashaparak}
    \delta\alpha=\frac{m}{b}\sin^2(\theta_I)\cos(\theta_I)
    =\frac{m}{2r_o}\sin(2\theta_I),
\end{equation}
where in the last term we have made the replacement $b=r_o\sin(\theta_I)$, with $r_o$ the radial distance between the observer and the lens. In the Fig\eqref{anglecompS-A} we have plotted both the Shapiro expression (with $\gamma=1$) and the Arakida expression for $r_o=1\,\text{AU}$, and also their difference $\delta\alpha$ as given by \eqref{eq:difdektashaparak}. At elongation angles $\theta_I$ equals to $45\degree$ or $135\degree$ the difference is as big as 1mas. Let us remember that instruments like GAIA can measure variations in angular position of the stars with a resolution as small as $1\mu\text{as}$ at elongation angles which vary between $\theta_I\approx 45\degree$ to $180\degree$. Even for planets like Jupiter, this $1\mu\text{as}$ light deflection is reached at elongation angles of $90\degree$ or at $17\degree$ for Saturn\cite{2010IAUS..261..306M,Crosta:2005ch}.

We will show below how the expression \eqref{eq:shapward} and also other more general results can be successfully recovered from the Ishihara \emph{et. al.} definition. 

All these discussed issues give us confidence in the expression defined by Ishihara \emph{et.al.} in \cite{Ishihara:2016vdc} and given by
\eqref{eq:asad} or its equivalent versions 
\eqref{eqmain} and \eqref{eqmain2}. 

Note that even when \eqref{def1} and \eqref{eqmain} 
are equivalent to the original result found by Ishihara \emph{et.al.} (\eqref{eq:asad}), they were never before explicitly presented in the literature. In particular, \eqref{def1} has a 
clear geometrical meaning.\footnote{In references 
\cite{Ishihara:2016vdc,Ishihara:2016sfv,Ono:2017pie,Ono:2018ybw} 
there is an alternative presentation for the deflection angle 
$\alpha$ as a particular sum of three angles, which is written in terms of 
two geometrical angles that $\gamma_\ell$ form with the radial curves 
and a coordinate angle $\varphi_{RS}$, but as the authors claim, this 
definition seems to rely on a choice of the angular
coordinate $\varphi_{RS}$, however they show that this definition is 
equivalent to the geometrical invariant version 
\eqref{eq:asad}. More details about the comparison between \eqref{def1} and the angular definition in terms of the sum of three angles\cite{Ishihara:2016vdc} are presented in the last part of Appendix 
\eqref{Ap}.}
As a useful test, in Appendix 
\eqref{Ap} we will show using an explicit example how the original 
version \eqref{eq:asad}, or its equivalent new finite region versions 
\eqref{def1} and \eqref{eqmain} give the same result. Of course, 
Eq.\eqref{eq:asad} is more easy to use, because one does not need to 
compute the geodesic curvatures. Therefore, from now on we will 
continue using this last expression.

\subsubsection{Relation between the Ishihara et.al. definition of deflection angle at finite distance and some known expressions in the literature using a post-newtonian approach.}\label{subsubpreli}
The deflection angle for a Schwarzschild metric and for a 
Kottler spacetime were calculated using equation 
\eqref{eq:asad} in \cite{Ishihara:2016vdc} and using the 
version \eqref{def:Arak} in \cite{Arakida:2017hrm}. In 
particular, the possibility of observing these finite 
corrections for the bending angle for a Schwarzschild 
spacetime were discussed 
\cite{Ishihara:2016sfv,Ono:2017pie}. As previously mentioned, the computation of 
corrections at finite distances for the deflection angle has been done by different authors under more general 
situations and well discussed in textbooks for many 
years (Expression \eqref{eq:isexp} can be found for example in \cite{Will:1993ns}). Recently these calculations have been performed using 
post-Newtonian methods, solving explicitly the geodesic 
equation in particular spacetimes
\cite{1970ApJ...162..345W,Shapiro67,Richter:1982zz,DeLuca:2003zz,Misner1973,Will:1993ns,He:2016vxc,deFelice:2006xm,deFelice:2004nf,Azreg-Ainou:2017obt}. In fact, such 
expressions are needed in high-precision 
astrometry\cite{deFelice:1998wk,deFelice:2001wr,Crosta:2003zh}. Unfortunately, the authors of \cite{Ishihara:2016sfv} 
or \cite{Arakida:2017hrm} do not try to make a comparison 
with these different results.

In this work, we show that the deflection angle which follows from \eqref{eq:asad} is in complete agreement with known finite distance expressions even considering second order effects and more general metrics than the Schwarzschild solution. In particular, we are interested in the comparison of the finite distance expression for the deflection angle as computed by Richter and Matzner for a \emph{parametrized-post-Newtonian (PPN) metric}\cite{Richter:1982zz}. Of course, the equivalence between the Ishihara \emph{et.al.} definition with the well-known expressions also implies that the Arakida definition can not reproduce these results.

A detailed discussion of the PPN metric first requires some review of basic facts and assumptions. Let us recall the form of the general PPN metric that represents the exterior of a static and axially symmetric compact body with mass $m$ and multipole moments $J_n$. For this case the metric is described by an expression similar to the expression given by Eq.\eqref{eq:spherimetr}  but now with the associated metric functions $\tilde{A}$, $\tilde{B}$ and $\tilde{C}$ depending on the coordinates $r$ and $\vartheta$:
\begin{eqnarray}
\begin{aligned}
\tilde{A}(r,\vartheta)=& 1+2\tilde{U}(r,\vartheta)+2\beta U^{2}(r,\vartheta), \label{A}\\ 
\tilde{B}(r,\vartheta)=& 1-2\gamma \tilde{U}(r,\vartheta)+\frac{3}{2}\nu \tilde{U}^{2}(r,\vartheta),\label{B} \\ 
\tilde{C}(r,\vartheta)=& B(r,\vartheta) r^{2},\label{C}
\end{aligned}
\end{eqnarray}
where the potential $\tilde{U}$ reads
\begin{equation}
\tilde{U}(r,\vartheta)=-\frac{m}{r}\bigg[1-\sum^\infty_{n=2}(\frac{R}{r})^nJ_n P_n(\cos(\vartheta))\bigg], 
\end{equation}
with $P_n(x)$ the Legendre polynomials. Here $\beta$, $\gamma$ and $\nu$ are three parameters which take the value $1$ in the Einstein general relativity theory. In that case, if $J_n=0$, this metric represents the second order version of the Schwarzschild metric. {Let us also assume} that in addition to the mass $m$, the only non-vanishing {multipole is the quadrupole moment,} $J_2$.

\begin{figure}[H]
\centering
\includegraphics[clip,width=85mm]{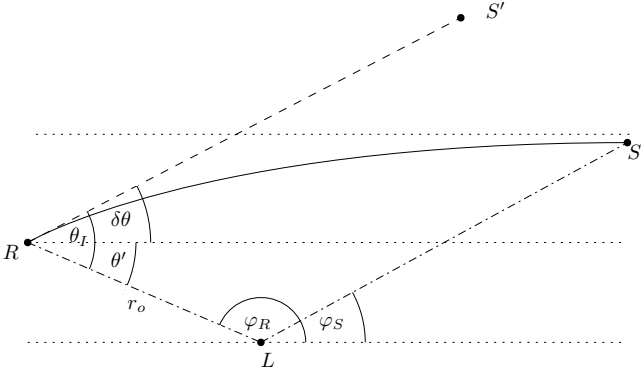}
\caption{A light ray travels from a distant source $S$ to the receiver $R$ through a region where a lens $L$ is present. The angle $\theta_I$ is defined by the angle between the lens, the receiver and the angular position of the image $S'$. The angle $\theta'$ is the angular position of the source in the absence of the lens as it would be seen by the receiver if the source were considered to be far away. The difference between these two angles is defined to be $\delta\theta$.}\label{finitedistance}
\end{figure}

Of course, this metric is not spherically symmetric. However if we restrict our study to the propagation of light rays in the plane defined by $\vartheta=\pi/2$, the PPN metric to this plane has an SO(2) symmetry and the metric functions are given by
\begin{eqnarray}
A(r)&=& \tilde{A}(r,\pi/2)= 1+2U(r)+2\beta U^{2}(r),\label{A1} \\ 
B(r)&=& \tilde{B}(r,\pi/2)=1-2\gamma U(r)+\frac{3}{2}\nu U^{2}(r),\label{A2} \\ 
C(r)&=&\tilde{C}(r,\pi/2)= B(r) r^{2},\label{A3}
\end{eqnarray}
with
\begin{equation}
U(r)=-\frac{m}{r}\bigg(1+\frac{R^{2}J_{2}}{2r^{2}}\bigg).
\end{equation}
Let a gravitational compact object be represented in the weak gravitational field region outside the object by the previous metric, and let us assume a lens $L$, receiver $R$ and a source $S$ configuration as shown in Fig.\eqref{finitedistance}. For the moment, we also assume that the source is far away from the lens  and we choose a new azimuth angular coordinate $\varphi$  such that $\varphi_S=0$, referring to Fig.\eqref{finitedistance}. More details between this angular coordinate and the previously defined $\hat\varphi$ will be given in Sec.\eqref{subsec:relationscoordinates} (see also Appendix\eqref{Bnueva}).  However, the receiver is assumed to be at a finite distance from the lens. In this case, the standard operational way to define the deflection angle is through the quantity 
\begin{equation}
\delta\theta=\theta_I-\theta',
\end{equation}
where $\theta_I$ is the angle between the image of the source as seen by the receiver and the receiver-lens axis, and $\theta'$ is the value that this angle should take if the lens were absent \cite{Richter:1982zz, DeLuca:2003zz,Will:1993ns,Misner1973}. If we were to assume that the receiver $R$ is at infinite distance from the lens, then $\delta\theta$ should agree with the asymptotic deflection angle $\alpha_\infty$. However, due to the finite distance location of the receiver there exists a disagreement between these two angles in general. 
Of course, $\delta\theta$ is not by itself directly observable if one uses a single observation; it must be measured using the Eddington procedure as explained above through two different observations at different times. We will use later this quantity to introduce another formula which takes into account the angular separation of the image of the source with respect to a reference object which is not necessarily the lens. Other observable quantities that can be computed from the deflection angle are the optical scalars known as shear and convergence (see for example \cite{PhysRevD.83.083007,PhysRevD.97.084010,Crisnejo:2018uyn} and references therein). However, in this discussion we will concentrate on the Eddington procedure and the associated differential astrometry technique (See for example \cite{Will:1993ns,Will2014,Poisson-2014,Turyshev:2008tm} and references therein for more details).
Different authors using different methods have computed the deflection angle $\delta\theta$ in terms of the parameters of the compact object (lens) and the observable angle $\theta_I$. These expressions can be found two separate ways: in terms of the impact parameter (which at finite distance is not an observable) or in terms of the radial coordinate $r_o$ between the receiver and the lens. If we consider only the computation of $\delta\theta$ at first order in $m$ and in $J_2$, the relation between the impact parameter and the radial coordinate $r_o$ is simply $b=r_o\sin(\varphi_R)$ which of course, must be corrected at second order.  

Before continuing, let us remark on the notation: even when $\delta\theta$ is a commonly used notation for the the deflection angle at finite distances, we will continue denoting $\delta\theta$ as $\alpha_{S_\infty}$ (we use the suffix ${S_\infty}$ in $\alpha$ as a reminder that the source is assumed to be placed at infinite distance from the lens). On the other hand in the case of infinite distances (for both, the receiver and the source) the deflection angle will be denoted as $\alpha_\infty$.
\begin{widetext}
More than three decades ago it was shown by Richter and Matzner in \cite{Richter:1982zz} that the deflection angle for the previous configuration of receiver, lens and source in a gravitational field represented by a PPN metric given by \eqref{A1}, \eqref{A2}, and \eqref{A3}, can be written in terms of the observable angle $\theta_I$ and the (non-observable)  impact parameter $b$ as
\begin{equation}    
\delta\theta\equiv\alpha_{{S_\infty}}=\alpha^{(1)}_{{S_\infty}}+\alpha^{(2)}_{{S_\infty}},
\end{equation}
where $\alpha^{(1)}_{S_\infty}$ and $\alpha^{(2)}_{S_\infty}$ are the linear and quadratic terms in the mass of the deflection angle:
\begin{eqnarray}
\alpha^{(1)}_{S_\infty}(b,\vartheta_I)&=&\frac{m}{b}( 1+\gamma)(1+\cos(\theta_{I}))\bigg[1+\frac{ J_{2}R^{2}}{2b^{2}}\bigg( 2+\cos(\theta_{I})-\cos^{2}(\theta_{I})\bigg)\bigg],\label{alpha1thetaI}\\
\alpha^{(2)}_{S_\infty}(b,\vartheta_I)&=&\frac{m^2}{b^2} (2-\beta+2\gamma+\frac{3}{4}\nu)(\pi-\theta_I+\sin (\theta_I) \cos (\theta_{I})).\label{alpha2thetaI}
\end{eqnarray}
\end{widetext}
In fact, a more general metric which admits rotation of the gravitational object and more general energy-momentum distributions has been studied \cite{Richter:1982zz}, but for our purposes it is sufficient to restrict the study to the considered case.

Equations \eqref{alpha1thetaI} and \eqref{alpha2thetaI} can also be rewritten in terms of the radial coordinate $r_o$ of the receiver which is related to $b$ (see \cite{Richter:1982zz}) by:
\begin{equation}\label{b-ro} 
\frac{1}{b}=\frac{1}{r_o\sin(\vartheta_I)}-(1+\gamma)\frac{m}{r^2_o\sin(\vartheta_I)}+\mathcal{O}(m^2,m J_2).
\end{equation}
In terms of $r_o$, the relations \eqref{alpha1thetaI} and \eqref{alpha2thetaI} read\cite{Richter:1982zz}
\begin{widetext}
\begin{eqnarray}
\alpha^{(1)}_{S_\infty}(r_o,\vartheta_I)&=&\frac{m}{r_o}( 1+\gamma)\bigg[\frac{(1+\cos(\theta_{I}))}{\sin(\vartheta_I)}+\frac{ J_{2}R^{2}}{2r^2_o\sin^3(\vartheta_I)}\bigg( 2+3\cos(\theta_{I})-\cos^{3}(\theta_{I})\bigg)\bigg],\label{alpha1thetaIv2}\\
\alpha^{(2)}_{S_\infty}(r_o,\vartheta_I)&=&\frac{m^2}{r^2_o} \bigg[(2-\beta+2\gamma+\frac{3}{4}\nu)\frac{\pi-\theta_I+\sin (\theta_I) \cos (\theta_{I})}{\sin^2(\vartheta_I)}-(1+\gamma)^2\frac{1+\cos(\vartheta_I)}{\sin(\vartheta_I)}\bigg].\label{alpha2thetaIv2}
\end{eqnarray} 
\end{widetext}
A natural question arises: Can these finite distance relations for the bending angle, \eqref{alpha1thetaI} and \eqref{alpha2thetaI} or their equivalent \eqref{alpha1thetaIv2} and \eqref{alpha2thetaIv2}, be recovered from the proposal given by the formula \eqref{eq:asad} of \cite{Ishihara:2016vdc} or the other nonequivalent alternative given by \eqref{def:Arak} of \cite{Arakida:2017hrm}?
 We will show in Section \eqref{subsection:thetaI} that the answer to this question is affirmative for the version given by the Ishihara \emph{et.al.} definition, giving us much more confidence to this geometrical way to compute the deflection angle at finite distances. Moreover, this non trivial result will follow as a particular case of the study of more general astrophysical situations, where the gravitational objects described by the PPN metric given by Eqs.\eqref{A1}, \eqref{A2} and \eqref{A3} are now surrounded by a plasma medium. That is, we will obtain expressions which can be used for a variety of spacetimes into the framework of gravitational metric theories which contains as a particular case the Einstein general relativity theory and which does not only represent the spacetime of a central spherical body mass but also it allows a the body with a nontrivial quadrupole moment and that can be immersed in a plasma environment. Of course, in that case, the light rays do not follow null geodesics of the PPN metric; however their dynamics is such that there exists an associated two-dimensional optical metric $g^{\text{opt}}_{ij}$ where the spatial orbits of the light rays in the physical metric can also be considered to be spatial geodesics of $g^{\text{opt}}_{ij}$ allowing us to use the Gibbons-Werner techniques.  
 
In particular, we will show for the first time that the relations \eqref{alpha1thetaI} and \eqref{alpha2thetaI} or \eqref{alpha1thetaIv2} and \eqref{alpha2thetaIv2} can be recovered and successfully derived from the simple and geometrical relation \eqref{eq:asad} and, what is more important, they can be generalized to more general scenarios taking into account the presence of a homogeneous plasma environment. As a byproduct we will also introduce for the first time new useful formulas for the separation angle between two sources.
However, let us first review the behavior of light rays in the presence of plasma and how they can be studied using the Gauss-Bonnet theorem.

\section{The optical metric and the Gauss-Bonnet theorem in a plasma environment}\label{section3}
\subsection{The optical metric associated to a plasma medium in an external gravitational field}
Let us consider a static spacetime $(\mathcal{M},g_{\alpha\beta})$ filled with a cold non-magnetized plasma described by the refractive index $n$ \cite{Bisnovatyi-Kogan:2015dxa,Perlick:2015vta},
\begin{equation}\label{refraction-index}
n^{2}(x,\omega(x))=1-\frac{\omega_{e}^{2}(x)}{\omega^{2}(x)},
\end{equation}
where $\omega(x)$ is the photon frequency measured by a static observer while $\omega_{e}(x)$ is the electron plasma frequency,
\begin{equation}
\omega^{2}_{e}(x)=\frac{4\pi e^{2}}{m_{e}} N(x)=K_e N(x),
\end{equation}
with $e$ and $m_{e}$ the charge of the electron and its mass, respectively; and $N(x)$ is the number density of electrons in the plasma. 

We are interested in the the deflection of the light path when rays travel through a gravitational {field in} a plasma {filled environment}. The dynamics of the light rays are usually described through the Hamiltonian \cite{Perlick:2015vta,Perlick-book}, 
\begin{equation}\label{Hamilt}
H(x,p)=\frac{1}{2}\bigg({g}^{\alpha \beta}(x)p_{\alpha}p_{\beta}+\omega_{e}^{2}(x)\bigg),
\end{equation}
where light rays are solutions of the Hamilton's equation
\begin{equation}\label{Hamil-eqn}
\ell^{\alpha}=\frac{d x^{\alpha}}{d\tilde{s}}=\frac{\partial H}{\partial p_{\alpha}}, \ \ \frac{d p_{\alpha}}{d\tilde{s}}=-\frac{\partial H}{\partial x^{\alpha}}; 
\end{equation}
with the constraint
\begin{equation}\label{Hamit-constraint}
H(x,p)=0,
\end{equation}
and $\tilde{s}$ is an curve parameter along the light curves.

From \eqref{Hamit-constraint} in can be shown that in general light rays do not follow timelike or null geodesics with respect to ${g}_{\alpha\beta}$. Instead, they describe timelike curves with the exception of a homogeneous plasma medium where light rays follow timelike geodesics of $g_{\alpha\beta}$. Note that only light rays with $\omega(x) > \omega_{e}(x)$ propagate through the plasma.

On the other hand, for the case of static spacetimes, even considering dispersive media one can use a Fermat-like principle\cite{book:75670}, where the spatial projections of the light rays on the slices $t=\text{constant}$ which solve the Hamilton's equations are also spacelike geodesics of the following Riemannian optical metric,
\begin{equation}\label{optical-metric}
g_{ij}^{\text{opt}}=-\frac{n^2}{{g}_{00}}{g}_{ij}.
\end{equation}
It was precisely this last fact that recently allowed us to study the deflection of light in plasma environments using the Gauss-Bonnet theorem \cite{Crisnejo:2018uyn}.

From now on, we will restrict our attention to static and axially symmetric metrics surrounded by a cold non-magnetized plasma, that is, the physical spacetime is assumed to be described by a metric of the form
\begin{equation}\label{eq:phys}
\begin{aligned}
ds^2 =& -\tilde{A}(r,\vartheta) dt^{2} + \tilde{B}(r,\vartheta) dr^{2} \\
&+\tilde{C}(r,\vartheta)(\Theta(r,\theta)d\vartheta^{2}+\sin^{2}\vartheta d\varphi^{2}),
\end{aligned}
\end{equation}
and with a dependence of the plasma frequency on the coordinates $r$ and $\vartheta$, $\omega_{e}=\omega_{e}(r,\vartheta)$. 
 Note that we are neglecting the self-gravitation of the plasma. We also assume asymptotic flatness and that the plasma medium is static with respect to observers following integral curves of the timelike Killing vector field $\xi^\alpha=(\frac{\partial}{\partial t})^\alpha$. 
Due to the gravitational redshift, the frequency of a photon at a given radial position $r$ is given by:
\begin{equation}
\omega(r,\vartheta)=\frac{\omega_{\infty}}{\sqrt{\tilde{A}(r,\vartheta)}},
\end{equation}
where $\omega_{\infty}$ is the photon frequency measured by an observer at infinity. 
Now we will  restrict to the study of light propagation in the plane defined by  $\vartheta=\pi/2$. If the spacetime under consideration is spherically symmetric this restriction does not constitute any loss of generality. However, for the axially-symmetric case we should keep in mind that our results will be only valid for light propagation on this plane. Restricted to $\vartheta=\pi/2$, 
all variables only have a radial dependence and the metric functions will be written without a tilde in a similar way as was done in \eqref{A1}, \eqref{A2}, \eqref{A3}.

As we are interested in the application of the Gauss-Bonnet theorem to the determination of the bending angle,  following our previous work\cite{Crisnejo:2018uyn}, we will make use of the associated 2-dimensional Riemannian manifold $\left(\mathcal{M}^{\text{opt}},{g}^\text{opt}_{ij}\right)$ with the $SO(2)$ optical metric \eqref{optical-metric} (restricted to the plane $\vartheta=\pi/2$), 
\begin{equation}\label{eq:opt}
d\sigma^{2}=g_{ij}^{\text{opt}} dx^{i}dx^{j}=\frac{n^{2}(r)}{A(r)} \bigg(B(r)dr^{2}+C(r)d\varphi^{2}
\bigg).
\end{equation}
This metric is conformally related to the induced metric on the spatial section $t=\text{constant}$, $\vartheta=\pi/2$, of the physical spacetime, and therefore it preserves the angles formed between two curves at a given point. 

The geodesic motion follows from the Lagrangian,
\begin{equation}\label{eq:Lmassive}
\mathcal{L}=\frac{1}{2}\bigg[\frac{n^{2}(r)}{A(r)} \bigg(B(r)\bigg(\frac{dr}{d\sigma}\bigg)^{2}+C(r)\bigg(\frac{d\varphi}{d\sigma}\bigg)^{2}\bigg)\bigg],
\end{equation}
with the constraint:
\begin{equation}\label{constraint}
\frac{n^{2}(r)}{A(r)} \bigg[B(r)\bigg(\frac{dr}{d\sigma}\bigg)^{2}+C(r)\bigg(\frac{d\varphi}{d\sigma}\bigg)^{2}\bigg]=1.
\end{equation}
In the case of a homogeneous plasma ($\omega_e=\text{constant}$, see below), it follows
from \eqref{eq:Lmassive} and \eqref{constraint}, the orbital equation is given by\cite{Crisnejo:2018uyn},
\begin{equation}\label{orbit}
\bigg(\frac{dr}{d\varphi}\bigg)^{2}=\frac{C(r)}{B(r)}\bigg[\frac{C(r)\,n^{2}(r)}{A(r)\,n_{0}^{2}\,b^{2}}-1\bigg],
\end{equation}
where $n_{0}^{2}=1-\frac{\omega_{e}^{2}}{\omega_{o}^{2}A(r_{o})}$, 
and with $\omega_o$ the frequency of the light ray measured by the receiver at $r_o$ (related to $\omega_\infty$ by $\omega_\infty=\omega_o\sqrt{A(r_o)}$).

Defining $u=\frac{1}{r}$, the above equation reduces to,
\begin{equation}\label{orbit-eq-u}
\bigg( \frac{du}{d\varphi} \bigg)^{2}=
\frac{u^{4} C(u)}{B(u)}\bigg(\frac{C(u)n^{2}(u)}{n_{0}^{2}(u_{0})b^{2}A(u)}-1 \bigg).
\end{equation}

In terms of the curvature tensor associated with the optical metric, the Gaussian curvature $\mathcal{K}$ can be computed from
\begin{equation}\label{KR}
\mathcal{K}= \frac{R_{r\varphi r\varphi}(g^{\text{opt}})}{\text{det}(g^{\text{opt}})}.
\end{equation}

\subsection{About the measurement of light deflection propagating into a plasma medium}\label{subsection:twoimages}
A standard procedure to measure the bending in light rays which propagate in a region with a lens present consists in observing how the relative angle $\Theta$ between two far away sources (one of them chosen as a reference $S_r$) changes when our Sun (or a different star or a planet) pass near the line of sight of the source $S$\cite{Will2014}. Therefore, in order to study how these relative angular positions depend of the location of the observer, lens, source and the plasma, a general formula for this angle is needed. As the light rays in plasma media follow in general timelike curves we need an expression for the relative angle for these kind of curves. The general expression for causal curves was recently derived  by Lebedev and Lake in \cite{Lebedev-2013,Lebedev-2016}. Their result is the following. Let $U^a$ be a 4-velocity of an observer and $K$, $W$ be two arbitrary future causal vectors with $\bar{K}$, $\bar{W}$ their spatial projections into the local frame of the observer. Then, the angle between $\bar{K}$, $\bar{W}$ in the position of the observer, which is a measurable quantity, is given by:
\begin{equation}\label{costheta}
     \cos(\Theta) = \frac{K_\alpha W^\alpha+(U_\alpha K^\alpha)(U_\beta W^\beta)}{\sqrt{K_\alpha K^\alpha + (U_\alpha K^\alpha)^2}\sqrt{W_\alpha W^\alpha + (U_\alpha W^\alpha)^2}}.
\end{equation}
It is important to note that \eqref{costheta} is an explicitly gauge invariant quantity.

As a corollary from the above expression we proof now the following useful result.
\begin{theorem}
Let $\gamma$ and $\gamma'$ be two light rays with 4-momentum $p^\alpha$ and $p'^{\alpha}$ propagating not necessarily at the same frequency into a cold plasma medium in a static and asymptotically flat spherically symmetric spacetime of the form \eqref{eq:spherimetr}. Then, at the position of an observer with 4-velocity $U^a$ the angle 
\eqref{costheta}  between the two rays takes the form,
\begin{equation}\label{costheta2}
    \cos(\Theta)=\frac{1}{n(p)n(p^\prime)}\bigg( 1+\frac{p_\alpha p^{\prime\alpha}}{(p_\beta U^\beta)(p^\prime_\beta U^\beta)} \bigg),
\end{equation} 
where $n(p)$ and $n(p')$ are the respective refractive indices which can be written in terms of the associated frequencies as observed in the position of the observer as $n^2(p(\omega))=1-\omega^2_e(r)/\omega^2(r)$ and $n^2(p^\prime(\omega^\prime))=1-\omega^2_e(r)/\omega^{\prime 2}(r)$.
\end{theorem}

Even though this is a very simple formula, we have no knowledge of a previous presentation of it. We note that this expression reduces to the well-known pure gravity expression when $n(p)=n(p')=1.$ (See for example \cite{Will:1993ns}). Note also that this expression can be used to compute the change in the relative position angle between the two images due to pure gravitational and plasma lensing effects, and additionally to aberration effects depending of the relative velocity between the observer and the lens (codified in the factors $p_\beta U^\beta$ and $p^\prime_\beta U^\beta$ which depend of $U^a$). However, even when the last effects can be easily introduced in the general formulae, for simplicity we will restrict here to the case that the observer is considered to be static with respect to the lens.\\

{\bf Proof:}
From the Hamiltonian given by \eqref{Hamilt} it follows that the tangent vectors to the light rays are parallel to the 4-momentum and therefore we can compute their relative angle using Eq.\eqref{costheta} with the vectors $K$ and $W$ identified with $p^\alpha$ and $p'^\alpha$.
On the other hand, for two photons with 4-momentum $p^\alpha$ and $p^{\prime\alpha}$ propagating into a plasma medium the respective refractive indices read  \cite{book:75670,BisnovatyiKogan:2010ar},
\begin{equation}
    n^2(p)=1+\frac{p_\alpha p^\alpha}{(p_\beta U^\beta)^2}, \ \ n^2(p^\prime)=1+\frac{p^\prime_\alpha p^{\prime\alpha}}{(p^\prime_\beta U^\beta)^2},
\end{equation} which can be re-expressed as
\begin{eqnarray}
    \sqrt{p_\alpha p^\alpha+(p_\alpha U^\alpha)^2}&=&n(p)(p_\alpha U^\alpha)\\
    \sqrt{p^\prime_\beta p^{\prime\beta}+(p^\prime_\beta U^\beta)^2}&=&n(p^\prime)(p^\prime_\beta U^\beta).
\end{eqnarray}
By replacing the last expressions into eq.\eqref{costheta} the result follows.\\

For a static plasma medium one can see from the Hamilton equations that,
\begin{equation}
    p^\alpha=\omega(x^i)\bigg(U^\alpha+n(x^i) \hat{e}^\alpha\bigg)
\end{equation}
where $\omega(x^j)=-p_\alpha U^\alpha$, $n^2(x^i)=1-\frac{w^2_e(x^i)}{w^2(x^i)}$ and $\hat{e}^\alpha$ is a spatial-like vector orthogonal to $U^\alpha$ and tangent to the 3-dimensional hypersurfaces $\Sigma_t$ defined by $t=\text{constant}$, They  are normalized to $1$ with respect to the metric \eqref{eq:spherimetr}. $U^\alpha$ is the 4-velocity of the observers orthogonal to the slices $\Sigma_t$, that is $U^\alpha=\frac{t^\alpha}{\sqrt{-{\bf g}(t^\alpha,t^\alpha)}}$, with $t^\alpha=(1,0,0,0)$. We have a similar expression for $p^{\prime\alpha}$, with respective frequency $\omega^\prime(x^i)$.

In order to construct from \eqref{costheta2} a relationship that can be easily tested by observations and also that takes into account the presence of the plasma we will follow the procedure given by Poisson and Will in \cite{Poisson-2014} (See also \cite{Will:1993ns}). As we are interested in the first order change of the relative angle $\Theta$ between two given sources (one of them used as a reference source $S_\text{ref}$) in the presence of a lens as compared with the relative angle $\Theta'$ without its presence, we must take into account that the spatial direction of the photon 4-momentum of each light ray will change compared with its unperturbed flat value.

\begin{figure}[H]
\centering
\includegraphics[clip,width=75mm]{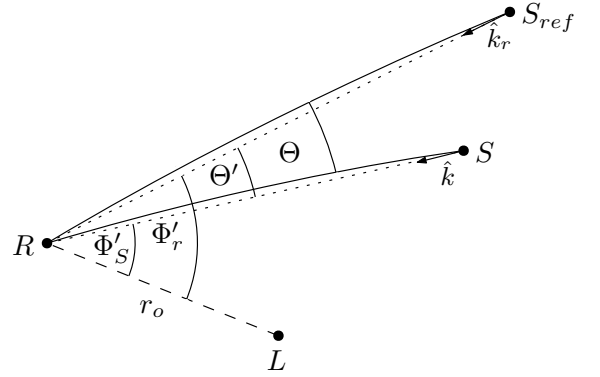}
\caption{$R$, $L$, $S$ and  $S_{\text{ref}}$ denote observer, lens, source and reference source positions. Solid lines mean the real paths of photons while the dot ones indicate the photon trajectory in the background. All angular quantities as well as lens-observer position $r_o$ are observable quantities.}
 \label{obs1}
\end{figure}

Let us consider a 
photon with 4-momentum $p^\alpha$ starting at the position of a source $S$, and another photon with 4-momentum $p^{\prime\alpha}$ starting from a reference source $S_{ref}$, both of them reaching the observer at $R$ (see Fig. \ref{obs1}). We consider that both sources are far away from the lens, which is in general the more common situation. We also assume that at linear order the metric is written in a cartesian and isotropic coordinate system as $g_{\alpha\beta}=\eta_{\alpha\beta}+\epsilon h_{\alpha\beta}$ with the only non-vanishing components of $h_{\alpha\beta}$ given by $h_{\alpha\beta}=(h_{00},h_{rr}\delta_{ij})$, where $\epsilon$ is a bookkeeper parameter which measures the deviation of the Minkowski spacetime. The unperturbed spatial directions of the light rays are given by $\hat{k}^i$ for the source $S$ and by $\hat{k}^i_r$ for the reference source (which are unit vectors with respect to the euclidean metric). Under these assumptions, at first order in $\epsilon$ the spatial vector $\hat{e}^\alpha=(0,\hat{e}^i)$ associated to the 4-momentum $p^\alpha$ of the light ray coming from $S$ and the spatial vector $\hat{e}^{\prime\alpha}=(0,\hat{e}^{\prime i})$ associated to the 4-momentum $p^{\prime}$ of the ray coming from $S_{ref}$ will be given by,
\begin{eqnarray}
    \hat{e}^i= (1-\frac{\epsilon}{2} h_{rr})\hat{k}^i-\alpha_S \hat{b}^i_S+\mathcal{O}(\epsilon^2),\\
    \hat{e}^{\prime i}=(1-\frac{\epsilon}{2} h_{rr})\hat{k}^i_r-\alpha_r \hat{b}^i_r+\mathcal{O}(\epsilon^2),
\end{eqnarray}
where we have taking into account that $\alpha_S$ and $\alpha_r$ are $\mathcal{O}(\epsilon)$ quantities representing the deviation angles associated to $S$ and $S_r$ respectively and $\hat{b}^i_S=\frac{\vec{b}^i_S}{b_S}$ and $\hat{b}^i_r=\frac{\vec{b}^i_r}{b_r}$ are the unit vectors (with respect to the euclidean metric) in the direction of the respective impact parameters vectors.

Now we compute the angle $\Theta$ using the formula $\eqref{costheta2}$ at first order in the perturbation of the flat background. After a simple computation we arrive to the following expression:
\begin{equation}\label{will-2}
    \cos(\Theta)=\hat{k}\cdot\hat{k}_r - \alpha_S (\hat{k}_{r}\cdot\hat{b}_S)  - \alpha_r (\hat{k}\cdot\hat{b}_r)+\mathcal{O}(\epsilon^2),
\end{equation}
where ``$\cdot$" means scalars product with respect to the background euclidean metric. Note that this expression has formally the same form that Eq. $10.74$ in \cite{Poisson-2014} even when in that reference it was deduced from the particular case of \eqref{costheta2} when $n(p)=n(p^\prime)=1$. In contrast, in our more general setting the light rays can follow timelike curves and the deflection angles are allowed to take into account the presence of the plasma.

\begin{figure}[H]
\centering
\includegraphics[clip,width=75mm]{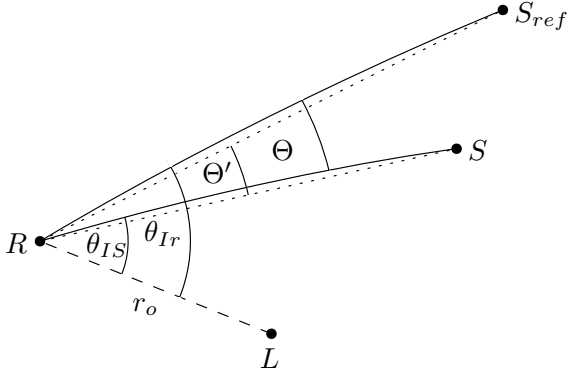}
\caption{From the position of the receiver different observable angles can be measured. $\Theta$ is the angle between $S_\text{ref}$ and $S$, $\theta_{IS}$ is the angle between $S$, $R$ and $L$, and $\theta_{Ir}$ is the angle between $S_\text{ref}$, $R$ and $L$.  }
 \label{obs2}
\end{figure}

Following \cite{Poisson-2014}, 
it is straightforward to see that one can rewrite \eqref{will-2} as
\begin{equation}\label{will-3}
\begin{aligned}
    \cos(\Theta)=&\cos(\Theta^\prime)- \alpha_S\bigg( \frac{\cos\Phi^\prime_r-\cos\Phi^\prime_S\cos\Theta^\prime}{\sin\Phi^\prime_S} \bigg) \\
    &- \alpha_r\bigg( \frac{\cos\Phi^\prime_S-\cos\Phi^\prime_r\cos\Theta^\prime}{\sin\Phi^\prime_r} \bigg),
\end{aligned}    
\end{equation}
where the angles $\Phi'_S$ and $\Phi'_r$ are the unperturbed values of the angles between observer, lens and source, and between observer, lens and reference source respectively. Note that, despite the scenario shown in figure \eqref{obs1}, the receiver, source, lens and reference source are not assumed to be necessarily at the same plane.
For small departures from the background path we define $\Delta\Theta=\Theta-\Theta^\prime\ll 1$, and expand the left hand side in  \eqref{will-3} as,
\begin{equation}
    \cos\Theta\approx\cos\Theta^\prime-\sin\Theta^\prime\Delta\Theta.
\end{equation}
Therefore, replacing it in  \eqref{will-3}, we obtain 
\begin{equation}\label{angle-obs-1}
\begin{aligned}
    \Delta\Theta=&\alpha_S\bigg( \frac{\cos\Phi^\prime_r-\cos\Phi^\prime_S\cos\Theta^\prime}{\sin\Phi^\prime_S\sin\Theta^\prime} \bigg) \\
    &+ \alpha_r\bigg( \frac{\cos\Phi^\prime_S-\cos\Phi^\prime_r\cos\Theta^\prime}{\sin\Phi^\prime_r\sin\Theta^\prime} \bigg).
\end{aligned}    
\end{equation}
Even more, as we are considering first order corrections, we can replace the angles $\Phi'_S$, $\Phi'_r$, and $\Theta'$ by the respective observable angular positions $\theta_{IS}$ and $\theta_{Ir}$, and $\Theta$ (See Fig.\eqref{obs2}), namely:
\begin{equation}\label{angle-obs-obs}
\begin{aligned}
    \Delta\Theta=&\alpha_S\bigg( \frac{\cos\theta_{Ir}-\cos\theta_{IS}\cos\Theta}{\sin\theta_{IS}\sin\Theta} \bigg) \\
    &+ \alpha_r\bigg( \frac{\cos\theta_{IS}-\cos\theta_{Ir}\cos\Theta}{\sin\theta_{Ir}\sin\Theta} \bigg),
\end{aligned}    
\end{equation}
obtaining in this way a relationship between observable quantities that allows us to test our expressions for the deflection angle of different plasma density profiles and gravitational fields.

It is worthwhile to note that if the reference source is chosen as the lens, that is, if the direction of the reference source and the lens coincide, ($\theta_{Ir}=0$ and $\Theta=\theta_{IS}$),  eq.\eqref{angle-obs-obs} reduces to:
\begin{equation}
    \Delta\Theta=\alpha_S.
\end{equation}

As astronomical angular measurements are performed on the celestial sphere, it is convenient to express \eqref{angle-obs-obs} in terms of angles projected on the plane of the sky.
Then, \eqref{angle-obs-obs} can be expressed in term of the projected angles $A$ and $B$ as,
\begin{equation}\label{eq:deltatrigAB}
    \Delta\Theta = \alpha_S \cos(B)+\alpha_r \cos(A),
\end{equation}
where $A$ is the angle between $S_{ref}-S$ and $S_{ref}-L$, and $B$ the one between $S_{ref}-S$ and $S-L$ directions projected on the celestial sphere.
The relation \eqref{eq:deltatrigAB} follow from the well-known spherical trigonometric identities \cite{Poisson-2014}:
\begin{equation}
    \cos(\theta_{Ir})=\cos(\theta_{IS})\cos(\Theta)+\sin(\theta_{IS})\sin(\Theta)\cos(B),
\end{equation}
 \begin{equation}
     \cos(\theta_{IS})=\cos(\theta_{Ir})\cos(\Theta)+\sin(\theta_{Ir})\sin(\Theta)\cos(A).
\end{equation}
In particular, if the lens, observer, and the two sources are at the same plane, with both sources at the same side of the lens as seen by the observer, then $B=0$ and $A=\pi$, and the variation in the separation angle reduces to
\begin{equation}\label{eq:mostsimple}
\Delta\Theta=\alpha_S-\alpha_r,
    \end{equation}
a result that also follows from a simple inspection of Fig.\eqref{obs2}.
Relations like \eqref{eq:deltatrigAB} with $\alpha_S$ and $\alpha_r$ given by the Shapiro formula (eq.\eqref{eq:shapward}), or the more general case given by the Richter and Matzner expression \eqref{alpha1thetaI} are of the common use in astronomical observations of the deflection angle, not only produced by the Sun but also big planets like Jupiter\cite{Crosta:2005ch}. The most simple situation follows from the use of \eqref{eq:mostsimple} for a pure monopole gravitational field in which the deflection angles $\alpha_S$ and $\alpha_r$ are given by the Shapiro formula, eq.\eqref{eq:shapward}. In such a situation, eq.\eqref{eq:mostsimple} gives
\begin{equation}
    \Delta\Theta=(1+\gamma)\frac{2m}{r_o}\frac{\sin(\frac{\theta_{IS}-\theta_{Ir}}{2})}{\sin\frac{\theta_{IS}}{2}\sin\frac{\theta_{Ir}}{2}}.
\end{equation}
 This expression was also recently used by Turyshev\cite{Turyshev:2008tm} to estimate the deflection angles caused by the monopole aspect of the gravitational fields of different celestial bodies in the solar
system and their observability in future space interferometry missions by using absolute and differential astrometry measurements. In that reference, similar analysis was made by the author by using the Richter and Matzner formula and a generalization of it in order to also estimate the observability of the quadrupole and octupole graviational fields contribution to the shift in angular positions of far away objects.

As we have shown here, the formal expressions \eqref{eq:deltatrigAB}, \eqref{angle-obs-obs} and \eqref{eq:mostsimple} can also be used for a more general situation where light rays are propagating in a plasma medium, with of course a corresponding expression for $\alpha_S$ and $\alpha_r$ that generalize the Shapiro expression in those environments. In the following sections we will give expressions for the deflection angles derived from the use of the Gauss-Bonnet theorem that not only allow to recover the Shapiro or Richter and Matzner formulas as particular cases, but also generalize them to more general astrophysical situations where a plasma medium is present.

\section{Finite distance corrections for the light deflection in a homogeneous plasma medium: PPN metric}\label{section4}

{Let us} consider a gravitational lens surrounded by a homogeneous plasma whose electron number density reads,
\begin{equation}
N(r,\vartheta)=N_{0}=\text{constant}.
\end{equation}

\subsection{Parametrized-post-Newtonian (PPN) metric}
As an initial example, we study the light propagation in the equatorial plane of a astrophysical object surrounded by an homogeneous plasma medium and whose gravitational field is described by Eqs. $\eqref{A1},\eqref{A2}$ and \eqref{A3}. In the following we assume that $J_2\ll 1$ such that we will also neglect terms of order  $\mathcal{O}(J_2\times m^2)$.

Due to the gravitational redshift and considering that both the source and the observer at a finite distance from the lens, the refractive index reads
\begin{equation}
n(r)=\sqrt{1-\frac{\omega_{e}^{2} A(r)}{\omega_{o}^{2}A(r_{o})}},
\end{equation}
where $r_{o}$ is the radial position of the observer from the lens.
The associated optical metric at the considered order follows from using the relation \eqref{eq:opt} and reads,
\begin{equation}\label{eq:optmetppn}
d\sigma^2=\Omega^2(dr^2+r^2d\varphi^2);
\end{equation}
with
\begin{equation}
\begin{aligned}
\Omega^2=&\frac{\omega_{o}^2-\omega_{e}^2}{\omega_{o}^2}+\frac{m}{\omega_{o}^2 r^3 r_{o}^3} \bigg[(\gamma+1) \omega_{o}^2 r_{o}^3 (J_2 R^2+2 r^2)\\
&-\omega_{e}^2 \bigg(J_2 R^2 (\gamma r_{o}^3+r^3)+2 r^2 r_{o}^2 (\gamma r_{o}+r)\bigg)\bigg]\\
&+\frac{m^2}{2 \omega_{o}^2 r^2 r_{o}^2}\bigg[\omega_{o}^2 r_{o}^2 \bigg(8 \gamma-4 \beta+3 \nu+8\bigg)\\
&+\omega_{e}^2 \bigg(4 (\beta-2) r^2-8 \gamma r r_{o}-3 \nu r_{o}^2\bigg)\bigg]\\
&+\mathcal{O}(m^3,m^2\times J_2).
\end{aligned}
\end{equation}
In order to implement the method described in Section \eqref{sec:remark} for calculating the bending angle at finite distances, first we need to solve the equation \eqref{orbit-eq-u}. As we are interested in second order correction in $m$ for the bending angle we only need to solve \eqref{orbit-eq-u} at first order, which explicitly reads,
\begin{equation}
\begin{aligned}
\bigg(\frac{du}{d\varphi}\bigg)^2=&\frac{1}{b^2}-u^2+\frac{m \, u}{b^2}(2+J_2 R^2 u^2) \bigg( \gamma+\frac{1}{1-\omega_{e}^2/\omega_{o}^2} \bigg),
\end{aligned}
\end{equation}
with the asymptotic condition,
\begin{equation}\label{orbit-eq-u-id}
\lim_{\varphi\to 0}u(\varphi)=0 .\end{equation}
Then, assuming a solution of the form,
\begin{equation}
 u(\varphi)=\frac{1}{b}[\sin(\varphi)+m u_{1}(\varphi)],
\end{equation}
we obtain at first order in $m$,
\begin{equation}\label{uppn}
\begin{aligned}
u(\varphi)=&\frac{\sin(\varphi)}{b} + \frac{m(1-\cos(\varphi))}{2b^4} \bigg( \gamma+\frac{1}{1-\omega_{e}^{2}/\omega_{o}^{2}} \bigg)\\
&\times \bigg( 2b^2+J_{2}R^{2}(1-\cos(\varphi)) \bigg)+\mathcal{O}(m^2).
\end{aligned}
\end{equation}
For completeness, in eq. \eqref{uppn} we have written explicitly terms of order $\mathcal{O}(m\times J_2)$. From this expression it is worthwhile to emphasize that since the Gaussian curvature is order $\mathcal{O}(m)$ (see \eqref{eq:Kppn} below), it follows that terms of the form $\mathcal{O}(m\times J_2)$ in $u$ will contribute to the deflection angle with corrections of order $\mathcal{O}(m^2\times J_2)$ which are of higher order than considered. Therefore, they are not necessary in the computation of the bending angle.  

In order to compute the bending angle using \eqref{eq:asad} we must integrate the Gaussian curvature $\mathcal{K}$ over ${}^\infty_R\square^\infty_S$. For this, we need to compute $\mathcal{K}$ using the relation \eqref{KR} for the optical metric \eqref{eq:optmetppn} at second order in $m$ neglecting terms of order $\mathcal{O}( m^2\times J_2)$. The result reads:
\begin{equation}\label{eq:Kppn}
\begin{aligned}
\mathcal{K}=&\frac{m(2r^{2}+9J_{2}R^{2})}{2r^{5}} \frac{\omega_{o}^{2}(\gamma \omega_{e}^{2}-(\gamma+1)\omega_{o}^{2})}{(\omega_{o}^{2}-\omega_{e}^{2})^{2}} \\
&+ \frac{m^{2}\omega_{o}^{2}}{r_{o}r^{4}(\omega_{o}^{2}-\omega_{e}^{2})^{3}} \bigg\{r_{o}\omega_{o}^{4}\bigg[4\beta-2+4\gamma+6\gamma^{2}-3\nu\bigg]\\
&-2\omega_{o}^{2}\omega_{e}^{2}\bigg[(2+\gamma)r+2r_{o}(\beta-2+\gamma+3\gamma^{2})-3r_{o}\nu\bigg] \\ &+\omega_{e}^{4}\bigg[2\gamma r+6\gamma^{2}r_{o}-3\nu r_{o}\bigg]\bigg\}+\mathcal{O}(m^3,m^2\times J_2).
\end{aligned}
\end{equation}
The two-form $\mathcal{K}dS$ reads:
\begin{equation}
\begin{aligned}
\mathcal{K}dS=&\bigg\{-\frac{m (9 J_2 R^2+2 r^2) (\omega_{o}^2(\gamma+1) -\gamma \omega_{e}^2)}{2 r^4 (\omega_{o}^2-\omega_{o}^2)}\\
&+\frac{m^2}{r^3 r_{o} (\omega_{o}^2-\omega_{e}^2)^2} \bigg[\omega_{o}^4 r_{o} \bigg(4 (\beta+\gamma^2-1)-3 \nu\bigg)\\
&-2 \omega_{o}^2 \omega_{e}^2 \bigg(r_{o} (2 \beta+4 \gamma^2-3 \nu-4)+r\bigg)\\
&+r_{o} \omega_{e}^4 (4 \gamma^2-3 \nu)\bigg]\bigg\}drd\varphi+\mathcal{O}(m^3,m^2\times J_2).
\end{aligned}
\end{equation}

Finally, after doing the corresponding integral \eqref{eq:asad}, the deflection angle follows:
\begin{equation}
\alpha=-\int^{\varphi_R}_{\varphi_S}\int^{\infty}_{r_{\gamma_\ell}}\mathcal{K}dS=\alpha^{(1)}+\alpha^{(2)},
\end{equation}
where
\begin{equation}
\begin{aligned}
r_{\gamma_\ell}=&\frac{1}{u(\varphi)}=\frac{b}{\sin(\varphi)}-\frac{1-\cos(\varphi)}{\sin^2(\varphi)}\bigg(\gamma+\frac{1}{1-{\omega^2_e}/{\omega^2_o}}\bigg)m\\
&+\mathcal{O}(m^2,m\times J_2),
\end{aligned}
\end{equation}
and with
\begin{equation}\label{alpha1-homo}
\begin{aligned}
\alpha^{(1)}=&\frac{m}{b}\bigg(\cos(\varphi_{S})-\cos(\varphi_{R})\bigg)\bigg( \gamma+\frac{1}{1-\omega_{e}^{2}/\omega_{o}^2} \bigg) \\
&\times \bigg[1+\frac{J_{2}R^{2}}{4b^{2}} \bigg( 4-\cos(2\varphi_{S})-\cos(\varphi_{R}-\varphi_{S})\\
&-\cos(2\varphi_{R})-\cos(\varphi_{S}+\varphi_{R}) \bigg)\bigg],
\end{aligned}
\end{equation}
the linear term in $m$ and the second order correction,
\begin{equation}\label{alpha2-homo}
\begin{aligned}
\alpha^{(2)}=&\frac{m^{2}}{4b^{2}(\omega_{o}^{2}-\omega_{e}^{2})^{2}}\bigg\{(\varphi_{S}-\varphi_{R})(\omega_{o}^{2}-\omega_{e}^{2}) \\
&\times \bigg( \omega_{o}^{2}(4\beta-8-8\gamma-3\nu)+3\nu\omega_{e}^{2} \bigg) \\
&+4 (\omega_{o}^{2}(1+\gamma)-\omega_{e}^{2}\gamma)^{2}\bigg(\sin(\varphi_{S})-\sin(\varphi_{R})\bigg) \\
&+\frac{1}{2}\bigg[ \omega_{o}^{4}\bigg( 4(\beta-1+\gamma^{2})-3\nu \bigg)+\omega_{e}^{4}(4\gamma^{2}-3\nu) \\
&-2\omega_{o}^{2}\omega_{e}^{2}(2\beta+4\gamma^{2}-3\nu) \bigg]\bigg( \sin(2\varphi_{R})-\sin(2\varphi_{S})\bigg) \\
&+ 8\omega_{o}^{2}\omega_{e}^{2}\cos(\varphi_{S})\bigg(\sin(\varphi_{R})-\sin(\varphi_{S})\bigg)\bigg\}.
\end{aligned}
\end{equation}
In \eqref{alpha2-homo}.we have used the approximation $r_{o}\approx b/\sin(\varphi_{R})$ which can be safely used at the considered order. 

Expressions \eqref{alpha1-homo} and \eqref{alpha2-homo} generalize previous known results in several ways. For the bending angle in a plasma environment these expressions take into account finite distance corrections, as well as second order effects in the mass and linear in the quadrupole moment. We are not aware of any previous derivations of these general expressions. 

Now, we will study some special cases of the above expressions which help to test their  validity and to give new relevant formulas for describing the lensing effects of the astrophysical objects under consideration.
\subsection{Special cases of \eqref{alpha1-homo} and \eqref{alpha2-homo}}
\subsubsection{Infinite distances case}
Let us consider the limit where the source and the observer are far away from the lens. In such a situation {we may take}
\begin{equation}\label{eq:tendencias}
\varphi_{R}\to \pi+\alpha^{(1)}_\infty\text{ and }  \varphi_{S}\to 0.
\end{equation}
{In principle, we should proceed as follows: first, as $\alpha^{(1)}_\infty$ is already $\mathcal{O}(m)$ we could compute it from \eqref{alpha1-homo} by taking $\varphi_R=\pi$ and $\varphi_S=0$. After that, as a second step, we should replace this obtained value for $\alpha^{(1)}_\infty$ in \eqref{eq:tendencias} in order to use again the Eq.\eqref{alpha1-homo} to obtain extra $\mathcal{O}(m^2)$ corrections that should be added to $\alpha^{(2)}_\infty$. However in the practice it is not necessary, because we have the following expansion in power of $m$ for the kind of trigonometric functions that appears in $\alpha^{(1)}$ (Eq. \eqref{alpha1-homo}): $\cos(n(x+\alpha^{(1)}_\infty))=\cos(nx)-n\sin(nx)\alpha^{(1)}_\infty+\mathcal{O}(m^2)$ with $n$ an integer number. In particular in our case we have $\cos(\varphi_R)=\cos(\pi+\alpha^{(1)}_\infty)=-1+\mathcal{O}(m^2)$. Therefore, the produced corrections will be $\mathcal{O}(m^3)$, which are not taken into account in our approximation. 
From the previous considerations, we see that we can replace $\varphi_R=\pi$ and $\varphi_S=0$}
into \eqref{alpha1-homo} and \eqref{alpha2-homo}, such that
the deflection angle for an astrophysical object described in the weak gravitational field for the PPN metric which takes into account the monopole and quadrupole gravitational moments is given by Eqs.\eqref{A1}, \eqref{A2} and \eqref{A3} reduces to:
\begin{widetext}
\begin{equation}\label{eqalphapp}
\begin{aligned}
\alpha=&\frac{2m(b^2+J_{2} R^{2})}{b^3}\bigg(\gamma+\frac{1}{1-\omega_{e}^{2}/\omega_{o}^{2} }\bigg)+ \frac{\pi m^{2}}{b^{2}}\bigg(\frac{2-\beta+2\gamma}{1-\omega_{e}^{2}/\omega_{o}^{2}}+\frac{3}{4}\nu\bigg).
\end{aligned}
\end{equation}
\end{widetext}
Despite the simplicity of this expression, it generalizes many recent results.
We have no knowledge of a previous presentation of this general formula. 

In particular, in the absence of plasma ($\omega_{e}=0$ or equivalently $\omega_e/\omega_{o}\ll 1$) the previous equation reduces to,
\begin{equation}
\alpha=2(\gamma+1)\frac{m}{b}+\pi (2-\beta+2\gamma+\frac{3}{4}\nu)\frac{m^{2}}{b^{2}},
\end{equation}
which coincides with the expression found in \cite{Epstein-1980,PhysRevD.97.084010}.

On the other hand, even considering the presence of the plasma, if the object under study is {a}  spherical mass ($J_2=0$) and the gravitational field is described by the Einstein general relativity theory ($\gamma=\nu=\beta=1$), then the equation \eqref{eqalphapp} reduces to
\begin{equation}
\alpha=\frac{2m}{b}\bigg(1+\frac{1}{1-\omega_{e}^{2}/\omega_{o}^{2}}\bigg)+\frac{3\pi}{4}\bigg(1+\frac{4}{1-\omega_{e}^{2}/\omega_{o}^{2}}\bigg)\frac{m^{2}}{b^{2}}.
\end{equation}
The first term of the previous expression agrees with the formula obtained for the first time by Bisnovatyi-Kogan and Tsupko in \cite{BisnovatyiKogan:2008yg,BisnovatyiKogan:2010ar}) and including the second term coincides with the result recently found by us in \cite{Crisnejo:2018uyn}.

\subsubsection{Schwarzschild metric at finite distances}\label{subsec:relationscoordinates}
The finite distance contributions for the bending angle in the presence of an homogeneous plasma in a Schwarzschild background 
follows by setting $\gamma=\beta=\nu=1$ and $J_2=0$ into equations \eqref{alpha1-homo} and \eqref{alpha2-homo}:
\begin{equation}
\begin{aligned}
\alpha^{(1)}=&\frac{m}{b}\bigg(1+\frac{1}{1-\omega_{e}^{2}/\omega_{o}^{2}}\bigg) \bigg(\cos(\varphi_S)-\cos(\varphi_R)\bigg);
\end{aligned}
\end{equation}
\begin{equation}
\begin{aligned}
\alpha^{(2)}=& \frac{m^2}{8b^2(\omega_{o}^2-\omega_e^2)^2}\bigg[ 6(\varphi_R-\varphi_S)(5\omega_{o}^{4}-6\omega_{o}^{2}\omega_{e}^{2}+\omega_{e}^{4}) \\
&+16\omega_{o}^{2}\omega_{e}^{2}\cos(\varphi_S)\sin(\varphi_R)-(\omega_{o}^{2}+\omega_{e}^{2})^{2} \sin(2\varphi_S)\\
&+ 8(\omega_{e}^{2}-2\omega_{o}^{2})^{2}\bigg( \sin(\varphi_{S})-\sin(\varphi_{R}) \bigg)\\
&+(\omega_{o}^{4}-6\omega_{o}^{2}\omega_{e}^{2}+\omega_{e}^{4})\sin(2\varphi_R) \bigg].
\end{aligned}
\end{equation}

These expressions generalize the  relations describing light deflection in a vacuum Schwarzschild spacetime recently found by Ishihara \emph{et.al.} in \cite{Ishihara:2016vdc} at first order in the mass and extended to second order by Ono \emph{et.al.} in \cite{Ono:2017pie}. In particular, in the absence of plasma or where the effect of the plasma enviornment is negligible ($\omega_e/\omega_o\ll 1$) these expressions reduce to the following vacuum values:
\begin{equation}
\begin{aligned}\label{aldas1}
\alpha^{(1)}_{\text{vac}}=&\frac{2m}{b} \bigg(\cos(\varphi_S)-\cos(\varphi_R)\bigg);\\
\end{aligned}
\end{equation}
\begin{equation}\label{aldas2}
\begin{aligned}
\alpha^{(2)}_{\text{vac}}=&\frac{m^2}{8b^2}\bigg[ 30(\varphi_R-\varphi_S) +\sin(2\varphi_R)-\sin(2\varphi_S)\\
&+32\bigg(\sin(\varphi_{S})-\sin(\varphi_{R})\bigg)\bigg].
\end{aligned}
\end{equation}
The expression \eqref{aldas1} is in complete agreement with the first order computation of the deflection angle  derived in reference \cite{Ishihara:2016vdc}. The analogous expression of \eqref{aldas2} has been computed in the appendix of reference \cite{Ono:2017pie}. However, note that even when there is perfect agreement between our first order expression \eqref{aldas1}
and the corresponding formula given by the authors in \cite{Ishihara:2016vdc}, it seems on first sight that there is an inconsistency between our second order correction as given by \eqref{aldas2} and the expression from the appendix of the article \cite{Ono:2017pie} which for the convenience of the reader and in order to differentiate from our expression \eqref{aldas2} we reproduce here under the alternative name of $\hat\alpha^{(2)}_{\text{vac}}$ and also with a $\hat{}$ in their angular variable $\varphi$:
\begin{equation}\label{ladeasad}
\begin{aligned}
\hat\alpha^{(2)}_{\text{vac}}=&\frac{m^2}{8b^2}\bigg[ 30(\hat\varphi_R- \hat\varphi_S)+\sin(2\hat\varphi_R)-\sin(2\hat\varphi_S)\bigg].
\end{aligned}
\end{equation}
It seems that an apparent discrepancy between \eqref{ladeasad} and \eqref{aldas2} appears, because of the following missing terms in \eqref{ladeasad}:
\begin{equation}
\delta=32\bigg(\sin(\varphi_{S})-\sin(\varphi_{R})\bigg);
\end{equation}
which is however present in \eqref{aldas2}. The difference is only apparent because the angular coordinate $\hat\varphi$ used by the authors of \cite{Ono:2017pie} is related to our $\varphi$ by
\begin{equation}\label{changephi}
\hat{\varphi}=\varphi-\frac{\alpha_\infty}{2}\approx \varphi-\frac{2m}{b}+\mathcal{O}(m^2).
\end{equation}
The transformation \eqref{changephi} follows from the fact that we have chosen the polar axis such that the orbit followed by a light ray which reaches the asymptotic region $r\to\infty$ 
(or, equivalently $u\to 0$) has the following angular coordinate behavior in this limit: $\varphi\to 0$ or  $\varphi\to\pi+\alpha_\infty$ (as can be seen from Eq.\eqref{uppn} with $\gamma=1$ and $\omega_e=0$). On the other hand, the authors of \cite{Ono:2017pie} choose the polar axis such that the closest approach of the light ray to the lens occurs when their angular coordinate $\hat\varphi$ takes the value $\hat\varphi=\pi/2$, resulting in a corresponding orbit which is symmetric with respect to the radial direction defined by $\hat\varphi=\pi/2$. As the total deflection angle at infinite distance is $\alpha_\infty$, the asymptotic points of the orbit occur when $\hat\varphi\to -\alpha_\infty/2$ (the position of an asymptotic source) or when $\hat\varphi\to \pi+\alpha_\infty/2$  (the position of an asymptotic receiver). Note that the difference between $\varphi$ and $\hat\varphi$ is $\mathcal{O}(m)$, and therefore $\alpha^{(2)}$ as given by Eq. \eqref{aldas2} preserves its form in terms of $\hat\varphi$. However, it also follows from the relation $\eqref{changephi}$ that at first order in $m$ we have
\begin{equation}\label{costrans}
\cos(\varphi)\approx\cos(\hat\varphi)-\frac{2m}{b}\sin{\hat\varphi}+\mathcal{O}(m^2).
\end{equation}
Hence, if we replace Eq.\eqref{costrans} into Eq.\eqref{aldas1}, it can be seen that new quadratic terms in $m$ appear as functions of the variable $\hat\varphi$ which  exactly cancel the apparent discrepant terms $\delta$ present in $\alpha^{(2)}_{\text{vac}}$. Therefore, when our expressions for the deflection angle are written in terms of the the angular coordinate $\hat\varphi$ of Ono \emph{et.al.} \cite{Ono:2017pie} the relation \eqref{ladeasad} is recovered.

\subsection{ Deflection angle in terms of the observable $\theta_{I}$ and comparison with previous particular known expressions}\label{subsection:thetaI}
Let us now compare between our finite distance results and the well known expressions from the literature \cite{Richter:1982zz,DeLuca:2003zz}. 
In order to do that we will assume the source is at infinite distance from the lens. In this case the deflection angle $\alpha^{(1)}$ and $\alpha^{(2)}$ take the following limits:
\begin{widetext}
\begin{equation}
\begin{aligned}
\alpha^{(1)}_{S_\infty}(b,\varphi_R):=&\lim_{\varphi_S\to 0}\alpha^{(1)}=\frac{m}{b}\bigg(1-\cos(\varphi_{R})\bigg)\bigg( \gamma+\frac{1}{1-\omega_{e}^{2}/\omega_{o}^2} \bigg)\times \bigg[1+\frac{J_{2}R^{2}}{4b^{2}} \bigg( 3-2\cos(\varphi_{R})
-\cos(2\varphi_{R})\bigg)\bigg],\label{eq:alf1lim}\\
\end{aligned}
\end{equation}
\begin{equation}\label{eq:alf2lim}
\begin{aligned}
\alpha^{(2)}_{S_\infty}(b,\varphi_R):=&\lim_{\varphi_S\to 0}\alpha^{(2)}=\frac{m^{2}}{4b^{2}(\omega_{o}^{2}-\omega_{e}^{2})^{2}}\bigg\{\varphi_{R}(\omega_{e}^{2}-\omega_{o}^{2}) 
\bigg( \omega_{o}^{2}(4\beta-8-8\gamma-3\nu)+3\nu\omega_{e}^{2} \bigg)\\
&-4 \sin(\varphi_{R})(\omega_{o}^{2}(1+\gamma)-\omega_{e}^{2}\gamma)^{2}+\frac{1}{2}\bigg[ \omega_{o}^{4}\bigg( 4(\beta-1+\gamma^{2})-3\nu \bigg)+\omega_{e}^{4}(4\gamma^{2}-3\nu) \\
&-2\omega_{o}^{2}\omega_{e}^{2}(2\beta+4\gamma^{2}-3\nu) \bigg]\sin(2\varphi_{R})+ 8\omega_{o}^{2}\omega_{e}^{2}\sin(\varphi_{R})\bigg\}.
\end{aligned}
\end{equation}
\end{widetext}
As seen from Fig.\eqref{finitedistance}, the following relation follows between the angular position of the receiver $\varphi_R$ and the angles $\theta_I$ and $\delta\theta$: 
\begin{equation}\label{vardelta}
\begin{aligned}
\varphi_{R}&=\pi-\theta_{I}+\delta\theta\\
&
=\pi-\theta_{I}+\alpha^{(1)}_{S_\infty}+\mathcal{O}(m^2)
.
\end{aligned}
\end{equation}
Finally, by replacing this relation into Eqs.\eqref{eq:alf1lim} and \eqref{eq:alf2lim} we obtain
\begin{widetext}
\begin{equation}\label{alpha1-homo-thetaI}
\begin{aligned}
\alpha^{(1)}_{S_\infty}(b,\theta_I)=&\frac{m}{b}\bigg(1+\cos(\theta_{I})\bigg)\bigg( \gamma+\frac{1}{1-\omega_{e}^{2}/\omega_{o}^2} \bigg) \bigg[1+\frac{ J_{2}R^{2}}{2b^{2}} \bigg( 2+\cos(\theta_{I})-\cos^{2}(\theta_{I}) \bigg)\bigg],
\end{aligned}
\end{equation}
 \begin{equation}\label{alpha2-homo-thetaI}
 \begin{aligned}
 \alpha^{(2)}_{S_\infty}(b,\theta_I)=&\frac{m^2}{8b^2(\omega_{o}^2-\omega_{e}^2)^2}\bigg[16 \omega_{o}^2\omega_{e}^2\sin(\theta_{I})+(\omega_{o}^2-\omega_{e}^2)\bigg(2(\pi -\theta_{I})+´\sin(2\theta_I) \bigg)\\
 &\times\bigg(\omega_{o}^2(-4 \beta+8\gamma+3\nu+8)- 3\nu\omega_{e}^2+8\sin (2\theta_{I})\omega_{o}^2\omega_{e}^2\bigg)\bigg].
 \end{aligned}
\end{equation}
\end{widetext}
Equations \eqref{alpha1-homo-thetaI}, and \eqref{alpha2-homo-thetaI} are the generalization of the relations \eqref{alpha1thetaI} and \eqref{alpha2thetaI} to the case of a PPN spacetime surrounded by a homogeneous plasma.

Alternatively, if we take into account the following relation between the impact parameter $b$ and the coordinate $r_o$ which follows from \eqref{uppn} and \eqref{vardelta} and generalizes the relation \eqref{b-ro}:
\begin{equation}\label{uppninv}
\begin{aligned}
\frac{1}{b}=&\frac{1}{r_o\sin(\theta_I)} - \frac{m}{r^2_o\sin(\theta_I)} \bigg( \gamma+\frac{1}{1-\omega_{e}^{2}/\omega_{o}^{2}} \bigg)\\
&+\mathcal{O}(m\times J_2,m^2),
\end{aligned}
\end{equation}
then Eqs. \eqref{alpha1-homo-thetaI}, and \eqref{alpha2-homo-thetaI} can be rewritten as:
\begin{equation}\label{alpha1-homo-thetaIv2}
\begin{aligned}
\alpha^{(1)}_{S_\infty}(r_o,\theta_I)=&\frac{m}{r_o}\frac{1+\cos(\theta_{I})}{\sin(\theta_I)}\bigg( \gamma+\frac{1}{1-\omega_{e}^{2}/\omega_{o}^2} \bigg) \\
&\times \bigg[1+\frac{ J_{2}R^{2}}{2r^{2}_o} \frac{ 2+\cos(\theta_{I})-\cos^{2}(\theta_{I})}{\sin^2(\vartheta_I)}\bigg],
\end{aligned}
\end{equation}
\begin{equation}\label{alpha2-homo-thetaIv3}
\begin{aligned}
\alpha^{(2)}_{S_\infty}(r_o,\theta_I)=&\frac{m^2}{r^2_o}\bigg\{\frac{1}{8(\omega_{o}^2-\omega_{e}^2)^2\sin^2(\theta_I)}\bigg[16 \omega_{o}^2\omega_{e}^2\sin(\theta_{I})\\
&+(\omega_{o}^2-\omega_{e}^2)\bigg(2(\pi -\theta_{I})+´\sin(2\theta_I) \bigg)\\
&\times\bigg(\omega_{o}^2(-4 \beta+8\gamma+3\nu+8)- 3\nu\omega_{e}^2\\
&+8\sin (2\theta_{I})\omega_{o}^2\omega_{e}^2\bigg)\bigg]\\
&-\frac{1+\cos(\theta_{I})}{\sin(\theta_I)}\bigg( \gamma+\frac{1}{1-\omega_{e}^{2}/\omega_{o}^2} \bigg)^2\bigg\}.
\end{aligned}
\end{equation}
In particular, {it} is easy to check that if $\omega_e=0$, or alternatively $\omega_e/\omega_0\ll 1$ then the equations \eqref{alpha1-homo-thetaI}, and \eqref{alpha2-homo-thetaI} or their alternative versions \eqref{alpha1-homo-thetaIv2}, \eqref{alpha2-homo-thetaIv3} reduce to the known expressions \eqref{alpha1thetaI} and \eqref{alpha2thetaI} by Richter and Matzner \cite{Richter:1982zz}. The advantage of relations \eqref{alpha1-homo-thetaIv2} and \eqref{alpha2-homo-thetaIv3} is that they are written in terms of physical observables.

It is very nice to see how starting with an 
elegant, geometrical and compact expression for 
the deflection angle as given by 
Eq.\eqref{eq:asad} well known formulas like 
\eqref{alpha1thetaI} and \eqref{alpha2thetaI} 
can be recovered. Moreover, we have not only 
confirmed for the first time 
the success of the Gauss-Bonnet formula 
\eqref{eq:asad} to recover known results of the 
angle deflection at finite distances (giving us confidence in that definition),  but also we have been able to generalize these results to more general astrophysical environments. In particular, from \eqref{alpha1-homo-thetaIv2} we see that the correction produced by a homogeneous plasma in the deflection angle even considering finite distances, is given by  a global factor $\gamma+\frac{1}{1-\omega_{e}^{2}/\omega_{o}^2}.$ This peculiar characteristic however does not remain if we consider the second order terms, in which case the plasma contribution is much more complicated. In particular, neglecting the quadrupole moment, and considering the validity of the Einstein equations we obtain that reduces to
\begin{equation}\label{eqobse}
\alpha^{(1)}_{S_\infty}(r_o,\theta_I)=\frac{m}{r_o}\frac{1+\cos(\theta_{I})}{\sin(\theta_I)}\bigg( 1+\frac{1}{1-\omega_{e}^{2}/\omega_{o}^2} \bigg).
\end{equation}
This expression can be compared with a similar relation obtained for the first time by Bisnovatyi-Kogan and Tsupko\cite{BisnovatyiKogan:2008yg,BisnovatyiKogan:2010ar} which reads
\begin{equation}\label{bys}
\alpha^{(1)}=\frac{2m}{b}\bigg( 1+\frac{1}{1-\omega_{e}^{2}/\omega_{o}^2} \bigg).
\end{equation}
The formula \eqref{bys} was obtained under the more common assumption of infinite distances. The advantage of \eqref{eqobse}, is that it is written in terms of the observable quantity $\theta_I$ and the coordinate distance $r_o$.

\section{Inhomogeneous plasma medium}\label{section5}
In this section, we focus on finite distance corrections to the deflection angle for light rays propagating in a non-uniform plasma. In this case, the steps that lead to the final expression for the deflection angle are basically the same which we applied in our previous article\cite{Crisnejo:2018uyn}, therefore we skip the intermediate computations and only present the essential steps. 

Let us consider an asymptotically flat and spherically symmetric gravitational lens surrounded by an inhomogeneous plasma whose electron number density $N(r)$ is a decreasing function of the radial coordinate $r$, and such that its radial derivative $N'(r)$ is also decreasing and smaller than $N(r)$. In isotropic coordinates, the components of the metric in the physical spacetime are codified in the following expressions:
\begin{equation}
A(r)=1-\mu h_{00}(r), \ B(r)=1+\epsilon h_{rr}(r), \ C(r)=r^2 B(r).
\end{equation}
The refractive index reads,
\begin{equation}
n(r)=\sqrt{1-\frac{\omega_{e}^{2} (1-\mu h_{00}(r)) }{\omega_{\infty}^{2}}},
\end{equation}
where as before, $\omega_\infty$ is related to the detected frequency by a receiver by
$\omega_\infty=\omega_o\sqrt{A(r_o)}$.

The associated optical metric is given by,
\begin{equation}\label{h00-inhom}
d\sigma^{2}=\bigg(\frac{(1+\epsilon h_{rr})(\omega_{\infty}^{2}-\omega_{e}^{2}+\mu \omega_{e}^{2} h_{00})}{\omega_{\infty}^{2}(1-\mu h_{00})}\bigg)(dr^{2}+r^{2}d\varphi^{2}).
\end{equation}
In general, the change in the deflection angle due to the presence of the refractive index is smaller than the main part due to the purely gravitational effect. We will assume as in \cite{BisnovatyiKogan:2010ar,Crisnejo:2018uyn} that the deflection angle is small and therefore as a first approximation the path followed by the light ray can be taken as the straight line geodesic of the flat euclidean space. We also neglect all higher order terms of the form $\mathcal{O}(N'^2,\mu N',\mu N'',\epsilon N'^2,\epsilon N'')$.
                      
Working at linear order in $\mu$ and $\epsilon$, and following the same steps explained in detail in \cite{Crisnejo:2018uyn}, we obtain for $\mathcal{K}dS$ (expressed in terms of the detected frequency $\omega_o$ by the receiver):
\begin{equation}
\mathcal{K}dS= \frac{1}{2}\bigg[\frac{K_{e}(rN{'}){'}}{\omega_{o}^{2}-\omega_{e}^{2}}  - \frac{\omega_{o}^{2}(rh_{00}{'}){'} }{\omega_{o}^{2}-\omega_{e}^{2}} \mu - (r h_{rr}{'}){'} \epsilon\bigg]dr d\varphi.
\end{equation}
By inserting this expression into Eq.\eqref{eq:asad}, we find that the deflection angle in this approximation is given by
\begin{widetext}
\begin{equation}\label{alpha-inhomo-GEN}
\begin{aligned}
\alpha&\approx -\lim_{R\to\infty}\int\int_{D_r} \mathcal{K}dS=-\int^{\varphi_R}_{\varphi_S}\int^\infty_{b/\sin\varphi} \frac{1}{2}\bigg[\frac{K_{e}(rN{'}){'}}{\omega_o^{2}-\omega_{e}^{2}}  - \frac{\omega_{o}^{2}(rh_{00}{'}){'} }{\omega_{o}^{2}-\omega_{e}^{2}} \mu - (r h_{rr}{'}){'} \epsilon\bigg]dr d\varphi.
\end{aligned}
\end{equation}
\end{widetext}
Using integration by parts in the first two terms of the radial integral and neglecting terms of order $\mathcal{O}(N'^2,h_{00}N')$, we obtain the final expression:
\begin{equation}\label{angletsu}
\alpha\approx\int^{\varphi_R}_{\varphi_S} \frac{1}{2}\bigg[\frac{K_{e}(rN{'})}{\omega_{o}^{2}-\omega_{e}^{2}} -  \frac{\omega_{o}^{2}(rh_{00}{'})}{\omega_{o}^{2}-\omega_{e}^{2}} \mu -(r h_{rr}{'})\epsilon\bigg]\bigg|_{r=b/\sin\varphi}d\varphi.
\end{equation}
This equation gives us a general formula to compute the deflection angle in a spherically symmetric spacetime when an inhomogeneous plasma medium is present taking into account finite distance corrections. Note that this expression can also be derived with the technique used by Bisnovatyi-Kogan and Tsupko in \cite{BisnovatyiKogan:2010ar}, where they find the deflection angle considering infinite distances by solving the Hamilton equations perturbatively for a not necessarily spherically symmetric metric of the form $g_{\alpha\beta}=\eta_{\alpha\beta}+h_{\alpha\beta}$.

  By assuming the condition $\omega_e/\omega_o\ll 1$ and motivated by the decomposition presented by Bisnovatyi-Kogan and Tsupko in \cite{BisnovatyiKogan:2010ar} for the deflection angle, Eq. \eqref{angletsu} can be decomposed in terms of the form:
\begin{equation}\label{alphadecomp}
\alpha = \alpha_{1} + \alpha_{2} + \alpha_{3} + \alpha_{4},
\end{equation}
where,
\begin{eqnarray}\label{alpha1-inhomo}
\alpha_{1}&=&-\frac{1}{2} \int_{\varphi_{S}}^{\varphi_{R}} r_\varphi\bigg( h'_{rr}(r_{\varphi})  \epsilon+  h_{00}^{{'}}(r_{\varphi})  \mu\bigg) \ d\varphi, \\
\label{alpha2-inhomo}
\alpha_{2}&=&-\frac{\mu}{2\omega_{o}^{2}}\int_{\varphi_{S}}^{\varphi_{R}} r_{\varphi}  h_{00}^{{'}}(r_{\varphi})  \omega_{e}^{2}(r_{\varphi}) d\varphi,\\
\label{alpha3-inhomo}
\alpha_{3}&=&\frac{K_{e}}{2\omega_{o}^{2}} \int_{\varphi_{S}}^{\varphi_{R}} r_{\varphi} N{'}(r_{\varphi}) d\varphi,\\
\label{alpha4-inhomo}
\alpha_{4} &=&\frac{K_{e}}{2\omega_{o}^{4}}\int_{\varphi_{S}}^{\varphi_{R}} r_{\varphi} N{'}(r_{\varphi}) \omega_{e}^{2}(r_{\varphi}) d\varphi,
\end{eqnarray}
and $r_{\varphi}=b/\sin(\varphi)$. 
These expressions are the finite distance counterparts of the expressions found by \cite{BisnovatyiKogan:2010ar}. In particular the first term $\alpha_1$ is the pure gravitational deflection angle; the second term $\alpha_2$ is a correction of the first due to the presence of the plasma, the third term is the pure refractive angle (present even without a gravitational field), and the last term is a correction  to the third term. 
As explained by Bisnovatyi-Kogan and Tsupko in \cite{BisnovatyiKogan:2010ar} in general astrophysical situations the first and the third terms in \eqref{alphadecomp} make the main contribution to the deflection angle, where in general $\alpha_3<\alpha_1$.

\subsection{Diverging lensing due solely to an inhomogeneous plasma medium}
An interesting example arises when we consider the limit in which gravity does not affect the deflection at all, $m\rightarrow 0$. In this case, spacetime is Minkowski and the perturbed components of the metric $h_{ij}$ vanish, which causes Eqs. \eqref{alpha1-inhomo} and \eqref{alpha2-inhomo} to vanish. This leaves only $\alpha_3$ and $\alpha_4$ to contribute to the deflection angle. These contributions are in the opposite sense to the gravitational deflection from $\alpha_1$ and $\alpha_2$, and therefore the lensing effect due to inhomogeneous plasma in the absence of gravitation is diverging, rather than the usual converging behavior of a gravitational lens \cite{2018MNRAS.475..867E}.

Lensing due to plasma is relevant to the observation of radio sources through the intervening interstellar medium (ISM), which can contain inhomogeneities in the electron density. These density perturbations can act as diverging lenses, causing frequency-dependent dimming of radio sources observed through the ISM. A number of astrophysical phenomena are associated with this type of lensing, including extreme scattering events \cite{AR_TUNTSOV16,AR_ROMANI,AR_FIEDLER,AR_CLEGGFEY} as well as pulsar scintillation \cite{AR_PEN1,AR_PEN2}. Furthermore, it has recently been suggested that plasma lensing may play a role in the mechanism responsible for generating fast radio bursts \cite{AR_CORDES}.

In terms of the derivation presented here, the inhomogeneous plasma case is particularly interesting due to the fact that it depends only on the Minkowski metric. In this case, the effect of gravitation is totally removed from the problem, which ultimately illustrates the elegant utility of the Gauss-Bonnet method when coupled to a Riemann optical metric representation.

\subsection{Schwarzschild spacetime with plasma density profile of the form $N(r)=N_or^{-h}$}

Now we will apply this general result to the case of Schwarzschild spacetime with the density profile
\begin{equation}
N(r)=N_or^{-h}, \ \ \ h>0.
\end{equation}
For this, we make the following identification,
\begin{equation}
\epsilon = \mu = m, \ \
h_{00}=h_{rr}=\frac{2}{r}.
\end{equation}
For completeness we write the expression for each individual term \eqref{alpha1-inhomo}, \eqref{alpha2-inhomo}, \eqref{alpha3-inhomo}, \eqref{alpha4-inhomo} but, as discussed, the main contribution to the deflection angle is given by $\alpha_1$ and $\alpha_3$. Explicitly, these terms read:
\begin{widetext}
\begin{equation}\label{a1F}
\begin{aligned}
\alpha_{1}=&\frac{2m}{b}\bigg(\cos(\varphi_{S})-\cos(\varphi_{R})\bigg),
\end{aligned}
\end{equation}
\begin{equation}\label{a2F}
\alpha_{2}=\frac{mK_{e}N_{o}}{\omega_{o}^{2}b^{h+1}}\bigg[\cos(\varphi_{S}) \ {}_{2}F_{1}\bigg(\frac{1}{2},-\frac{h}{2};\frac{3}{2};\cos^{2}(\varphi_{S})\bigg)-\cos(\varphi_{R}) \ {}_{2}F_{1}\bigg(\frac{1}{2},-\frac{h}{2};\frac{3}{2};\cos^{2}(\varphi_{R})\bigg)\bigg],
\end{equation}
\begin{equation}\label{a3F}
\alpha_{3}=-\frac{K_{e}N_{o}h}{2\omega_{o}^{2}b^{h}}\bigg[\cos(\varphi_{S}) \ {}_{2}F_{1}\bigg(\frac{1}{2},\frac{1-h}{2};\frac{3}{2};\cos^{2}(\varphi_{S})\bigg)-\cos(\varphi_{R}) \ {}_{2}F_{1}\bigg(\frac{1}{2},\frac{1-h}{2};\frac{3}{2};\cos^{2}(\varphi_{R})\bigg)\bigg],
\end{equation}
\begin{equation}\label{a4F}
\alpha_{4}=-\frac{K_{e}^{2}N_{o}^{2}h}{2\omega_{o}^{4}b^{2h}}\bigg[\cos(\varphi_{S}) \ {}_{2}F_{1}\bigg(\frac{1}{2},\frac{1-2h}{2};\frac{3}{2};\cos^{2}(\varphi_{S})\bigg)-\cos(\varphi_{R}) \ {}_{2}F_{1}\bigg(\frac{1}{2},\frac{1-2h}{2};\frac{3}{2};\cos^{2}(\varphi_{R})\bigg)\bigg],
\end{equation}
\end{widetext}
with ${}_2F_1(a,b;c;x)$ the ordinary hypergeometric function\cite{book:112863}. 

These analytic and closed expressions generalize the known equivalent formulas for the infinite distance case. In particular we can also recover the expressions for the infinite distance case by taking the limit of the previous expressions in the limit of $\varphi_S\to\ 0$ and $\varphi_R\to \pi+\mathcal{O}(m)$. In this case, they reduce to:
\begin{equation}
a_1^{\infty}=\frac{4m}{b},
\end{equation}
\begin{equation}
\alpha_2^{\infty}=\frac{\sqrt{\pi } m  K_e N_o \Gamma \left(\frac{h}{2}+1\right)}{b^{h+1}\omega_{o}^2 \Gamma \left(\frac{h+3}{2}\right)},
\end{equation}
\begin{equation}\label{alp3inf}
\alpha_3^{\infty}=-\frac{\sqrt{\pi }  K_e N_o \Gamma \left(\frac{h+1}{2}\right)}{b^{h}\omega_{o}^2 \Gamma \left(\frac{h}{2}\right)},
\end{equation}
\begin{equation}
\alpha_4^{\infty}=-\frac{\sqrt{\pi }  K_e^2 N_o^2 b^{-2 h} \Gamma \left(h+\frac{1}{2}\right)}{2 \omega_{o}^4 \Gamma (h)},
\end{equation}
where $\Gamma(x)$ is the Gamma function. Expression \eqref{alp3inf} was found by first time by Giampieri\cite{giampieri1996} and rederived by Bisnovatyi-Kogan and Tsupko in \cite{BisnovatyiKogan:2010ar}.
As an application, we will study in the next subsection a particular charge density profile which describes the plasma in our solar system.

\subsubsection{A particular plasma model for solar corona}
Let us consider the following electronic density profile particular model for the solar plasma neglecting latitude variations \cite{Tyler-1977,giampieri1996,Bertotti-1998,Turyshev-2018-b},
\begin{equation}
    N(r)=\bigg[ C_2 \bigg(\frac{R_\odot}{r}\bigg)^{2}+C_6 \bigg(\frac{R_\odot}{r}\bigg)^{6}+C_{16} \bigg(\frac{R_\odot}{r}\bigg)^{16}\bigg]\text{cm}^{-3},
\end{equation}
with
\begin{eqnarray}
C_2 &=& 3.44\times 10^5, \\
C_6 &=& 1.55\times 10^8, \\
C_{16} &=& 2.99\times 10^8,
\end{eqnarray}
and $r\geq R_\odot$. The value of coefficients $C_i$ have been empirically determined in the past. We will only take into account the main contribution of the plasma, that is, $\alpha_3$. Considering the case in which the source is far away from the lens and taking $\varphi_R=\pi-\theta_I$, the main contribution of the plasma is given by (see also Appendix \eqref{Bp}),
\begin{widetext}
\begin{equation}\label{alpha3-sun-2}
\begin{aligned}
\alpha_3 =&-\frac{1}{(2\pi)^2}\bigg(\frac{1\text{Hz}}{f}\bigg)^2\bigg[ \frac{\tilde{C}_2}{2}\bigg(\frac{R_\odot}{r_o}\bigg)^2 \ \frac{\pi-\theta_I+\cos(\theta_I)\sin(\theta_I)}{\sin^2(\theta_I)}+\frac{\tilde{C}_6}{64} \bigg(\frac{R_\odot}{r_o}\bigg)^6 \ \bigg(60(\pi-\theta_I)+45\sin(2\theta_I)-9\sin(4\theta_I)\\
&+\sin(6\theta_I)\bigg)\times\frac{1}{\sin^6(\theta_I)}+\frac{\tilde{C}_{16}}{458752} \bigg(\frac{R_\odot}{r_o}\bigg)^{16}\bigg( 720720(\pi-\theta_I)+640640\sin(2\theta_I)-224224\sin(4\theta_I)+81536\sin(6\theta_I)\\
&-25480\sin(8\theta_I)
+6272\sin(10\theta_I)-1120\sin(12\theta_I)+128\sin(14\theta_I)-7\sin(16\theta_I) \bigg) \times \frac{1}{\sin^{16}(\theta_I)}
\bigg],
\end{aligned}    
\end{equation}
\end{widetext}
where
\begin{eqnarray}
\tilde{C}_2 &=& (5.64\times 10^4)^2 \ C_2=1.09\times 10^{15} \ , \\
\tilde{C}_6 &=& (5.64\times 10^4)^2 \ C_6=4.93\times 10^{17} \ , \\
\tilde{C}_{16} &=& (5.64\times 10^4)^2 \ C_{16}=9.51\times 10^{17} \ .
\end{eqnarray}
In this way, the main plasma contribution given by \eqref{alpha3-sun-2} is only expressed in terms of observable quantities: the elongation angle $\theta_I$ and the distance Sun-Earth $r_o$. It is easy to check that if we consider observations from the Earth at small impact parameters of the order of a few solar radius (which implies that $\Theta_I\approx 0$ and the infinite distance expression is a good approximation), the formula \eqref{alpha3-sun-2} reduces to the expression given by Giampieri\cite{giampieri1996} (see also \cite{Bertotti-1998}) and recently rederived by Turyshev and Toth in \cite{Turyshev-2018-b} using the full Maxwell equations. More precisely, in such an approximation, eq.\eqref{alpha3-sun-2} reduces to the following expression which agrees with eq.(185) of reference \cite{Turyshev-2018-b} {(see also eq. (123) of \cite{Turyshev-2018-bc} which is the published version of \cite{Turyshev-2018-b} and also \cite{Turyshev:2018gjj,Turyshev:2018dzp} 
of the same authors where the same 
expression appears but with a factor of 2 
difference due to their treatment of the 
plasma as two one-way contributions: a) on a way in the solar system and b) on a way 
out of the solar system, and therefore 
they consider a one-way deflection angle. 
In particular in \cite{Turyshev:2018gjj} 
the authors also describe the phase 
evolution of a plane
wave propagating in the vicinity of a 
massive body in the presence of 
plasma)}. In order to 
make the comparison more simple we have re-written $\alpha_3$ in terms of the 
wavelength $\lambda=c/f$ and the impact 
parameter $b=r_o\sin\theta_I$, 
\begin{widetext}
\begin{equation}\label{alpha3-sun-lambda}
\begin{aligned}
\alpha_{3\infty} =&-\bigg(\frac{1\text{Hz}}{f}\bigg)^2\bigg[ \frac{\tilde{C}_2}{8\pi}\bigg(\frac{R_\odot}{b}\bigg)^2+\frac{15\tilde{C}_6}{64\pi} \bigg(\frac{R_\odot}{b}\bigg)^6+\frac{6435\tilde{C}_{16}}{16384\pi} \bigg(\frac{R_\odot}{b}\bigg)^{16}
\bigg]\\
=&-\bigg(\frac{\lambda}{1\mu\text{m}}\bigg)^2\frac{1}{(2.99792458\times 10^{14})^2}\bigg[ \frac{\tilde{C}_2}{8\pi}\bigg(\frac{R_\odot}{b}\bigg)^2+\frac{15\tilde{C}_6}{64\pi} \bigg(\frac{R_\odot}{b}\bigg)^6+\frac{6435\tilde{C}_{16}}{16384\pi} \bigg(\frac{R_\odot}{b}\bigg)^{16}\bigg]\\
=&-\bigg(\frac{\lambda}{1\mu\text{m}}\bigg)^2\bigg[ 4.82\times 10^{-16}\bigg(\frac{R_\odot}{b}\bigg)^2+4.09\times 10^{-13} \bigg(\frac{R_\odot}{b}\bigg)^6+1.32\times 10^{-12} \bigg(\frac{R_\odot}{b}\bigg)^{16}
\bigg],
\end{aligned}    
\end{equation}
\end{widetext}

\begin{figure}[H]
\centering
\includegraphics[clip,width=85mm]{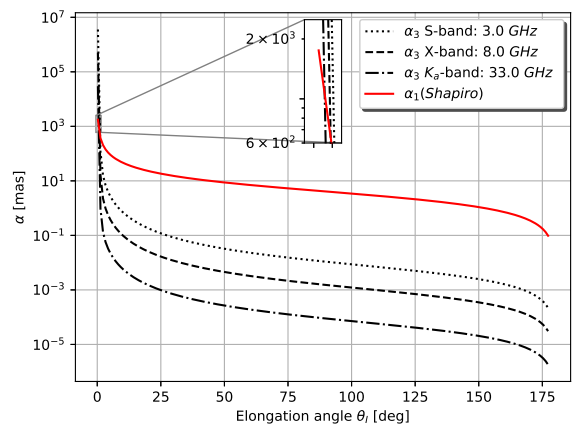}
\caption{Plasma contribution to the deflection angle (with opposite sign) as function of the elongation angle $\Theta_I$ for three different frequencies at different bands. For completeness the pure gravity contribution given by the Shapiro expression with the Eddington parameter $\gamma=1$ is also shown. In the zoomed region is shown the values of the deflection angles in the neighborhood of the Sun surface. }\label{solarangle}
\end{figure}

Coming back to the expression \eqref{alpha3-sun-2}, in Fig. \eqref{solarangle} we plot its contribution (with opposite 
sign) as well as the deflection angle purely due to gravity $\alpha_1$ for a range of different frequencies. Depending on the 
frequency-band, the plasma contribution can be of the 
order of $1\mu\text{as}$ even for large elongation angles, 
and of the order of $1\text{mas}$ for frequencies in the S-band up to elongation angles as large as $25\degree$. Of course, 
at smaller frequencies the contribution is more significant. 
Finally, let us remark that we can also use the same 
expression \eqref{alpha3-sun-2} to compute the deflection 
angle of a reference source different to our Sun if it is 
contained in the ecliptic plane (in order that the 
electronic charge density be a good model), and therefore 
to use the relations discussed in Section 
\eqref{subsection:twoimages} (in particular eq.\eqref{eq:mostsimple}) to compute the separation 
angle between the sources and its variation at different 
times depending of the relative position between these 
sources and our Sun as seen from the Earth in their orbit 
around it. 

\section{Final remarks}
In conclusion, our calculations achieve three main goals. First, by carefully constructing a finite quadrilateral region to apply the Gibbons-Werner method, we have resolved an apparent contradiction in the literature when they are applied to static and spherically symmetric asymptotically flat spacetimes.  Second, our results are derived in terms of observable quantities that facilitate comparison with previous, well-studied cases in the literature. Third, by making use of the Gibbons-Werner approach of coupling a Riemann optical metric to the Gauss-Bonnet method, we have expanded on well-known cases in the literature. For example, by including the effects of the PPN expansion and a possible quadrupole moment into the case of a homogeneous plasma in a gravitational field, as well as including the corrections arising from consideration of the finite distance between source, lens and observer. This work demonstrates the utility and elegance of the Gauss-Bonnet theorem and the Gibbons-Werner method and their relevance for all forms of lensing - both gravitational (converging) and plasma (diverging).

\section*{Acknowledgments}
We thanks to Slava G. Turyshev for helpful comments that allowed to us to improve our paper. E.G. acknowledges support from CONICET and SeCyT-UNC. A.R. acknowledges support from Dr. Samar Safi-Harb at the University of Manitoba.

\appendix

\section{}\label{Ap}
\subsection{Explicit comparison between formulas \eqref{eq:asad}, \eqref{def1} and \eqref{eqmain}}
In this Appendix we will use a particular example to illustrate how the alternative expressions for the deflection angle given by Eqs.\eqref{eq:asad}, \eqref{def1} and \eqref{eqmain} give the same results.
Let us focus on a Schwarzschild metric written in isotropic coordinates: 
\begin{equation}
\begin{aligned}
ds^2=&-\bigg(\frac{1-\frac{m}{2r}}{1+\frac{m}{2r}}\bigg)^2 dt^2+\bigg(1+\frac{m}{2r}\bigg)^4\\&
\times\bigg[dr^2+r^2\bigg(d\vartheta^2+\sin^2(\vartheta)d\varphi^2\bigg)\bigg]. 
\end{aligned}
\end{equation}
Let us focus on the plane defined by $\vartheta=\pi/2$. We shall calculate the deflection angle to second order precision in the mass $m$. However, for the moment we will write the exact relationships. The associated optical metric is given by:
\begin{equation}\label{eq:optschiso}
d\sigma^2=\frac{(1+\frac{m}{2r})^6}{(1-\frac{m}{2r})^2}\bigg(dr^2+r^2d\varphi^2\bigg);
\end{equation}
and the associated Gaussian curvature reads:
\begin{equation}
\mathcal{K}=-\frac{128mr^3(4r^2-2rm+m^2)}{(2r+m)^8}.
\end{equation}
The surface element is
\begin{equation}
dS=\sqrt{\det(g^{\text{opt}}_{ij})}drd\varphi=
\frac{(2r+m)^{6}}{16r^3(2r-m)^2}drd\varphi.
\end{equation}
Therefore at second order in the mass we obtain
\begin{equation}
\mathcal{K}dS=\bigg(-\frac{2m}{r^2}+\frac{m^2}{r^3}\bigg)drd\varphi+\mathcal{O}(m^3),
\end{equation}
which can be rewritten in terms of the variable $u=1/r$ as
\begin{equation}
\mathcal{K}dS=\bigg({2m}-{m^2}{u}\bigg)dud\varphi+\mathcal{O}(m^3).
\end{equation}
Now, we will use equation \eqref{eqmain} to compute the deflection angle. To do that, we must integrate over a region $\tilde{D}_r$ bounded by the radial curves $\gamma_R$, $\gamma_S$, the geodesic $\tilde{\gamma}_{\ell}$ and the arc of circle segment $\gamma_C$. $\gamma_R$ and $\gamma_S$ are given by $\varphi=\varphi_R$ and $\varphi=\varphi_S$, respectively. In terms of the coordinate $u$, $\gamma_C$ is defined by $u=1/r_C=\text{constant}$, and finally the spatial geodesic $\tilde{\gamma}_{\ell}$ describing the orbit of a light ray between $S$ and $R$ is given by:
\begin{equation}\label{uint}
\begin{aligned}
u_\ell=&\frac{\sin(\varphi)}{b}+\frac{2m(1-\cos(\varphi))}{b^2}\\
&+\frac{15m^2\cos(\varphi)(\tan(\varphi)-\varphi)}{4b^3}+\mathcal{O}(m^3).
\end{aligned}
\end{equation}
This expression for $u_\ell$ follows by solving Eq.\eqref{orbit-eq-u} at second order in the mass,
\begin{equation}\label{eq-orb-apA}
\bigg(\frac{du_\ell}{d\varphi}\bigg)^2=\frac{1}{b^2}-u_\ell^2+\frac{4 m u_\ell}{b^2}+ \frac{15 m^2 u_\ell^2}{2b^2},
\end{equation}
with the asymptotic condition,
\begin{equation}\label{orbit-eq-u-id2}
\lim_{\varphi\to 0}u_\ell(\varphi)=0.
\end{equation}
Note however, that in order to compute the integral of the Gaussian curvature we only need to consider the first two terms in \eqref{uint} because $\mathcal{K}dS$ is already order $m$.
Hence we obtain for the first term of $\eqref{eqmain}$:

\begin{widetext}
\begin{equation}\label{eq:Kdscalciso}
\begin{aligned}
-\int\int_{\tilde{D}_r}\mathcal{K}dS=&\int^{\varphi_R}_{\varphi_S}\int^{\frac{\sin(\varphi)}{b}+\frac{2m(1-\cos(\varphi))}{b^2}}_{1/r_C}\bigg({2m}-{m^2}{u}\bigg)dud\varphi+\mathcal{O}(m^3)\\
=&\frac{2m}{b}\bigg[\cos(\varphi_S)-\cos(\varphi_R)+\frac{b}{r_C}(\varphi_S-\varphi_R)\bigg]\\
&+\frac{m^2}{8b^2}\bigg[30(\varphi_R-\varphi_S)+\sin(2\varphi_R)-\sin(2\varphi_S)+32(\sin(\varphi_S)-\sin(\varphi_R))+\frac{4b^2}{r^2_C}(\varphi_R-\varphi_S)\bigg]+\mathcal{O}(m^3).
\end{aligned}
\end{equation}

We must also compute the other two terms of \eqref{eqmain}. The last term is computed in the Euclidean metric and it is simply given by Eq.\eqref{eq:kappaeuc}. In order to compute the second term we need first to compute the geodesic curvature of $\tilde{\gamma}_C$ defined by $r=r_C=\text{constant}$. The exact value of this curvature in the optical metric \eqref{eq:optschiso} is given by
\begin{equation}
\tilde{\kappa}_{\tilde{\gamma}_C}=\frac{4r_C[4r_C(r_C-2m)+m^2]}{(2r_C+m)^4}
\end{equation}
and therefore we obtain for the second term of \eqref{eqmain}
\begin{equation}\label{eq:kapisotcal}
\begin{aligned}
-\int_{\tilde{\gamma}_{C (S\to R)}}  \tilde\kappa\;d\tilde\sigma=& \int^{\varphi_S}_{\varphi_R}  \tilde\kappa_{\tilde{\gamma}_C}\;\sqrt{g^{\text{opt}}_{\varphi\varphi}}d\varphi=\int^{\varphi_S}_{\varphi_R} \bigg(1-\frac{2}{r_C}+\frac{m^2}{2r^2_C}\bigg)d\varphi+\mathcal{O}(m^3)\\
=&\varphi_S-\varphi_R+\frac{2m}{r_C}(\varphi_R-\varphi_S)+\frac{m^2}{2r^2_C}(\varphi_R-\varphi_S)+\mathcal{O}(m^3)
\end{aligned}
\end{equation}
Taking cognizance of Eqs.\eqref{eq:kappaeuc}, \eqref{eq:Kdscalciso} and \eqref{eq:kapisotcal}, and replacing these expressions into the formula \eqref{eqmain} we obtain:
\begin{equation}\label{eq:disfint1}
\begin{aligned}
\alpha=&\frac{2m}{b}\bigg[\cos(\varphi_S)-\cos(\varphi_R)\bigg]+\frac{m^2}{8b^2}\bigg[30(\varphi_R-\varphi_S)+\sin(2\varphi_R)-\sin(2\varphi_S)+32(\sin(\varphi_S)-\sin(\varphi_R))\bigg]+\mathcal{O}(m^3),
\end{aligned}
\end{equation}
\end{widetext}
which agrees with our previous expression given by \eqref{aldas1} and \eqref{aldas2} obtained directly using Eq.\eqref{eq:asad}.

Now we will repeat the computation, but using the expression given by \eqref{def1}. To do that we must compute the sum of angles in the regions $D_r$ on the Euclidean space, and also in the region $\tilde{D}_r$ of the optical space. 

In the Euclidean space, it is easy to see that the sum of the interior angles is:
\begin{equation}\label{eq:epsiiso}
\sum_i\epsilon_i=2\pi+\varphi_R-\varphi_S.
\end{equation}
Instead, for the region $\tilde{D}_r$ in the optical metric, we have
\begin{equation}\label{eq:epsitildeiso}
\sum_i\tilde\epsilon_i=\pi+\tilde\epsilon_1+\tilde\epsilon_2;
\end{equation}
with $\tilde\epsilon_1$ and $\tilde\epsilon_2$ the angles formed by the curve $\tilde{\gamma}_{\ell}$ and the radial curves $\tilde{\gamma}_R$ and $\tilde\gamma_{S}$ respectively.
As we wish to compute the deflection angle to second order in the mass, we need to use the expression for  orbital equation which describes $\tilde\gamma_{\ell}$ with all the terms that appear in Eq.\eqref{uint}.
\begin{widetext}
The angle $\tilde{\epsilon}_1$ can be computed from the following relation (see Fig.\eqref{grafico-no-flat} )
\begin{equation}\label{eq:A1}
\begin{aligned}
\tan{\tilde\epsilon_1}=&-\bigg[\frac{\sqrt{g^{\text{opt}}_{\varphi\varphi}}}{\sqrt{g^{\text{opt}}_{rr}}}\frac{d\varphi}{dr}\bigg]\bigg|_{\tilde\gamma_{\ell}(\varphi_R)}=&-\bigg[r\frac{d\varphi}{dr}\bigg]\bigg|_{\tilde\gamma_{\ell}(\varphi_R)}=
u(\varphi_R)  
{\frac{d\varphi}{du}\bigg|_{\varphi=\varphi_R}}\end{aligned},  
\end{equation}
and similarly, $\tilde{\epsilon}_2=\pi-\tilde{\chi}_2$ with $\tilde\chi_2$ the supplementary angle to $\tilde{\epsilon}_2$ which satisfies: 
\begin{equation}\label{eq:A2}
\begin{aligned}
\tan{\tilde\chi_2}=&-\bigg[\frac{\sqrt{g^{\text{opt}}_{\varphi\varphi}}}{\sqrt{g^{\text{opt}}_{rr}}}\frac{d\varphi}{dr}\bigg]\bigg|_{\tilde\gamma_{\ell}(\varphi_S)}=&-\bigg[r\frac{d\varphi}{dr}\bigg]\bigg|_{\tilde\gamma_{\ell}(\varphi_S)}=
u(\varphi_S)  
{\frac{d\varphi}{du}\bigg|_{\varphi=\varphi_S}}.
\end{aligned}    
\end{equation}
Using \eqref{uint}, we obtain that 
\begin{equation}
\tan\tilde{\epsilon}_1=u(\varphi_R)  
{\frac{d\varphi}{du}\bigg|_{\varphi=\varphi_R}}
=\tan(\varphi_R)-\frac{2m}{b}\frac{1-\cos(\varphi_R)}{\cos^2(\varphi_R)}+\frac{m^2}{8b^2}\frac{15(\sin(2\varphi_R)-2\varphi_R)\cos(\varphi_R)-16\sin(2\varphi_R)+32\sin(\varphi_R)}{\cos^3(\varphi_R)}. 
\end{equation}
\end{widetext}
Hence, to the considered order we obtain:
\begin{equation}
\begin{aligned}
\tilde{\epsilon}_1=&\varphi_R-\frac{2m}{b}(1-\cos(\varphi_R))\\&
-\frac{m^2}{8b^2}(30\phi_R+\sin(2\varphi_R)-32\sin(\varphi_R))+\mathcal{O}(m^3).
\end{aligned}
\end{equation}
Repeating for $\tilde{\chi}_2$, which is obtained from \eqref{eq:A2} we find:
\begin{equation}
\begin{aligned}
\tilde{\chi}_2=&\varphi_S-\frac{2m}{b}(1-\cos(\varphi_S))\\&
-\frac{m^2}{8b^2}(30\phi_S+\sin(2\varphi_S)-32\sin(\varphi_S))+\mathcal{O}(m^3).
\end{aligned}
\end{equation}
Therefore, we arrive at
\begin{equation}\label{eq:sumepsiti}
\begin{aligned}
\sum_i\tilde{\epsilon}_i=&2\pi+\tilde{\epsilon}_1-\tilde{\chi}_2\\
=&2\pi+\varphi_R-\varphi_S+\frac{2m}{b}(\cos(\varphi_R)-\cos(\varphi_S))\\
&+\frac{m^2}{8b^2}\bigg[30(\varphi_S-\varphi_R)+\sin(2\varphi_S)-\sin(2\varphi_R)\\
&+32(\sin(\varphi_R)-\sin(\varphi_S))\bigg].
\end{aligned}
\end{equation}
Finally, by replacing \eqref{eq:epsiiso} and \eqref{eq:sumepsiti} into the angular definition for the deflection angle (given by \eqref{def1}), we recover the expression \eqref{eq:disfint1}.

As a final comment let us note that Ishihara \emph{et.al} also express the angle deflection in terms of  two angles $\Psi_S$ and $\Psi_R$ and the coordinate angle $\varphi_{RS}=\varphi_R-\varphi_S$. Their expression reads\cite{Ishihara:2016vdc}:
\begin{equation}\label{def:asang}
\alpha=\Psi_R-\Psi_S+\varphi_{RS}.
\end{equation}
On the other hand, the inner angles $\tilde\epsilon_1$ and $\tilde\epsilon_2$ used in this work are related to the angles $\Psi_S$ and $\Psi_R$  by:
\begin{eqnarray}
\tilde\epsilon_1&=&\pi-\Psi_R,\\
\tilde\epsilon_2&=&\Psi_S. 
\end{eqnarray}
Then, taking into account the relations 
\eqref{eq:epsiiso} and \eqref{eq:epsitildeiso}, it 
is easy to see that the definition \eqref{def1} 
which was based in the sum of the inner angles of 
finite quadrilateral regions agrees with the 
expression as \eqref{def:asang} (which does not make mention to any region). 
\section{}\label{Bnueva}
\subsection{Relationship between different angular coordinates and the elongation angle}
In 
this appendix we make some general 
discussion about the expressions for the 
deflection angle, the orbit equation and 
the relation with the elongation angle 
$\theta_I$ when one uses different 
azimuth angular coordinates systems (with the restriction that all of them define 
the same rotational Killing vector and 
therefore they can only differ in a 
constant $\delta$). 

{
Consider the 
Fig.\eqref{grafico-polar}, which is 
basically the same figure as 
Fig.\eqref{grafico-no-flat} but with the 
difference than now we have introduced 
two different angular coordinates systems with respective polar axis $A_{\text{axis}}$ and $\delta_{\text{axis}}$.
\begin{figure}[H]
\centering
\includegraphics[clip,width=75mm]{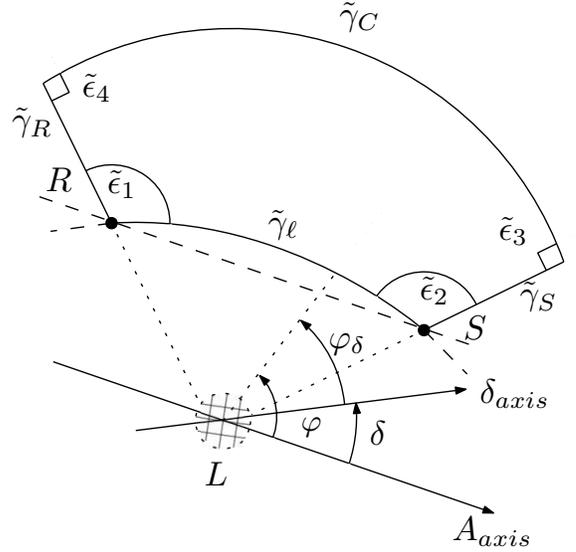}
\caption{Relation between two arbitrary azimuth angular coordinates in terms of the angular separation $\delta$ between the respective polar axis.}
\label{grafico-polar}
\end{figure}
These coordinates are named $\varphi$ and $\varphi_\delta$ respectively. As it can be seen from the Fig.\eqref{grafico-polar}, they are related by
\begin{equation}\label{eq:coordelta}
    \varphi=\varphi_\delta+\delta.
\end{equation}
Let us assume that the source is far 
 away from the lens and the observer. 
 Then, each of these azimuth angular 
 coordinates can be identified by the 
 angle $\varphi_S$ that the light ray 
 connecting $S$ with $R$ takes in this 
 limit. We will denote this angle by 
 $\Delta$, that is 
 $\lim_{S\to\infty}\varphi_{\delta\, S}=\Delta$, 
 where the limit is taken along the light curve connecting the source with the 
 receiver. For example, the azimuth 
 angular coordinate $\hat\varphi$ used by Ishihara \emph{et.al} and also by 
 Arakida is such that the closest 
 approach of the light ray to the lens is at $\hat\varphi_{\text{min}}=\pi/2$, or 
 equivalently, because of the symmetry  
 of the light ray trajectory around the  
 radial direction defined by $\hat\varphi_{\text{min}}$  we can codify the same information by taking
 $\Delta=-\alpha_\infty/2$. Another natural possibility was considered in the main 
 body text were we took a new angular 
 variable such that for a far away source $\lim_{S\to\infty}\varphi_S=0$.
 }
 
 {We wish to write 
 expressions for the deflection and the
 elongation angle which remain valid in 
 any of the azimuth angular coordinates 
 defined by different $\Delta$. One way 
 to do that in a very easy way is to 
 start by choosing a particular  
 $A_{\text{axis}}$ polar axis where the 
 corresponding angular coordinate 
 $\varphi$ is such that   
 $\lim_{S\to\infty}\varphi_S=0$. 
 Therefore, as we said before, any other $\delta_{\text{axis}}$ polar axis can be identified by the angle $\Delta$ (See 
 Fig.\eqref{grafico-polar2} which is 
 basically the same graphic as in 
 Fig.\eqref{grafico-polar} but  "rotated" in such a way that now the 
 $A_{\text{axis}}$ polar axis is plotted 
 as "horizontal").
 \begin{figure}[H]
\centering
\includegraphics[clip,width=75mm]{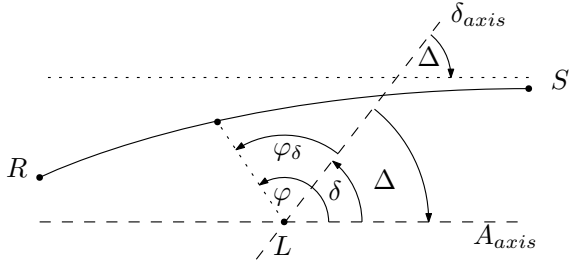}
\caption{A particular choice for the $\varphi$ coordinate such that for a far away source S, $\varphi_S\to 0$. Any other azimuth angular coordinate $\phi_\delta$ such that $\varphi_{\delta\,S}\to \Delta$ for a far away source is related to $\varphi$ by $\delta=-\Delta$. }
\label{grafico-polar2}
\end{figure}
Note also that, by 
 construction the $\delta$ shift between 
 these azimuth angular coordinates is 
 simply given by $\delta=-\Delta$, and as we know the expressions for the orbit 
 (in isotropic coordinates), for the 
 deflection angle and for the elongation 
 angle in the coordinates associated to 
 the $A_{\text{axis}}$ polar axis, we can directly write expressions for the same quantities in the other coordinates systems related by  \eqref{eq:coordelta}.
 }
 
 {Let us start with the elongation angle $\theta_I$. It is given in the $\varphi$ coordinates of Fig.\eqref{grafico-polar2} (see also Fig.\eqref{finitedistance}) by $\theta_I=\pi-\varphi_R+\delta\theta$, and therefore in terms of the azimuth coordinate $\varphi_\delta$ it will be given by
 \begin{equation}\label{eq:thetaIgeneraldelta}
     \theta_I=\pi-(\varphi_{\delta\,R}-\Delta)+\delta\theta.
 \end{equation}
 In particular, if the polar axis $\delta_{\text{axis}}$ is chosen such that $\Delta=-\alpha_\infty/2$ (the Ishihara \emph{et.al.} and Arakida choice) we have the following result:
 \begin{equation}
\begin{aligned}
    \Theta_I &=\pi-\hat\varphi_R-\overbrace{\frac{\alpha_\infty}{2}+\delta\theta}^{\mathcal{O}(m)}\\
    &=\pi-\hat\varphi_R+\mathcal{O}(m);
\end{aligned}
\end{equation}
a relation that was used by us in Sec.\eqref{sec:remark} in order to relate the expression for the deflection angle computed by these authors with the elongation angle.
In a similar way, we can proceed with the orbit equation and the deflection angle. 
In particular, for a Schwarzschild 
spacetime written in isotropic 
coordinates the orbit equation at linear 
order in the mass is given by the first 
two terms of  eq.\eqref{uint}, and 
therefore in any other choice for the 
azimuth coordinate we have that $u$ is 
given by (keeping all the other coordinates the same)
\begin{equation}\label{uint2bis}
u_\ell(\varphi_\delta)=\frac{\sin(\varphi_\delta-\Delta)}{b}+\frac{2m(1-\cos(\varphi_\delta-\Delta))}{b^2}+\mathcal{O}(m^2).
\end{equation}
Similarly, as we know the expression for 
the deflection angle in the coordinates 
associated to $A_\text{axis}$ 
(\eqref{eq:disfint1}), we can compute the deflection angle in any other associated angular coordinate system (for simplicity we continue writing our expressions for far away sources, namely $\varphi_S=0$ and $\varphi_{\delta S}=\Delta$): 
\begin{widetext}
\begin{equation}\label{eq:disfintDelta}
\begin{aligned}
\alpha=&\frac{2m}{b}\bigg[1-\cos(\varphi_
{\delta\,R}-\Delta)\bigg]+\frac{m^2}{8b^2}\bigg[
30(\varphi_{\delta\,R}-\Delta)+\sin(2(\varphi_{\delta\,R}-\Delta))-32(\sin(\varphi_{\delta\,R}-\Delta))\bigg]+\mathcal{O}(
m^3).
\end{aligned}
\end{equation}
\end{widetext}
This expression is valid for any $\Delta$, but in the case of the Ishihara \emph{et.al.} the $\hat\varphi$ coordinate was chosen such that $\Delta=-\alpha_\infty/2$, a quantity already $\mathcal{O}(m)$ and therefore the expression can be re-expanded as we explained in Section \eqref{subsec:relationscoordinates}.
}

{Of course, even when we obtained all these expressions in a very easy way from the known expressions in the $A_{\text{axis}}$ related coordinates, they can also be explicitly obtained from the orbit equation in isotropic coordinates and the definition of the deflection angle. Namely, we directly  solve \eqref{eq-orb-apA} with the asymptotic condition
\begin{equation}\label{eq:asympcondphidelta}
\lim_{\varphi_\delta\to \Delta}u(\varphi_\delta)=0. 
\end{equation}
To do that, we choose the ansatz
\begin{equation}
u(\varphi_\delta)=\frac{\sin(\varphi_\delta-\Delta)}{b}+\frac{2m u_1(\varphi_\delta)}{b}+\mathcal{O}(m^2),
\end{equation}
where the first term describe the unperturbed flat orbit in this coordinate system. 
From \eqref{eq-orb-apA} we obtain that at first order in $m$, $u_1(\varphi_\delta)$ must satisfy the following differential equation: 
\begin{equation}
b\cos(\phi_\delta-\Delta)\frac{du_1}{d\phi_\delta}+\sin(\phi_\delta-\Delta)(bu_1-1)=0.
\end{equation}
The general solution of this equation is given by
\begin{equation}
    u_1(\varphi_\delta)=\frac{1}{b}+C\cos(\varphi_\delta-\Delta);
\end{equation}
with $C$ an integration constant that is
fixed by requiring the asymptotic  condition \eqref{eq:asympcondphidelta} (resulting in $C=-1/b$). Therefore at linear order in $m$ the orbit equation reads 
\begin{equation}
u(\varphi_\delta)=\frac{\sin(\varphi_\delta-\Delta)}{b}+\frac{2m(1-\cos(\varphi_\delta-\Delta))}{b^2}+\mathcal{O}(m^2);
\end{equation}
which agrees with the expression obtained before.
Using this expression for the orbit and repeating the same procedure as in the eq.\eqref{eq:Kdscalciso} of Appendix \ref{Ap} we obtain the following expression for the deflection angle
\begin{widetext}
\begin{equation}
\begin{gathered}
\alpha=\frac{2m}{b}\bigg[\cos(\varphi_{\delta\,S}-\Delta)-\cos(\varphi_{\delta\,R}-\Delta)\bigg]+\frac{m^2}{8b^2}\bigg[30(\varphi_{\delta\,R}-\varphi_{\delta\,S})+\sin(2(\varphi_{\delta\,R}-\Delta))-\sin(2(\varphi_{\delta\,S}-\Delta))\\
+32(\sin(\varphi_{\delta\,S}-\Delta)-\sin(\varphi_{\delta\,R}-\Delta))\bigg]+\mathcal{O}(m^3);
\end{gathered}
\end{equation}
\end{widetext}
which for a far away source ($\phi_{\delta\,S}=\Delta$ agrees with the expression \eqref{eq:disfintDeltafinal}).
}

{ Finally, let us write the deflection angle in terms of $\theta_I$. From the expression \eqref{eq:disfintDelta} and \eqref{eq:thetaIgeneraldelta} it follows that at linear order in $m$, $\alpha$ is given as expected by the Shapiro-Ward formula \eqref{eq:shapward}(with $\gamma=1$). In order to compute the deflection angle from \eqref{eq:disfintDelta} at second order in $\theta_I$ we must take into account that 
\begin{equation}
\delta\theta=\alpha^{(1)}_{S_\infty}+\mathcal{O}(m^2)=\frac{2m}{b}\bigg[1+\cos(\theta_I)\bigg]+\mathcal{O}(m^2).
\end{equation}
Therefore by doing the replacement 
\begin{equation}
    \varphi_{\delta\,R}-\Delta=\pi-\theta_I+\delta\theta,
\end{equation}
in Eq.\eqref{eq:disfintDelta} and taking into account that
\begin{equation}
\begin{aligned}
    \cos(\varphi_{\delta\,R}-\Delta)=&\cos(\pi-\theta_I+\delta\theta)\\=&-\cos(\theta_I)-\frac{m}{b}\bigg[2\sin(\theta_I)+\sin(2\theta_I)\bigg]\\&+\mathcal{O}(m^2),
    \end{aligned}
\end{equation} 
and that the second term in \eqref{eq:disfintDelta} expressed in terms of $\theta_I$ reads
\begin{equation}
    \frac{m^2}{8b^2}\bigg[30(\pi-\theta_I)-\sin(2\theta_I)-32\sin(\theta_I)\bigg];
\end{equation}
we can write the deflection angle in terms of $\theta_I$ as
\begin{equation}\label{eq:disfintDeltafinal}
\begin{aligned}
\alpha=&\frac{2m}{b}\bigg[1+\cos(\theta_I)\bigg]+\frac{15m^2}{4b^2}\bigg[
\pi-\theta_I+\sin(\theta_i)\cos(\theta_I)\bigg]\\&
+\mathcal{O}(
m^3);
\end{aligned}
\end{equation}
which agrees with the more general relation given by the expressions \eqref{alpha1thetaI} and \eqref{alpha2thetaI} for the particular case of all the PPN parameters set to 1 and $J_2=0$. 
}

\section{}\label{Bp}
\subsection{Schwarzschild spacetime with plasma density profile of the form $N(r)=N_{oh}r^{-h}$. Particular cases}

In order to compute the main contribution to the deflection angle due to the presence of the plasma for the Sun, we need to calculate the explicit form of $\alpha_3$ for the particular cases $h=2$, $6$, $16$, and with respective numerical constants $N_{o2}$,  $N_{o6}$ and $N_{o16}$. In terms of standard trigonometric functions they read,
\begin{widetext}

\begin{equation}
\begin{aligned}
 h=2: \    \alpha_3=&\frac{K_{e}N_{o2}}
{2b^{2}\omega_{o}^{2}}\bigg[\varphi_{S}-
\varphi_{R}+\cos(\varphi_{R})\sin(\varphi_{R}) -
\cos(\varphi_{S})\sin(\varphi_{S})\bigg],\\
h=6: \ \alpha_3=&\frac{K_{e}N_{o6}}{64b^{6}\omega_{o}^{2}}\bigg[60(\varphi_{S}-\varphi_{R})+45\bigg(\sin(2\varphi_{R})-\sin(2\varphi_{S})\bigg)+9\bigg(\sin(4\varphi_{S})-\sin(4\varphi_{R})\bigg)+\sin(6\varphi_{R})-\sin(6\varphi_{S})\bigg], \\
h=16: \ \alpha_3=&\frac{K_e N_{o16}}{458752 b^{16}\omega_o^2}\bigg[ 720720(\varphi_S-\varphi_R)+640640\bigg(\sin(2\varphi_R)-\sin(2\varphi_S) \bigg)
        +224224\bigg( \sin(4\varphi_S)-\sin(4\varphi_R) \bigg) \\
        &+ 81536 \bigg( \sin(6\varphi_R)-\sin(6\varphi_S) \bigg)
        +25480 \bigg( \sin(8\varphi_S)-\sin(8\varphi_R) \bigg) + 6272\bigg( \sin(10\varphi_R)-\sin(10\varphi_S) \bigg)\\
        &+1120\bigg( \sin(12\varphi_S)-\sin(12\varphi_R) \bigg)+ 128\bigg( \sin(14\varphi_R)-\sin(14\varphi_S) \bigg)
        +7\bigg(\sin(16\varphi_S)-\sin(16\varphi_R) \bigg) \bigg].
\end{aligned}
\end{equation}

\end{widetext}


\begin{thebibliography}{100}
	
	\bibitem{Hoekstra:2013via}
	Henk Hoekstra, Matthias Bartelmann, Haakon Dahle, Holger Israel, Marceau
	Limousin, and Massimo Meneghetti.
	\newblock {Masses of galaxy clusters from gravitational lensing}.
	\newblock {\em Space Sci. Rev.}, 177:75--118, 2013.
	
	\bibitem{2015IAUS..311...86M}
	R.~{Mandelbaum}.
	\newblock {Galaxy Halo Masses from Weak Gravitational Lensing}.
	\newblock In M.~{Cappellari} and S.~{Courteau}, editors, {\em Galaxy Masses as
		Constraints of Formation Models}, volume 311 of {\em IAU Symposium}, pages
	86--95, 2015.
	
	\bibitem{Giocoli:2013tga}
	Carlo Giocoli, Massimo Meneghetti, R.~Benton Metcalf, Stefano Ettori, and Lauro
	Moscardini.
	\newblock {Mass and Concentration estimates from Weak and Strong Gravitational
		Lensing: a Systematic Study}.
	\newblock {\em Mon. Not. Roy. Astron. Soc.}, 440(2):1899--1915, 2014.
	
	\bibitem{Lewis20061}
	Antony Lewis;~Anthony Challinor.
	\newblock Weak gravitational lensing of the {CMB}.
	\newblock {\em Physics Reports}, 429, 2006.
	
	\bibitem{AR-EHT}
	Avery~E. Broderick, Vincent~L. Fish, Michael~D. Johnson, Katherine Rosenfeld,
	Carlos Wang, Sheperd~S. Doeleman, Kazunori Akiyama, Tim Johannsen, and
	Alan~L. Roy.
	\newblock {Modeling Seven Years of Event Horizon Telescope Observations with
		Radiatively Inefficient Accretion Flow Models}.
	\newblock {\em Astrophys. J.}, 820(2):137, 2016.
	
	\bibitem{AR-CLUSTER}
	A.~{Cava}, D.~{Schaerer}, J.~{Richard}, P.~G. {P{\'e}rez-Gonz{\'a}lez},
	M.~{Dessauges-Zavadsky}, L.~{Mayer}, and V.~{Tamburello}.
	\newblock {The nature of giant clumps in distant galaxies probed by the anatomy
		of the cosmic snake}.
	\newblock {\em Nature Astronomy}, 2:76--82, January 2018.
	
	\bibitem{Collett:2018gpf}
	Thomas~E. Collett, Lindsay~J. Oldham, Russell~J. Smith, Matthew~W. Auger,
	Kyle~B. Westfall, David Bacon, Robert~C. Nichol, Karen~L. Masters, Kazuya
	Koyama, and Remco van~den Bosch.
	\newblock {A precise extragalactic test of General Relativity}.
	\newblock 2018.
	
	\bibitem{AR-PSR1}
	Peter Goldreich and William~H. Julian.
	\newblock {Pulsar electrodynamics}.
	\newblock {\em Astrophys. J.}, 157:869, 1969.
	
	\bibitem{AR-PSR2}
	J.~Pétri.
	\newblock {Theory of pulsar magnetosphere and wind}.
	\newblock {\em J. Plasma Phys.}, 82(5):635820502, 2016.
	
	\bibitem{AR-HITOMI}
	Felix Aharonian et~al.
	\newblock {Temperature structure in the Perseus cluster core observed with
		Hitomi}.
	\newblock {\em Publ. Astron. Soc. Jap.}, 70(2):11, 2018.
	
	\bibitem{Solar-Radio}
	D.~O. Muhleman and I.~D. Johnston.
	\newblock Radio propagation in the solar gravitational field.
	\newblock {\em Phys. Rev. Lett.}, 17:455--458, Aug 1966.
	
	\bibitem{Muhleman-1970}
	D.~O. Muhleman, R.~D. Ekers, and E.~B. Fomalont.
	\newblock Radio interferometric test of the general relativistic light bending
	near the sun.
	\newblock {\em Phys. Rev. Lett.}, 24:1377--1380, Jun 1970.
	
	\bibitem{1977ApJ...211..943M}
	D.~O. {Muhleman}, P.~B. {Esposito}, and J.~D. {Anderson}.
	\newblock {The electron density profile of the outer corona and the
		interplanetary medium from Mariner-6 and Mariner-7 time-delay measurements}.
	\newblock {\em \apj}, 211:943--957, February 1977.
	
	\bibitem{Tyler-1977}
	G.~Leonard Tyler, Joseph~P. Brenkle, Thomas~A. Komarek, and Arthur~I.
	Zygielbaum.
	\newblock {The Viking Solar Corona Experiment}.
	\newblock {\em Journal of Geophysical Research}, 82:4335--4340, 1977.
	
	\bibitem{Turyshev2010}
	Slava~G. Turyshev and Viktor~T. Toth.
	\newblock The pioneer anomaly.
	\newblock {\em Living Reviews in Relativity}, 13(1):4, Sep 2010.
	
	\bibitem{Breuer-1980}
	J.~Breuer, R. A.;~Ehlers.
	\newblock Propagation of high-frequency electromagnetic waves through a
	magnetized plasma in curved space-time. i.
	\newblock {\em Proceedings Mathematical Physical \& Engineering Sciences}, 370,
	03 1980.
	
	\bibitem{Breuer-1981a}
	J.~Breuer, R. A.;~Ehlers.
	\newblock Propagation of high-frequency electromagnetic waves through a
	magnetized plasma in curved space-time. ii - application of the asymptotic
	approximation.
	\newblock {\em Proceedings of the Royal Society of London Series A}, 01 1981.
	
	\bibitem{Breuer-1981b}
	J.~Breuer, R. A.;~Ehlers.
	\newblock Propagation of electromagnetic waves through magnetized plasmas in
	arbitrary gravitational fields.
	\newblock {\em Astronomy \& Astrophysics}, 03 1981.
	
	\bibitem{Perlick-book}
	Volker Perlick.
	\newblock {\em Ray Optics, Fermat{'}s Principle, and Applications to General
		Relatively}.
	\newblock Lecture Notes in Physics 61. Springer-Verlag Berlin Heidelberg, first
	edition, 2000.
	
	\bibitem{1991CuSc...60..106S}
	G.~{Swarup}.
	\newblock {An International Telescope for Radio Astronomy}.
	\newblock {\em Current Science, Vol.~60, NO.2/JAN25, P.106, 1991}, 60:106,
	January 1991.
	
	\bibitem{2013AA...556A...2V}
	M.~P. {van Haarlem}, M.~W. {Wise}, A.~W. {Gunst}, and et. al.
	\newblock {LOFAR: The LOw-Frequency ARray}.
	\newblock {\em Astron. Astrophys}, 556:A2, August 2013.
	
	\bibitem{2009IEEEP..97.1497L}
	C.~J. {Lonsdale}, R.~J. {Cappallo}, {Morales}, and et. al.
	\newblock {The Murchison Widefield Array: Design Overview}.
	\newblock {\em IEEE Proceedings}, 97:1497--1506, August 2009.
	
	\bibitem{Budianu-2015}
	A.~Budianu, A.~Meijerink, M.J. Bentum, and D.~A.~M. Alessandro.
	\newblock {Swarm-to-Earth communication in {OLFAR}}.
	\newblock {\em Elsevier Science. Acta Astron{\'a}utica}, 107:14--19, 2015.
	
	\bibitem{Bentum:2016ekl}
	Mark~J. Bentum, Luca Bonetti, and Alessandro D. A.~M. Spallicci.
	\newblock {Dispersion by pulsars, magnetars, fast radio bursts and massive
		electromagnetism at very low radio frequencies}.
	\newblock {\em Adv. Space Res.}, 59:736--747, 2017.
	
	\bibitem{BisnovatyiKogan:2008yg}
	G.~S. Bisnovatyi-Kogan and O.~{\relax Yu}. Tsupko.
	\newblock {Gravitational radiospectrometer}.
	\newblock {\em Grav. Cosmol.}, 15:20--27, 2009.
	
	\bibitem{BisnovatyiKogan:2010ar}
	G.~S. Bisnovatyi-Kogan and O.~{\relax Yu}. Tsupko.
	\newblock {Gravitational lensing in a non-uniform plasma}.
	\newblock {\em Mon. Not. Roy. Astron. Soc.}, 404:1790--1800, 2010.
	
	\bibitem{Tsupko:2013cqa}
	Oleg~Yu Tsupko and Gennady~S. Bisnovatyi-Kogan.
	\newblock {Gravitational lensing in plasma: Relativistic images at homogeneous
		plasma}.
	\newblock {\em Phys. Rev.}, D87(12):124009, 2013.
	
	\bibitem{Tsupko:2014sca}
	O.~{\relax Yu}. Tsupko and G.~S. Bisnovatyi-Kogan.
	\newblock {Influence of Plasma on Relativistic Images of Gravitational
		Lensing}.
	\newblock {\em Nonlin. Phenom. Complex Syst.}, 17(4):455--457, 2014.
	
	\bibitem{Tsupko:2014lta}
	O.~{\relax Yu}. Tsupko and G.~S. Bisnovatyi-Kogan.
	\newblock {Gravitational lensing in the presence of plasmas and strong
		gravitational fields}.
	\newblock {\em Grav. Cosmol.}, 20(3):220--225, 2014.
	
	\bibitem{Perlick:2015vta}
	Volker Perlick, Oleg~{\relax Yu}. Tsupko, and Gennady~S. Bisnovatyi-Kogan.
	\newblock {Influence of a plasma on the shadow of a spherically symmetric black
		hole}.
	\newblock {\em Phys. Rev.}, D92(10):104031, 2015.
	
	\bibitem{Bisnovatyi-Kogan:2015dxa}
	G.~S. Bisnovatyi-Kogan and O.~{\relax Yu}. Tsupko.
	\newblock {Gravitational Lensing in Plasmic Medium}.
	\newblock 2015.
	\newblock [Plasma Phys. Rep.41,562(2015)].
	
	\bibitem{Perlick:2017fio}
	Volker Perlick and Oleg~{\relax Yu}. Tsupko.
	\newblock {Light propagation in a plasma on Kerr spacetime: Separation of the
		Hamilton-Jacobi equation and calculation of the shadow}.
	\newblock {\em Phys. Rev.}, D95(10):104003, 2017.
	
	\bibitem{Bisnovatyi-Kogan:2017kii}
	Gennady Bisnovatyi-Kogan and Oleg Tsupko.
	\newblock {Gravitational Lensing in Presence of Plasma: Strong Lens Systems,
		Black Hole Lensing and Shadow}.
	\newblock {\em Universe}, 3(3):57, 2017.
	
	\bibitem{2013ApSS.346..513M}
	V.~S. {Morozova}, B.~J. {Ahmedov}, and A.~A. {Tursunov}.
	\newblock {Gravitational lensing by a rotating massive object in a plasma}.
	\newblock {\em Astrophysics and Space Science}, 346:513--520, 2013.
	
	\bibitem{2016ApSS.361..226A}
	A.~{Abdujabbarov}, B.~{Juraev}, B.~{Ahmedov}, and Z.~{Stuchl{\'{\i}}k}.
	\newblock {Shadow of rotating wormhole in plasma environment}.
	\newblock {\em Astrophysics and Space Science}, 361:226, 2016.
	
	\bibitem{2017IJMPD..2650051A}
	A.~{Abdujabbarov}, B.~{Toshmatov}, Z.~{Stuchl{\'{\i}}k}, and B.~{Ahmedov}.
	\newblock {Shadow of the rotating black hole with quintessential energy in the
		presence of plasma}.
	\newblock {\em International Journal of Modern Physics D}, 26:1750051--239,
	2017.
	
	\bibitem{2017IJMPD..2641011A}
	A.~{Abdujabbarov}, B.~{Toshmatov}, J.~{Schee}, Z.~{Stuchl{\'{\i}}k}, and
	B.~{Ahmedov}.
	\newblock {Gravitational lensing by regular black holes surrounded by plasma}.
	\newblock {\em International Journal of Modern Physics D}, 26:1741011--187,
	2017.
	
	\bibitem{2017PhRvD..96h4017A}
	A.~{Abdujabbarov}, B.~{Ahmedov}, N.~{Dadhich}, and F.~{Atamurotov}.
	\newblock {Optical properties of a braneworld black hole: Gravitational lensing
		and retrolensing}.
	\newblock {\em Phys. Rev. D.}, 96(8):084017, 2017.
	
	\bibitem{2018arXiv180203293T}
	B.~{Turimov}, B.~{Ahmedov}, A.~{Abdujabbarov}, and C.~{Bambi}.
	\newblock {Gravitational lensing by magnetized compact object in the presence
		of plasma}.
	\newblock {\em ArXiv: 1802.03293}, 2018.
	
	\bibitem{Rogers:2015dla}
	Adam Rogers.
	\newblock {Frequency-dependent effects of gravitational lensing within plasma}.
	\newblock {\em Mon. Not. Roy. Astron. Soc.}, 451(1):17--25, 2015.
	
	\bibitem{Rogers:2016xcc}
	Adam Rogers.
	\newblock {Escape and Trapping of Low-Frequency Gravitationally Lensed Rays by
		Compact Objects within Plasma}.
	\newblock {\em Mon. Not. Roy. Astron. Soc.}, 465(2):2151--2159, 2017.
	
	\bibitem{Rogers:2017ofq}
	Adam Rogers.
	\newblock {Gravitational Lensing of Rays through the Levitating Atmospheres of
		Compact Objects}.
	\newblock {\em Universe}, 3:3, 2017.
	
	\bibitem{2018MNRAS.475..867E}
	X.~{Er} and A.~{Rogers}.
	\newblock {Two families of astrophysical diverging lens models}.
	\newblock {\em Mon.Not.Roy.Astron.Soc.}, 475:867--878, March 2018.
	
	\bibitem{Crisnejo:2018uyn}
	Gabriel Crisnejo and Emanuel Gallo.
	\newblock {Weak lensing in a plasma medium and gravitational deflection of
		massive particles using the Gauss-Bonnet theorem. A unified treatment}.
	\newblock {\em Phys. Rev.}, D97(12):124016, 2018.
	
	\bibitem{Er:2013efa}
	Xinzhong Er and Shude Mao.
	\newblock {Effects of plasma on gravitational lensing}.
	\newblock {\em Mon. Not. Roy. Astron. Soc.}, 437(3):2180--2186, 2014.
	
	\bibitem{PhysRevD.83.083007}
	Emanuel Gallo and Osvaldo~M. Moreschi.
	\newblock Gravitational lens optical scalars in terms of energy-momentum
	distributions.
	\newblock {\em Phys. Rev. D}, 83:083007, 2011.
	
	\bibitem{Boero:2016nrd}
	Ezequiel~F Boero and Osvaldo~M Moreschi.
	\newblock Gravitational lens optical scalars in terms of energy-momentum
	distributions in the cosmological framework.
	\newblock {\em Mon. Not. Roy. Astron. Soc.}, 475(4):4683--4703, 2018.
	
	\bibitem{PhysRevD.97.084010}
	Gabriel Crisnejo and Emanuel Gallo.
	\newblock Expressions for optical scalars and deflection angle at second order
	in terms of curvature scalars.
	\newblock {\em Phys. Rev. D}, 97:084010, 2018.
	
	\bibitem{Gibbons:2008rj}
	G.~W. Gibbons and M.~C. Werner.
	\newblock {Applications of the Gauss-Bonnet theorem to gravitational lensing}.
	\newblock {\em Class. Quant. Grav.}, 25:235009, 2008.
	
	\bibitem{Weyl-1917}
	Hermann Weyl.
	\newblock Zur gravitationstheorie.
	\newblock {\em Annalen der Physik}, 359, 1917.
	
	\bibitem{Jusufi:2015laa}
	Kimet Jusufi.
	\newblock {Gravitational lensing by Reissner-Nordstr{\"o}m black holes with
		topological defects}.
	\newblock {\em Astrophys. Space Sci.}, 361(1):24, 2016.
	
	\bibitem{Jusufi:2016wiz}
	Kimet Jusufi.
	\newblock {Light Deflection with Torsion Effects Caused by a Spinning Cosmic
		String}.
	\newblock {\em Eur. Phys. J.}, C76(6):332, 2016.
	
	\bibitem{Jusufi:2016sym}
	Kimet Jusufi.
	\newblock {Quantum effects on the deflection of light and the Gauss-Bonnet
		theorem}.
	\newblock {\em Int. J. Geom. Meth. Mod. Phys.}, 14(10):1750137, 2017.
	
	\bibitem{Jusufi:2017gyu}
	Kimet Jusufi.
	\newblock {Deflection angle of light by wormholes using the Gauss-Bonnet
		theorem}.
	\newblock {\em Int. J. Geom. Meth. Mod. Phys.}, 14(12):1750179, 2017.
	
	\bibitem{Jusufi:2017hed}
	Kimet Jusufi, Izzet Sakalli, and Ali {\"O}vg{\"u}n.
	\newblock {Effect of Lorentz Symmetry Breaking on the Deflection of Light in a
		Cosmic String Spacetime}.
	\newblock {\em Phys. Rev.}, D96(2):024040, 2017.
	
	\bibitem{Jusufi:2017vta}
	Kimet Jusufi, Ali {\"O}vg{\"u}n, and Ayan Banerjee.
	\newblock {Light deflection by charged wormholes in Einstein-Maxwell-dilaton
		theory}.
	\newblock {\em Phys. Rev.}, D96(8):084036, 2017.
	\newblock [Addendum: Phys. Rev.D96,no.8,089904(2017)].
	
	\bibitem{Jusufi:2017drg}
	Kimet Jusufi, Nayan Sarkar, Farook Rahaman, Ayan Banerjee, and Sudan Hansraj.
	\newblock {Deflection of light by black holes and massless wormholes in massive
		gravity}.
	\newblock ArXiv:1712.10175 (2017).
	
	\bibitem{Jusufi:2017xnr}
	Kimet Jusufi, Farook Rahaman, and Ayan Banerjee.
	\newblock {Semiclassical gravitational effects on the gravitational lensing in
		the spacetime of topological defects}.
	\newblock {\em Annals Phys.}, 389:219--233, 2018.
	
	\bibitem{Jusufi:2018kmk}
	Kimet Jusufi, Ali {\"O}vg{\"u}n, Ayan Banerjee, and ˙~Izzet Sakalli.
	\newblock {Gravitational lensing by wormholes supported by electromagnetic,
		scalar, and quantum effects}.
	\newblock ArXiv:1802.07680 (2018).
	
	\bibitem{Jusufi:2018waj}
	Kimet Jusufi.
	\newblock {Conical Morris-Thorne Wormholes with a Global Monopole Charge}.
	\newblock Arxiv: 1803.02317 (2018).
	
	\bibitem{Sakalli:2018nug}
	Izzet Sakalli, Kimet Jusufi, and Ali {\"O}vg{\"u}n.
	\newblock {Analytical Solutions in a Cosmic String Born-Infeld-dilaton Black
		Hole Geometry: Quasinormal Modes and Hawking Radiation}.
	\newblock Arxiv: 1803.10583 (2018).
	
	\bibitem{Arakida:2017hrm}
	Hideyoshi Arakida.
	\newblock {Light Deflection and Gauss-Bonnet Theorem: Definition of Total
		Deflection Angle and its Applications}.
	\newblock {\em Gen. Rel. Grav.}, 50(5):48, 2018.
	
	\bibitem{Ishihara:2016vdc}
	Asahi Ishihara, Yusuke Suzuki, Toshiaki Ono, Takao Kitamura, and Hideki Asada.
	\newblock {Gravitational bending angle of light for finite distance and the
		Gauss-Bonnet theorem}.
	\newblock {\em Phys. Rev.}, D94(8):084015, 2016.
	
	\bibitem{Ishihara:2016sfv}
	Asahi Ishihara, Yusuke Suzuki, Toshiaki Ono, and Hideki Asada.
	\newblock {Finite-distance corrections to the gravitational bending angle of
		light in the strong deflection limit}.
	\newblock {\em Phys. Rev.}, D95(4):044017, 2017.
	
	\bibitem{Ovgun:2018xys}
	Ali {\"O}vg{\"u}n, Kimet Jusufi, and Izzet Sakalli.
	\newblock {Exact traversable wormhole solution in bumblebee gravity}.
	\newblock {\em arXiv:1804.09911}, 2018.
	
	\bibitem{Ovgun:2018ran}
	Ali {\"O}vg{\"u}n, Kimet Jusufi, and Izzet Sakalli.
	\newblock {Gravitational Lensing Under the Effect of Weyl and Bumblebee
		Gravities: Applications of Gauss-Bonnet Theorem}.
	\newblock {\em arXiv:1805.09431}, 2018.
	
	\bibitem{Jusufi:2018kry}
	Kimet Jusufi.
	\newblock {Gravitational Deflection of Massive Particles by Kerr Black Holes
		and Teo Wormholes Viewed as a Topological Effect}.
	\newblock {\em arXiv:1806.01256}, 2018.
	
	\bibitem{Ovgun:2018prw}
	Ali {\"O}vg{\"u}n, Galin Gyulchev, and Kimet Jusufi.
	\newblock {Weak Gravitational lensing by phantom black holes and phantom
		wormholes using the Gauss-Bonnet theorem}.
	\newblock {\em arXiv:1806.03719}, 2018.
	
	\bibitem{Ovgun:2018oxk}
	Ali {\"O}vg{\"u}n.
	\newblock {Maxwell's Fisheye Lensing Effect by Black holes and Gauss-Bonnet
		Theorem}.
	\newblock {\em arXiv:1806.05549}, 2018.
	
	\bibitem{Ovgun:2018fte}
	Ali {\"O}vg{\"u}n, Izzet Sakalli, and Joel Saavedra.
	\newblock {Weak gravitational lensing by Kerr-MOG Black Hole and Gauss-Bonnet
		theorem}.
	\newblock {\em arXiv:1806.06453}, 2018.
	
	\bibitem{Ono:2018ybw}
	Toshiaki Ono, Asahi Ishihara, and Hideki Asada.
	\newblock {Deflection angle of light for an observer and source at finite
		distance from a rotating wormhole}.
	\newblock {\em Phys. Rev.}, D98(4):044047, 2018.
	
	\bibitem{Werner:2012rc}
	M.~C. Werner.
	\newblock {Gravitational lensing in the Kerr-Randers optical geometry}.
	\newblock {\em Gen. Rel. Grav.}, 44:3047--3057, 2012.
	
	\bibitem{Jusufi:2017lsl}
	Kimet Jusufi, Marcus~C. Werner, Ayan Banerjee, and Ali {\"O}vg{\"u}n.
	\newblock {Light Deflection by a Rotating Global Monopole Spacetime}.
	\newblock {\em Phys. Rev.}, D95(10):104012, 2017.
	
	\bibitem{Jusufi:2017vew}
	Kimet Jusufi and Ali {\"O}vg{\"u}n.
	\newblock {Light Deflection by a Quantum Improved Kerr Black Hole Pierced by a
		Cosmic String}.
	\newblock Arxiv: 1707.02824 (2017).
	
	\bibitem{Jusufi:2017mav}
	Kimet Jusufi and Ali {\"O}vg{\"u}n.
	\newblock {Gravitational Lensing by Rotating Wormholes}.
	\newblock {\em Phys. Rev.}, D97(2):024042, 2018.
	
	\bibitem{Jusufi:2017uhh}
	Kimet Jusufi and Ali {\"O}vg{\"u}n.
	\newblock {Effect of the cosmological constant on the deflection angle by a
		rotating cosmic string}.
	\newblock {\em Phys. Rev.}, D97(6):064030, 2018.
	
	\bibitem{Jusufi:2018jof}
	Kimet Jusufi, Ali {\"O}vg{\"u}n, Joel Saavedra, Yerko Vásquez, and P. A.
	González.
	\newblock {Deflection of light by rotating regular black holes using the
		Gauss-Bonnet theorem}.
	\newblock {\em Phys. Rev.}, D97(12):124024, 2018.
	
	\bibitem{Ono:2017pie}
	Toshiaki Ono, Asahi Ishihara, and Hideki Asada.
	\newblock {Gravitomagnetic bending angle of light with finite-distance
		corrections in stationary axisymmetric spacetimes}.
	\newblock {\em Phys. Rev.}, D96(10):104037, 2017.
	
	\bibitem{1970ApJ...162..345W}
	W.~R. {Ward}.
	\newblock {General-Relativistic Light Deflection for the Complete Celestial
		Sphere}.
	\newblock {\em The Astrophysical Journal}, 162:345, October 1970.
	
	\bibitem{Shapiro67}
	Irwin Shapiro.
	\newblock {New Method for the Detection of Light Deflection by Solar Gravity}.
	\newblock {\em Science}, 18(3790):806--808, 1967.
	
	\bibitem{Richter:1982zz}
	Gary~W. Richter and Richard~A. Matzner.
	\newblock {Second-order contributions to gravitational deflection of light in
		the parametrized post-Newtonian formalism}.
	\newblock {\em Phys. Rev.D}, 26:1219--1224, 1982.
	
	\bibitem{DeLuca:2003zz}
	Sophie Pireaux.
	\newblock {\em {Light deflection experiments as a test of relativistic theories
			of gravitation}}.
	\newblock PhD thesis, Louvain U., 2002.
	
	\bibitem{Misner1973}
	C.~W. {Misner}, K.~S. {Thorne}, and J.~A. {Wheeler}.
	\newblock {\em Gravitation. San Francisco: W.H.~Freeman and Co.}
	\newblock 1973.
	
	\bibitem{Will:1993ns}
	C.~M. Will.
	\newblock {\em {Theory and experiment in gravitational physics. Cambridge, UK:
			Univ. Press}}.
	\newblock 1993.
	
	\bibitem{He:2016vxc}
	Guansheng He and Wenbin Lin.
	\newblock {Gravitational deflection of light and massive particles by a moving
		Kerr–Newman black hole}.
	\newblock {\em Class. Quant. Grav.}, 33(9):095007, 2016.
	\newblock [Addendum: Class. Quant. Grav.34,no.2,029401(2017)].
	
	\bibitem{deFelice:2006xm}
	Fernando de~Felice, Alberto Vecchiato, Maria~Teresa Crosta, Mario~G. Lattanzi,
	and Beatrice Bucciarelli.
	\newblock {A general relativistic model for the light propagation in the
		gravitational field of the Solar System: The dynamical case}.
	\newblock {\em Astrophys. J.}, 653:1552--1565, 2006.
	
	\bibitem{deFelice:2004nf}
	F.~de~Felice, M.~T. Crosta, Alberto Vecchiato, M.~G. Lattanzi, and
	B.~Bucciarelli.
	\newblock {A General relativistic model of light propagation in the
		gravitational field of the Solar System: The Static case}.
	\newblock {\em Astrophys. J.}, 607:580--595, 2004.
	
	\bibitem{PhysRevLett.75.1439}
	D.~E. Lebach, B.~E. Corey, I.~I. Shapiro, M.~I. Ratner, J.~C. Webber, A.~E.~E.
	Rogers, J.~L. Davis, and T.~A. Herring.
	\newblock Measurement of the solar gravitational deflection of radio waves
	using very-long-baseline interferometry.
	\newblock {\em Phys. Rev. Lett.}, 75:1439--1442, Aug 1995.
	
	\bibitem{Shapiro-2004}
	J.~L.; Lebach D. E.; Gregory J.~S. Shapiro, S. S.;~Davis.
	\newblock Measurement of the solar gravitational deflection of radio waves
	using geodetic very-long-baseline interferometry data, 1979-1999.
	\newblock {\em Physical Review Letters}, 92, 3 2004.
	
	\bibitem{Fomalont-1975}
	E.~B. {Fomalont} and R.~A. {Sramek}.
	\newblock {A confirmation of Einstein's general theory of relativity by
		measuring the bending of microwave radiation in the gravitational field of
		the sun}.
	\newblock {\em Astrophysical Journal}, 199:749--755, August 1975.
	
	\bibitem{Robertson91}
	D.~S. Robertson, W.~E. Carter, and W.~H. Dillinger.
	\newblock {New measurement of solar gravitational deflection of radio signals
		using VLBI}.
	\newblock {\em Nature}, 349:768--770, 1991.
	
	\bibitem{Lambert-2009}
	S.~B. Lambert and C.~Le~Poncin-Lafitte.
	\newblock {Determination of the relativistic parameter gamma using very long
		baseline interferometry}.
	\newblock {\em Astron. Astrophys.}, 499:331, 2009.
	
	\bibitem{2017arXiv170206647T}
	O.~{Titov}.
	\newblock {Testing of General Relativity with Geodetic VLBI}.
	\newblock {\em arXiv:1702.06647}, 2017.
	
	\bibitem{1997froeschle}
	M.~{Froeschle}, F.~{Mignard}, and F.~{Arenou}.
	\newblock {Determination of the PPN Parameter gamma with the HIPPARCOS Data}.
	\newblock In R.~M. {Bonnet}, E.~{H{\o}g}, P.~L. {Bernacca}, L.~{Emiliani},
	A.~{Blaauw}, C.~{Turon}, J.~{Kovalevsky}, L.~{Lindegren}, H.~{Hassan},
	M.~{Bouffard}, B.~{Strim}, D.~{Heger}, M.~A.~C. {Perryman}, and L.~{Woltjer},
	editors, {\em Hipparcos - Venice '97}, volume 402 of {\em ESA Special
		Publication}, pages 49--52, August 1997.
	
	\bibitem{2010hobbs}
	D.~{Hobbs}, B.~{Holl}, L.~{Lindegren}, F.~{Raison}, S.~{Klioner}, and
	A.~{Butkevich}.
	\newblock {Determining PPN {$\gamma$} with Gaia's astrometric core solution}.
	\newblock In S.~A. {Klioner}, P.~K. {Seidelmann}, and M.~H. {Soffel}, editors,
	{\em Relativity in Fundamental Astronomy: Dynamics, Reference Frames, and
		Data Analysis}, volume 261 of {\em IAU Symposium}, pages 315--319, January
	2010.
	
	\bibitem{deFelice:1998wk}
	Fernando de~Felice, M.~G. Lattanzi, A.~Vecchiato, and P.~L. Bernacca.
	\newblock {General relativistic satellite astrometry. I. A Nonperturbative
		approach}.
	\newblock {\em Astron. Astrophys.}, 332:1133--1141, 1998.
	
	\bibitem{deFelice:2001wr}
	Fernando de~Felice, B.~Bucciarelli, M.~Lattanzi, and A.~Vecchiato.
	\newblock {General relativistic satellite astronomy II. Modelling parallax and
		proper motion}.
	\newblock {\em Astron. Astrophys.}, 373:336--344, 2001.
	
	\bibitem{Crosta:2003zh}
	Maria~Teresa Crosta, Alberto Vecchiato, Fernando de~Felice, Mario~G. Lattanzi,
	and Beatrice Bucciarelli.
	\newblock {Some aspects of relativistic astrometry from within the Solar
		System}.
	\newblock {\em Celestial Mech.}, 87:209--218, 2003.
	
	\bibitem{kaplan115}
	George~H. Kaplan.
	\newblock High-precision algorithms for astrometry: A comparison of two
	approaches.
	\newblock {\em The Astronomical Journal}, 115(1):361, 1998.
	
	\bibitem{Turyshev:2008tm}
	Slava~G. Turyshev.
	\newblock {Relativistic gravitational deflection of light and its impact on the
		modeling accuracy for the Space Interferometry Mission}.
	\newblock {\em Astron. Lett.}, 35:215--234, 2009.
	
	\bibitem{book:75670}
	John~Lighton Synge.
	\newblock {\em Relativity: The general theory}.
	\newblock North Holland Publishing Co., 1960.
	
	\bibitem{refId0}
	{Titov, O.}, {Girdiuk, A.}, {Lambert, S. B.}, {Lovell, J.}, {McCallum, J.},
	{Shabala, S.}, {McCallum, L.}, {Mayer, D.}, {Schartner, M.}, {de Witt, A.},
	{Shu, F.}, {Melnikov, A.}, {Ivanov, D.}, {Mikhailov, A.}, {Yi, S.}, {Soja,
		B.}, {Xia, B.}, and {Jiang, T.}
	\newblock Testing general relativity with geodetic vlbi - what a single,
	specially designed experiment can teach us.
	\newblock {\em Astronomy and Astrophysics}, 618:A8, 2018.
	
	\bibitem{2010IAUS..261..306M}
	F.~{Mignard} and S.~A. {Klioner}.
	\newblock {Gaia: Relativistic modelling and testing}.
	\newblock In S.~A. {Klioner}, P.~K. {Seidelmann}, and M.~H. {Soffel}, editors,
	{\em Relativity in Fundamental Astronomy: Dynamics, Reference Frames, and
		Data Analysis}, volume 261 of {\em IAU Symposium}, pages 306--314, January
	2010.
	
	\bibitem{Ward-1970}
	W.~R. {Ward}.
	\newblock {General-Relativistic Light Deflection for the Complete Celestial
		Sphere}.
	\newblock {\em Astrophysical Journal}, 162:345, October 1970.
	
	\bibitem{Poisson-2014}
	Will~C.M. Poisson~E.
	\newblock {\em Gravity: Newtonian, Post-Newtonian, Relativistic}.
	\newblock CUP, 2014.
	
	\bibitem{ZHANG201639}
	Bo~Zhang, Jun-Jie Wei, He~Gao, and Xue-Feng Wu.
	\newblock Testing einstein's equivalence principle with multi-band very long
	baseline array measurements of agn core shifts.
	\newblock {\em Journal of High Energy Astrophysics}, 9-10:39 -- 45, 2016.
	
	\bibitem{gould93}
	A.~{Gould}.
	\newblock {Deflection of light by the earth}.
	\newblock {\em The Astrophysical Journal}, 414:L37--L40, September 1993.
	
	\bibitem{titovreduction}
	{Titov, O.} and {Girdiuk, A.}
	\newblock Applying the theory of general relativity to reducing geodetic vlbi
	data.
	\newblock {\em Astronomy and Astrophysics}, 574:A128, 2015.
	
	\bibitem{Crosta:2005ch}
	Maria~Teresa Crosta and F.~Mignard.
	\newblock {Micro-arcsecond light bending by jupiter}.
	\newblock {\em Class. Quant. Grav.}, 23:4853--4871, 2006.
	
	\bibitem{Azreg-Ainou:2017obt}
	Mustapha Azreg-Aïnou, Sebastian Bahamonde, and Mubasher Jamil.
	\newblock {Strong Gravitational Lensing by a Charged Kiselev Black Hole}.
	\newblock {\em Eur. Phys. J.}, C77(6):414, 2017.
	
	\bibitem{Will2014}
	Clifford~M. Will.
	\newblock The confrontation between general relativity and experiment.
	\newblock {\em Living Reviews in Relativity}, 17(1):4, 2014.
	
	\bibitem{Lebedev-2013}
	Dmitri Lebedev and Kayll Lake.
	\newblock {On the influence of the cosmological constant on trajectories of
		light and associated measurements in Schwarzschild de Sitter space}.
	\newblock 1308.4931 (2013).
	
	\bibitem{Lebedev-2016}
	Dmitri Lebedev and Kayll Lake.
	\newblock {Relativistic Aberration and the Cosmological Constant in
		Gravitational Lensing I: Introduction}.
	\newblock ArXiv:1609.05183 (2016).
	
	\bibitem{Epstein-1980}
	Irwin~I. Epstein, Reuben;~Shapiro.
	\newblock Post-post-{N}ewtonian deflection of light by the {S}un.
	\newblock {\em Physical Review D}, 22, 12 1980.
	
	\bibitem{AR_TUNTSOV16}
	Artem~V. Tuntsov, Mark~A. Walker, Leon V.~E. Koopmans, Keith~W. Bannister,
	Jamie Stevens, Simon Johnston, Cormac Reynolds, and Hayley~E. Bignall.
	\newblock Dynamic spectral mapping of interstellar plasma lenses.
	\newblock {\em The Astrophysical Journal}, 817(2):176, 2016.
	
	\bibitem{AR_ROMANI}
	R.~W. {Romani}, R.~D. {Blandford}, and J.~M. {Cordes}.
	\newblock {Radio caustics from localized interstellar medium plasma
		structures}.
	\newblock {\em \nat}, 328:324--326, July 1987.
	
	\bibitem{AR_FIEDLER}
	R.~L. {Fiedler}, B.~{Dennison}, K.~J. {Johnston}, and A.~{Hewish}.
	\newblock {Extreme scattering events caused by compact structures in the
		interstellar medium}.
	\newblock {\em \nat}, 326:675--678, April 1987.
	
	\bibitem{AR_CLEGGFEY}
	Andrew~W. Clegg, Alan~L. Fey, and T.~Joseph~W. Lazio.
	\newblock {The Gaussian plasma lens in astrophysics. Refraction}.
	\newblock {\em Astrophys. J.}, 496:253, 1998.
	
	\bibitem{AR_PEN1}
	R.~{Main}, I.-S. {Yang}, V.~{Chan}, D.~{Li}, F.~X. {Lin}, N.~{Mahajan}, U.-L.
	{Pen}, K.~{Vanderlinde}, and M.~H. {van Kerkwijk}.
	\newblock {Pulsar emission amplified and resolved by plasma lensing in an
		eclipsing binary}.
	\newblock {\em \nat}, 557:522--525, May 2018.
	
	\bibitem{AR_PEN2}
	Dana Simard and Ue-Li Pen.
	\newblock Predicting pulsar scintillation from refractive plasma sheets.
	\newblock {\em Mon. Not. R. Astron. Soc.}, 478(1):983--994, 2018.
	
	\bibitem{AR_CORDES}
	J.~M. Cordes, I.~Wasserman, J.~W.~T. Hessels, T.~J.~W. Lazio, S.~Chatterjee,
	and R.~S. Wharton.
	\newblock {Lensing of Fast Radio Bursts by Plasma Structures in Host Galaxies}.
	\newblock {\em Astrophys. J.}, 842(1):35, 2017.
	
	\bibitem{book:112863}
	John~Liu Murray~Spiegel, Seymour~Lipschutz.
	\newblock {\em Mathematical handbook of formulas and tables}.
	\newblock Schaum's Outline Series. McGraw-Hill, 3 edition, 2008.
	
	\bibitem{giampieri1996}
	G.~{Giampieri}.
	\newblock {Relativity Experiments in the Solar System}.
	\newblock In M.~{Carfora}, M.~{Cavagli{\`a}}, P.~{Fr{\'e}}, G.~{Pizzella},
	C.~{Reina}, and A.~{Treves}, editors, {\em General Relativity and
		Gravitational Physics}, page 181, 1996.
	
	\bibitem{Bertotti-1998}
	Bruno Bertotti;~Giacomo Giampieri.
	\newblock Solar coronal plasma in doppler measurements.
	\newblock {\em Solar Physics}, 178, 02 1998.
	
	\bibitem{Turyshev-2018-b}
	Slava~G. Turyshev and Viktor~T. Toth.
	\newblock {Scattering of light by plasma in the solar system}.
	\newblock ArXiv: 1805.00398v3 (2018).
	
	\bibitem{Turyshev-2018-bc}
	Slava~G. {Turyshev} and Viktor~T. {Toth}.
	\newblock {Diffraction of light by plasma in the solar system}.
	\newblock {\em Journal of Optics}, 21(4):045601, Apr 2019.
	
	\bibitem{Turyshev:2018gjj}
	Slava~G. Turyshev and Viktor~T. Toth.
	\newblock {Diffraction of light by the gravitational field of the Sun and the
		solar corona}.
	\newblock {\em Phys. Rev.}, D99(2):024044, 2019.
	
	\bibitem{Turyshev:2018dzp}
	Slava~G. Turyshev and Viktor~T. Toth.
	\newblock {Optical properties of the solar gravitational lens in the presence
		of the solar corona}.
	\newblock {\em Eur. Phys. J. Plus}, 134:63, 2019.
	
\end{thebibliography}
\end{document}